\documentclass[prl,onecolumn,superscriptaddress,nofootinbib]{revtex4}

\usepackage{graphicx}

\usepackage{enumerate}
\usepackage{epsfig,subfigure}
\usepackage{amsmath,amssymb,latexsym}
\usepackage{ascmac}
\usepackage{bm}
\usepackage{enumitem}

\def\U#1{{\rm #1}} 

\newtheorem{definition}{Definition}
\newtheorem{lemma}{Lemma}
\newtheorem{proposition}{Proposition}

\newcommand{\bra}[1]{\langle #1 |}
\newcommand{\ket}[1]{| #1 \rangle}
\newcommand{\expect}[1]{\left\langle #1 \right\rangle} 
\newcommand{\vir}{\U{vir}}

\newcommand{\nul}{\U{null}}

\def\Pr{\U{Pr}}
\def\tr{\U{tr}}
\def\({\left(}
\def\){\right)}

\begin{document}
\title
{Protocol-level description and self-contained security proof of decoy-state BB84 QKD protocol}
\author{Akihiro Mizutani}
\affiliation{Faculty of Engineering, University of Toyama, Gofuku 3190, Toyama 930-8555, Japan}
\author{Toshihiko Sasaki}
\affiliation{Quantinuum K.K. Otemachi Financial City Grand Cube 3F, Global Business Hub Tokyo 
1-9-2 Otemachi, Chiyoda-ku, Tokyo 100-0004 Japan}
\author{Go Kato}
\affiliation{Quantum ICT Laboratory, NICT, Koganei, Tokyo 184-8795, Japan}

\begin{abstract}
  {
  In this paper, we present a flowchart-based description of the decoy-state BB84 quantum key distribution (QKD) protocol and provide a step-by-step, self-contained information-theoretic security proof for this protocol within the universal composable security framework. As a result, our proof yields a key rate consistent with previous findings. Importantly, unlike all the prior security proofs, our approach offers a fully rigorous and mathematical justification for achieving the key rate with the claimed correctness and secrecy parameters, thereby representing a significant step toward the formal certification of QKD systems.
 } 
 \end{abstract}
\maketitle

\renewcommand\thefootnote{}
\footnotetext{
\textcopyright{} 2025 Quantum Key Distribution Technology Promotion Committee, Quantum Forum. All rights reserved.}
\renewcommand\thefootnote{\arabic{footnote}}

\tableofcontents

\section{Introduction}
Quantum key distribution (QKD) enables information-theoretically secure communication between distant parties, Alice and Bob, against an eavesdropper (Eve) with unlimited computational power~\cite{bb84,kiyoNphoto,feihureview,Pirandola2020}. QKD has already reached the stage of practical implementation~\cite{imple1,imple2,imple3,imple4,imple5,imple6}, 
and commercial QKD products are beginning to appear. 
However, for QKD to be widely deployed in society, it is necessary to establish certification system by a certification organization that can evaluate the security of implemented QKD protocols. 
In order to obtain such certification for a QKD protocol, at least the following two points need to be clearly addressed. 
\begin{enumerate}
\item
\label{req:r1}
An explicit description of the implemented QKD protocol
\item
\label{req:r2}
A self-contained security proof of the implemented QKD protocol
\end{enumerate}
Currently, the most practically important and de facto standard QKD protocol is 
the decoy-state BB84 protocol~\cite{decoy1,decoy2,decoy2005HoiKwong}. 
A typical experimental setup of this protocol, such as the one considered in this paper, is that 
Alice sends intended quantum states using a fully phase-randomized coherent light source, and Bob employs threshold detectors characterized by detection efficiency and dark count probability. 
So far, numerous security proofs have been given for this protocol~\cite{concise2014,nakayamahayashi2014,rusca2018,Tupkary2024,Wiesemann2024,kamin2024,Tupkary2025}, 
and the security proof with the typical experimental setup is considered to be well established. 
However, various theoretical techniques used in security proofs have been scattered across different papers, each adopting slightly different assumptions and protocols. 
As a result, it has remained unclear whether a self-contained security proof can be provided for the decoy-state BB84 protocol by combining these techniques. 
In particular, the classical post-processing steps performed by Alice and Bob—such as when they announce basis choices, bit values, or measurement outcomes, as well as their choices of bit error correction, error verification, and privacy amplification methods—have multiple options. 
Therefore, providing a self-contained security proof with specifying classical post-processing methods 
would be valuable for formal certification of QKD systems.

In this paper, we provide a flowchart-based description of the decoy-state BB84 protocol, explicitly specifying not only the quantum communication procedures but also classical post-processing procedures performed by Alice and Bob. 
To enable rigorous verification of the security proof, we minimized verbal explanations and instead presented a self-contained, step-by-step security proof based on mathematics. Specifically, every operation performed at each step of the protocol is expressed using completely positive trace-preserving (CPTP) maps, thereby providing an unambiguous mathematical representation of the protocol. 
This approach allows us to uniquely characterize the quantum state $\hat{\rho}_{ABE}$ among the secret keys held by Alice and Bob, and Eve's system. 
For the security criterion, we adopt the universal composable security framework~\cite{ben2005,renner2009,renner2022}, 
in which the trace distance between $\hat{\rho}_{ABE}$ and its ideal state can be upper-bounded by a small constant. 
This constant can be decomposed into the correctness and secrecy parameters~\cite{koashi2009njp,hoiknowng2003}, and 
we show through explicit step by step calculations, these two parameters can be derived from the actual protocol. 
We note that a closely related work aiming to provide a rigorous security proof is the one~\cite{tomamichel2017} based on the entropic uncertainty principle~\cite{EUP1} 
with an ideal single-photon source. In contrast, our proof assumes a practical coherent light source, and it is fully self-contained, offering a complete understanding of the security proof.

The key rate derived from our proof corresponds to the extraction of the secret key from detection events originating from Alice's single-photon emissions, and is in close agreement with previously known key rates~\cite{concise2014,nakayamahayashi2014,rusca2018,Tupkary2024,Wiesemann2024,kamin2024,Tupkary2025}. 
However, we stress that no prior work has rigorously answered—purely based on mathematical derivations—why the key rate 
with the claimed correctness and secrecy parameters can be obtained. 
Therefore, our result provides a complete and rigorous answer to the security of the de facto standard QKD protocol and could satisfy the requirement described above as Point~\ref{req:r2}. 
We believe that this result constitutes an important step toward future certification of QKD systems.

This paper is organized as follows. In Secs.~\ref{sec:yougoexplain}-\ref{sec:symbol}, we summarize the technical terms related to our decoy-state BB84 protocol, the assumptions on the protocol, and the symbols used throughout this paper. 
In Sec.~\ref{sec:flowchart}, we describe the procedures executed by Alice and Bob in our QKD protocol in a uniquely determined manner using a flowchart format. The overall flowchart consists of four sub-flowcharts: Flowchart 1 illustrates the exchange of information that must be shared between Alice and Bob before the start of the protocol; Flowchart 2 covers transmission and measurement of light pulses; Flowchart 3 describes the public disclosure of classical information during quantum communication; and Flowchart 4 details the operations conducted by Alice and Bob on classical computers to generate a secret key. In Sec.~\ref{sec:explicitPAform}, we present a formula of the amount of privacy amplification that determines how 
much reconciled keys should be shortened. 

In Sec.~\ref{chap:security}, we provide the security proof of the protocol described in the flowchart in Sec.~\ref{sec:flowchart}. 
Specifically, in Sec.~\ref{sec:defsymbols}, we summarize the symbols used throughout the proof. 
In Sec.~\ref{sec:mathematicalQKDprotocol}, we mathematically describe the operations performed by Alice and Bob in each step of the flowchart in Sec.~\ref{sec:flowchart}, as well as Eve's coherent attacks. 
In Sec.~\ref{sec:secdef}, we define the security criterion of our protocol and show that once the correctness parameter 
($\epsilon_c$)  and the secrecy parameter ($\epsilon_s$) are determined, the overall security parameter of the protocol is 
given by $\epsilon_c+\epsilon_s$. 
We also show that the correctness parameter can be directly obtained from the actual protocol. 
The secrecy parameter is rigorously derived in Sec.~\ref{sec:dervationsecrecy} based on the phase error correction approach~\cite{koashi2005,koashi2009njp}. 
In the derivation, we first introduce a virtual protocol in Sec.~\ref{subsec:virtualprotocol}, 
and estimate the number of phase errors in Alice's virtual qubits through a stochastic trial defined on the virtual protocol in 
Secs.~\ref{subsec:equivalence_actual_virtual} and \ref{subsec:mainprops}. 
This leads to the derivation of the amount of privacy amplification given in Sec.~\ref{sec:explicitPAform}, 
which ensures $\epsilon_s$-secrecy.

\section{QKD Protocol}
\label{chap:QKDprotocol}

\subsection{Technical terms}
\label{sec:yougoexplain}
We summarize in Table~\ref{table:yougo} the technical terms that appear in the following sections.

\begin{table}[ht]
\begin{center}
\begin{tabular}{|p{4cm}|p{12cm}|}
\hline 
Technical term & Explanation \\\hline\hline
Classical channel
& A communication channel that does not use quantum effects. It includes, e.g., Internet and e-mail. The information exchanged over this channel is perfectly available to an eavesdropper.
\\\hline
Quantum channel & A communication channel for transmitting quantum states.
\\\hline
Syndrome of linear code & 
The information used to correct bit errors in the sifted key. In this paper, the error correction code for bit-error correction is assumed to be any linear code that always output information indicating positions of bit errors. 
Note that the positions indicated are not always correct, and to confirm if bit errors are corrected, 
Alice and Bob use error verification (see the next item).
\\\hline
Hash function to check success or failure of error correction & 
A hash function used to check if the keys shared by Alice and Bob are the same sequence after bit error correction. 
This procedure is called error verification. The QKD protocol in this paper assumes that the 
surjective universal2 hash function is used for this. 
\\\hline
Surjective universal2 hash function & 
A universal2 hash function is a function probabilistically chosen from a family of functions designed to ensure that the probability of different inputs producing the same hash values is low. If the output of the universal2 hash function is $N_{\U{verify}}$ bits, the probability that any two different inputs result in the same hash values (i.e., the probability that a collision occurs) is at most $2^{-N_{\U{verify}}}$. A  surjective function with $N_{\U{verify}}$-bit output is a function 
such that at least one input exists for every $N_{\U{verify}}$-bit output.
\\\hline
Dual universal2 hash function~\cite{tsurumaru2010} & 
A linear, random, surjective hash function $f$, mapping $n$ input bits to $m$ output bits, is called dual universal2 if, for any non-zero input $y$, the probability that $y$ lies in the orthogonal space of the kernel of $f$ is upper bounded by $1/2^{n-m}$.
\\\hline
Privacy amplification&Privacy amplification is one of the key generation processes in the QKD protocol, performed by Alice and Bob using classical computers (as shown in Step~4 of Fig.~\ref{fig:flowKD2}). Specifically, it involves inputting a key that may have been partially leaked to eavesdroppers and shortening its length to generate a QKD key that is completely secure from eavesdroppers.
\\\hline
Hash functions for privacy amplification&The hash function used by Alice and Bob to obtain the QKD key. 
The QKD protocol in this paper assumes that the surjective dual universal2 hash function is used.
\\\hline
Mach-Zehnder interferometer&
The Mach-Zehnder interferometer is a device that uses optical interference to measure the phase difference of light. In this interferometer, light incident from the quantum channel is first split into two optical paths. After passing through these paths, 
the light is recombined, and the phase difference information of the light before recombination is obtained by detecting the 
light using photon detectors.
\\\hline
Dark count probability of a photon detector&
A dark count refers to the detection of a photon by a photon detector due to factors such as stray light, rather than light from the quantum channel. The probability of a dark count occurring during the detection of each optical pulse is called the dark count probability. This value (which takes between 0 and 1) can be any number for executing a QKD protocol, but a smaller value results in better protocol performance (allowing for a higher rate of key generation per unit of time). 
\\\hline
Detection efficiency of a photon detector& 
The detection efficiency of a photon detector refers to the fraction of incident photons that the detector successfully detects and converts into an electrical signal. Detection efficiency takes a value ranging from 0\% (no detection) to 100\% 
(perfect detection). While a QKD protocol can be executed with any value of detection efficiency, a higher detection 
efficiency results in better performance of the QKD protocol (i.e., a higher amount of QKD key generation per unit of time).
\\\hline
Block&
As explained in Fig.~\ref{fig:flowsift}, once Bob has finished receiving the $M$ double pulses, he announces the measurement outcomes of these pulses via a classical channel. Subsequently, Alice also announces information about the transmitted states of these pulses via a classical channel. These $M$ double pulses are called a block.
\\\hline
      \end{tabular}
\end{center}
\caption{Summary of technical terms in this paper.}
\label{table:yougo}
\end{table}

\subsection{ Assumptions for the protocol}
\label{sec:assumptions-of-actual-protocol}
We summarize in Table~\ref{table:assumption} the assumptions regarding the QKD protocol for which 
we provide the security proof in Sec.~\ref{chap:security}.

\begin{table}[ht]
\begin{center}
\begin{tabular}{|p{2.5cm}|p{2.5cm}|p{11cm}|}
\hline 
Name &Assumed entity &   Assumptions \\\hline\hline
Time ordering assumption & Time ordering & 
There is a method to guarantee time ordering. The method is available at certain points in the protocol.
\\\hline
Authenticated classical channel assumption&Classical channel&
Classical channels are authenticated and not disconnected. 
If we can assume the existence of a pre-shared QKD key, Wegman-Carter schemes~\cite{wegmancarter} using this key can achieve information-theoretic secure falsifiability. 
Alternatively, with the computationally secure authentication, we assume that falsifiability cannot be broken within the protocol execution time. 
\\\hline
Perfect state-preparation assumption&Equipment for the transmitter&
The state generated by the transmitter is two consecutive ideal single-mode coherent light pulses (double pulse), and the intensity of a particular, say the $i$th, double pulse sent from the transmitter is in a desired value dependent only on the $i$th intensity choice, which is $S$, $D$, or $V$ defined in Sec.~\ref{subsubsec:symbols}. 
The phase difference of the $i$th double pulse can be in a desired value dependent only on the $i$th bit choice and the basis choice.
\\\hline
Perfect phase randomization assumption &Equipment for the transmitter& 
The relative phase between different double pulses is perfectly randomized.
\\\hline
Ideal random number assumption&Equipment for the transmitter and the receiver&
Any random number generated is a true random number. 
\\\hline
Ideal phase modulation assumption (Receiver side) &Equipment for the receiver& 
The relative phase modulation between double pulses acting on the longer arm of the Mach-Zehnder interferometer, performed immediately before the photon detectors, is mode-independent, and ideally implemented.
\\\hline
Assumption of identical performance of photon detectors &
Equipment for the receiver& The performance of all photon detectors used by the receiver is identical.
\\\hline
Photon detector model &Equipment for the receiver& Photon detectors are modeled by the dark count probability and the detection efficiency. 
Photon detectors operate with these two parameters, independent of detection round. 
\\\hline
No side channel assumption &Eavesdroppers &
The eavesdropper can coherently modify and observe the quantum states of all the double-pulses transmitted by the transmitter, input arbitrary quantum states to the receiver instead of the original quantum states of the double-pulses sent from the transmitter, and arbitrarily eavesdrop on the contents of the classical channel. However, by any other means, it is impossible for eavesdroppers to obtain the internal information held by the transmitter or the receiver.
\\\hline
      \end{tabular}
\end{center}
\caption{Summary of assumptions we make on our QKD protocol}
\label{table:assumption}
\end{table}

\clearpage

\subsection{Symbols used in the protocol}
\label{sec:symbol}
Definitions of symbols that appear in the QKD protocol in Sec.~\ref{sec:flowchart} are summarized in the following six tables 
(Tables~\ref{table:talbe_specific_meaning}-\ref{table:talbe_MSN_data}). 
The last three tables 
(Tables~\ref{table:talbe_Alice_parameter}-\ref{table:talbe_MSN_data})
also show the information that can be obtained during the protocol and when that information is ready to be disclosed on a classical channel. The term ``disclosure" in this paper means that the information is transmitted from the transmitter to the receiver or from the receiver to the transmitter through a classical channel. Note that the information disclosed will be available also to an eavesdropper due to the use of a classical channel, but there is no security problem in it as this leaked information is properly taken into account in the security proof. 

\subsubsection{Symbols with specific meanings}
\label{subsubsec:symbols}
The following symbols have some specific meanings.

\begin{table}[ht]
\begin{center}
\begin{tabular}{|p{2cm}|p{11cm}|}
\hline 
Symbol & Definition \\\hline\hline
$S,D,V$ & They specify the intensity label of the double-pulse to be sent. 
(Specifically, $S, D$, and $V$ denote the signal, decoy, and vacuum (another decoy).) 
\\\hline
$Z,X$ & They identify the type of modulation of the double-pulse being sent.
(Specifically, $Z$ and $X$ denote the $Z$- and $X$-bases of the quantum state, respectively.) 
\\\hline
$\emptyset$ & 
It indicates that the receiver has not yet completed the measurement of the double-pulses to be received. 
The phrase ``the measurement outcome of the $i$th pulse is $\emptyset$" means that the current time is before the 
$i$th measurement time, which is defined prior to the start of the QKD protocol. 
\\\hline
Click event&It is the event where one or both detectors detect a photon(s) at the $i$th measurement time, which is defined before the start of the QKD protocol. 
Note that ``the $i$th double-pulse is detected" means that ``the $i$th measurement outcome is a click event".
\\\hline
No click event & It is the event where no detector detects a photon(s) at the $i$th measurement time, which is defined before the start of the QKD protocol.
\\\hline
      \end{tabular}
\end{center}
\caption{Summary of symbols with specific meanings}
\label{table:talbe_specific_meaning}
\end{table}

\subsubsection{Constants}
Constants that must be determined before the QKD protocol starts are summarized as follows. 
These constants need not to be secret from eavesdroppers. 
\begin{table}[ht]
\begin{center}
\begin{tabular}{|p{4cm}|p{9cm}|}
\hline 
Symbol & Definition \\\hline\hline
$N_{\U{block}}$ & Total number of blocks of the double-pulses sent by the transmitter.
\\\hline
$M$ & Total number of the double-pulses transmitted in one block.
\\\hline
$N$ & Total number of the double-pulses sent by the transmitter. ($N=MN_{\U{block}}$)
\\\hline
$p_{\omega}$ ($\omega\in\{S,D,V\}$)
& Probability of generating the double-pulses with the intensity specified by $\omega$.
\\\hline
$\mu_{\omega}$ ($\omega\in\{S,D,V\}$)
& The intensity of the double-pulses specified by $\omega$.
\\\hline
$p_{\alpha}$ ($\alpha\in\{Z,X\}$)
& Probability that the modulation type (basis) is $\alpha$.
\\\hline
$p_a=1/2$ ($a\in\{0,1\}$)
& Probability of the modulation bit value to be $a$.
\\\hline
$\theta_{a,\alpha}$ ($\alpha\in\{Z,X\},a\in\{0,1\}$)
& Relative phase between the individual pulses of the double-pulse for basis $\alpha$ and bit $a$. 
($\theta_{0,Z}=0$, $\theta_{1,Z}=\pi$, $\theta_{0,X}=\pi/2$, $\theta_{1,X}=3\pi/2$)
\\\hline
$p_{\beta}$ ($\beta\in\{Z,X\}$)
& Probability that the measurement basis is chosen to be $\beta$ (probability that the type of modulation to be applied to the double-pulse before the interference for photon detection is $\beta$).
\\\hline
$N_{\rm verify}$ & Bit length of the hash value to be used to verify that the reconciled keys are equal.
\\\hline
$e_{\rm bit}$ & 
Upper bound on the bit error rate (the fraction of bits with bit errors present in the sifted key) that is assumed and defined by Alice and Bob before the protocol starts.
An upper bound on the bit error rate is estimated by Alice and Bob before executing the protocol 
(the security is guaranteed for any choices of the values as described in Sec.~\ref{sec:explicitPAform}). 
\\\hline
$\epsilon_{\U{secrecy}}$
&The constant that appears in determining the amount of privacy amplification. 
This constant takes a value between 0 and 1, which quantifies the security of the QKD key. 
Note that a smaller value indicates higher security. 
\\\hline
      \end{tabular}
\end{center}
\caption{Summary of symbols regarding constants}
\label{table:talbe_constant}
\end{table}
\clearpage

\subsubsection{Parameters, etc}
Symbols representing parameters used to describe Alice's and Bob's procedures are defined as follows.

\begin{table}[ht]
\begin{center}
\begin{tabular}{|p{1cm}|p{15cm}|}
\hline 
Symbol & Definition \\\hline\hline
$i$ & Parameter specifying the $i$th double-pulse. $i\in\{1,2,...,N\}$
\\\hline
$j$ & Parameter specifying the $j$th block of the double-pulses.  $j\in\{1,2,...,N_{\U{block}}\}$
\\\hline
$S_j$ & Set of indices $i$ belonging to the $j$th block of the double-pulses. 
$S_j=\{(j-1)M+1,(j-1)M+2,...,jM\}$
\\\hline
      \end{tabular}
\end{center}
\caption{Summary of symbols regarding parameters, etc}
\label{table:talbe_parameter}
\end{table}

\subsubsection{Symbols for describing the double-pulses to be sent by the transmitter (Alice)}
Symbols for describing the quantum state of Alice's transmitted double-pulses are defined as follows. 
Here, definitions of $S^{\U{det}}_j$ and $S^{X,\U{det}}_j$ are given in Sec.~\ref{subsec:boblabel}.

\begin{table}[ht]
\begin{center}
\begin{tabular}{|p{1cm}|p{7cm}|p{8cm}|}
\hline 
Symbol & Definition & Timing of disclosure\\\hline\hline
$\omega_i$ & 
Random variable representing the intensity choice of the $i$th double-pulse. $\omega_i\in\{S,D,V\}$
&After the reception of all double-pulses belonging to the $j$th block and $y_i, \beta_i, b_i$ $(i\in S_j)$ 
transmitted by Bob are completed, $\omega_i$ $(i\in S_j^{\U{det}})$ is disclosed by Alice. 
(See the definition of $y_i, \beta_i, b_i$ in Sec.~\ref{subsubsec:symbolsBob}, 
and also see Note~1 in Sec.~\ref{subsec:INF}) 
\\\hline
$\alpha_i$ & 
Random variable representing the basis choice of the $i$th double-pulse. $\alpha_i\in\{Z,X\}$ 
&
After the reception of all double-pulses belonging to the $j$th block and $y_i, \beta_i, b_i$ $(i\in S_j)$ 
transmitted by Bob are completed, $\alpha_i$ $(i\in S_j^{\U{det}})$ is disclosed by Alice. 
(See Note~1 in Sec.~\ref{subsec:INF}) 
\\\hline
$a_i$ & Random variable representing the bit choice of the $i$th double-pulse. $a_i\in\{0,1\}$
& 
When $\alpha_i=Z$, $a_i$ is not disclosed. (See Note 2 in Sec.~\ref{subsec:INF}) 
\\
&&
When $\alpha_i=X$, after the reception of all double-pulses belonging to the $j$th block and $y_i, \beta_i, b_i$ 
$(i\in S_j)$ transmitted by Bob are completed, $a_i$ $(i\in S_j^{\U{det}})$ is disclosed by Alice.
\\\hline
      \end{tabular}
\end{center}
\caption{Summary of symbols regarding Alice's emitted states}
\label{table:talbe_Alice_parameter}
\end{table}

Only the receiver will know when a measurement is finished. Therefore, it must be verified in some way if the disclosure is indeed made after the measurement, i.e., the verification of the order of time. The protocols implemented utilize the 
time ordering assumption that there is a method to check the order of time.

\subsubsection{Symbols for describing measurement outcomes by the receiver (Bob)}
\label{subsubsec:symbolsBob}
Symbols for describing Bob's measurement outcomes are defined as follows.

\begin{table}[ht]
\begin{center}
\begin{tabular}{|p{1cm}|p{7.5cm}|p{7.5cm}|}
\hline 
Symbol & Definition & Timing of disclosure\\\hline\hline
$y_i$ & Random variable representing the measurement outcome for the $i$th 
double-pulse, in particular, whether it is a click event or not. 
$y_i\in\{\U{No~click,click}\}$& 
After the reception of all double-pulses belonging to the $j$th block is completed, 
$y_i$ $(i\in S_j)$  is disclosed by Bob. 
\\\hline
$\beta_i$ & 
Random variable representing the basis used in the measurement for the $i$th double-pulse. 
$\beta_i\in\{Z,X\}$
& After the reception of all double-pulses belonging to the $j$th block is completed, $\beta_i$ $(i\in S_j)$ 
is disclosed by Bob (See Notes~3 and 4 in Sec.~\ref{subsec:INF})
\\\hline
$b_i$ 
& 
Random variable representing the outcome of the measurement for the $i$th double-pulse. 
&
When $\beta_i=Z$, $b_i$ is not disclosed. (See Note 3 in Sec.~\ref{subsec:INF})
\\
& $b_i\in\{0,1,\U{No~click}\}$
& 
When $\beta_i=X$, after the reception of all double-pulses belonging to the $j$th block, $b_i$ $(i\in S_j)$ 
is disclosed by Bob. (See Note 4 in Sec.~\ref{subsec:INF}) 
\\\hline
      \end{tabular}
\end{center}
\caption{Summary of symbols regarding Bob's measurement outcomes}
\label{table:talbe_Bob_parameter}
\end{table}

\subsubsection{Sets of the double-pulse indices used for key generation}
\label{subsec:boblabel}
Symbols related to the sets representing in which time slots Bob detected a photon (Bob's detector clicked) 
are defined as follows.

\begin{table}[ht]
\begin{center}
\begin{tabular}{|p{2cm}|p{2.5cm}|p{11.5cm}|}
\hline 
Symbol & Definition & Timing of disclosure\\\hline\hline
$S_j^{Z,\U{det}}$ & $\{i\in S_j|\alpha_i=\beta_i=Z,y_i=\U{click}\}$
&
After the reception of all pulses belonging to the $j$th block and $y_i, \beta_i, b_i$ $(i\in S_j)$ transmitted by Bob are completed, $S_j^{Z,\U{det}}$ is disclosed by Alice. (See Note~1 in Sec.~\ref{subsec:INF})
\\\hline
$S_j^{X,\U{det}}$ & $\{i\in S_j|\alpha_i=\beta_i=X,y_i=\U{click}\}$&
After the reception of all pulses belonging to the $j$th block and $y_i, \beta_i, b_i$ $(i\in S_j)$ transmitted by Bob are completed, $S_j^{X,\U{det}}$ is disclosed by Alice. (See Note~1 in Sec.~\ref{subsec:INF})
\\\hline
$S_j^{\U{det}}$ & $S_j^{Z,\U{det}}\cup S_j^{X,\U{det}}$ &
After the reception of all pulses belonging to the $j$th block and $y_i, \beta_i, b_i$ $(i\in S_j)$ transmitted by Bob are completed, $S_j^{\U{det}}$ is disclosed by Alice. (See Note~1 in Sec.~\ref{subsec:INF})
\\\hline
\end{tabular}
\end{center}
\caption{Summary of symbols regarding the sets of the double-pulse indices}
\label{table:talbe_MSN_data}
\end{table}

\clearpage
\subsection{Flowcharts at each step of the QKD protocol}
\label{sec:flowchart}

\subsubsection{Overall Flowchart}
\begin{figure}[ht]
    \centering
    \includegraphics[width=10cm]{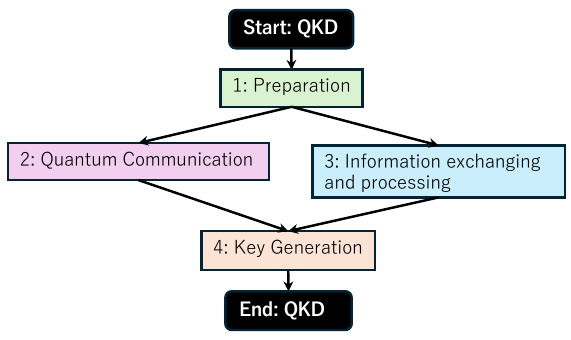}
    \caption{Overall flowchart of the QKD protocol.}
    \label{fig:zentai}
\end{figure}
Figure~\ref{fig:zentai} shows the overall flowchart of the QKD protocol. 
Immediately after the start of QKD, ``1: Preparation" is executed, and when ``1: Preparation" is completed, 
``2: Quantum communication" and ``3: Information exchanging and processing" are performed in parallel. 
When these two processes are completed, ``4: Key generation" is executed. After that, the QKD protocol ends. 
Each process in this flowchart will be described in a flowchart manner, which will be shown in the Secs. from \ref{sec:flowpreparation} to \ref{sec:keydistil}.

\clearpage
\subsubsection{Preparation Flowchart}
\label{sec:flowpreparation}
\begin{figure}[ht]
    \centering
    \includegraphics[width=9cm]{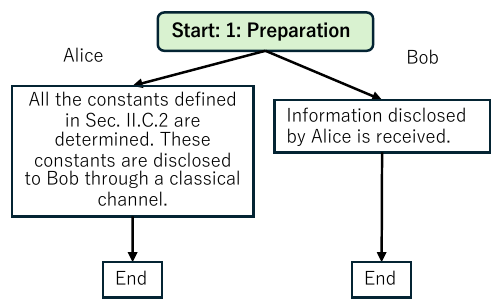}
    \caption{Flowchart for Preparation.}
    \label{fig:flow1}
\end{figure}

Figure~\ref{fig:flow1} shows the procedure for disclosing and exchanging information that the transmitter (Alice) and the receiver (Bob) follow before quantum communication starts. Here, the procedure ``End" for Alice (Bob) is defined as Alice 
(Bob) disclosing to Bob (Alice) that Alice (Bob) has finished the flowchart through a classical channel and that Alice (Bob) receives Bob's (Alice's) acceptance of the disclosed information. 
The same definition is used in Secs.~\ref{subsec:QCFLOW}, \ref{subsec:INF} and \ref{sec:keydistil}.

\label{sec:jizen}

\clearpage
\subsubsection{Quantum communication flowchart}
\label{subsec:QCFLOW}
\begin{figure}[ht]
    \centering
    \includegraphics[width=17.5cm]{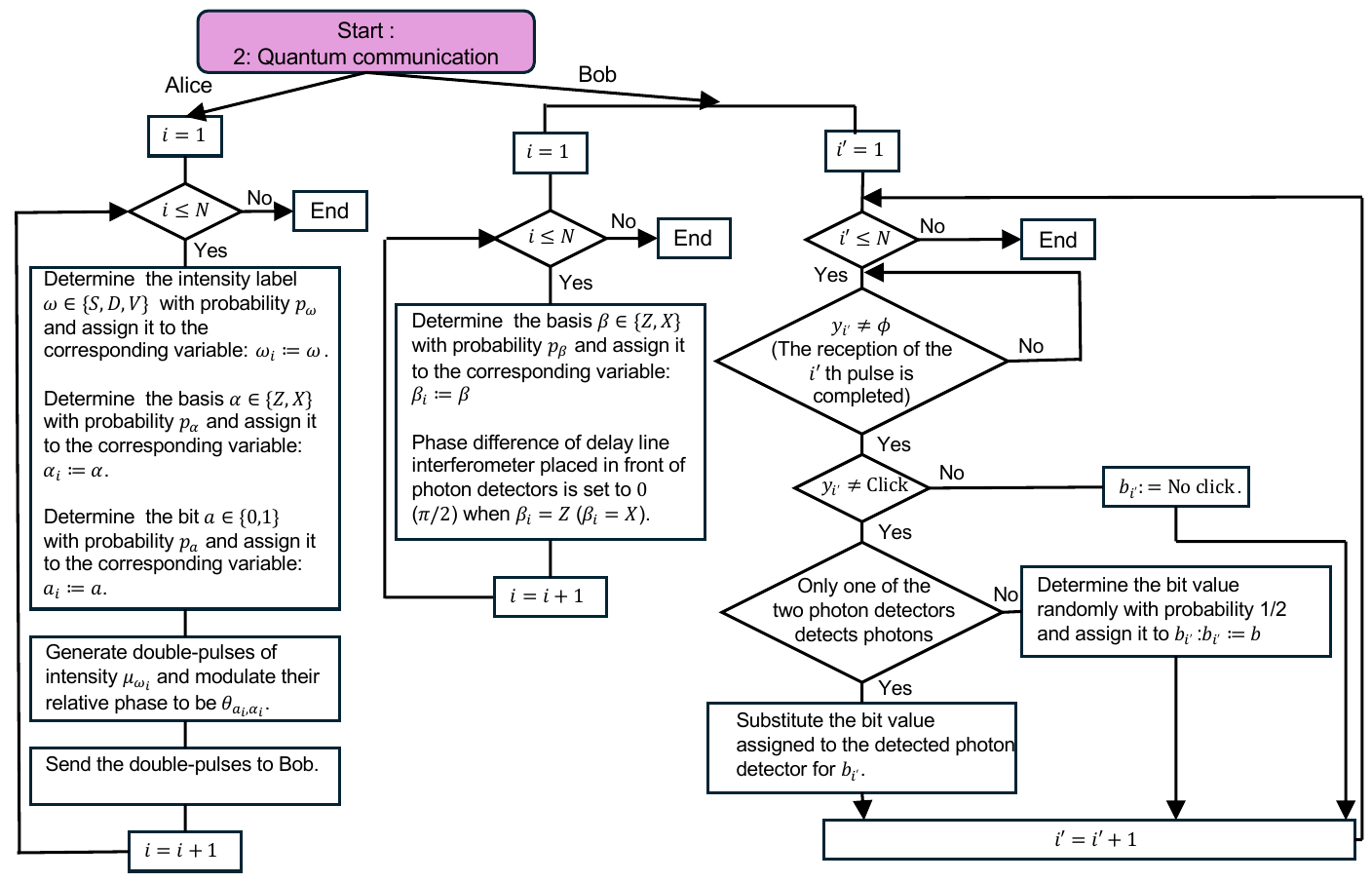}
    \caption{Flowchart for Quantum communication. 
    }
    \label{fig:flowQC}
\end{figure}

Figure~\ref{fig:flowQC} shows Alice's procedure for sending the double-pulses, 
    Bob's measurement, and the procedure for obtaining the measurement outcomes. 

\clearpage
\subsubsection{Information exchanging and processing flowchart}
\label{subsec:INF}
\begin{figure}[htbp]
    \centering
    \includegraphics[width=15cm]{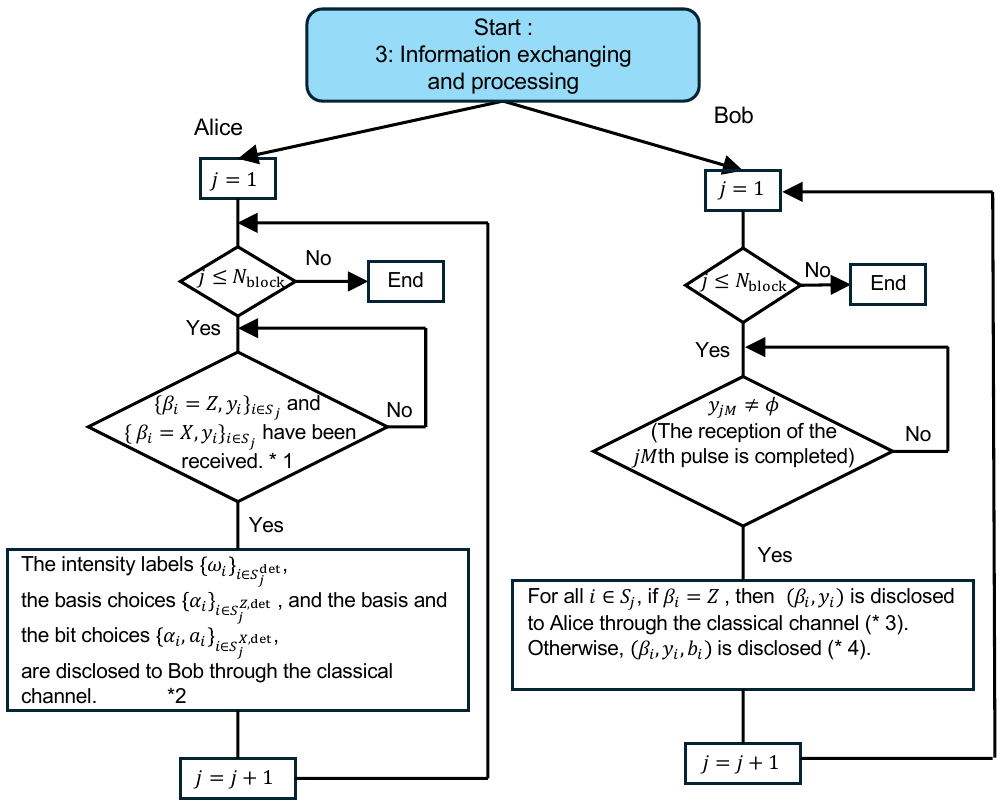}
    \caption{Flowchart for Information exchanging and processing.}
    \label{fig:flowsift}
\end{figure}

Figure \ref{fig:flowsift} shows how the transmitter and the receiver exchange information obtained through 
``2: Quantum communication" in Sec.~\ref{subsec:QCFLOW}.

{\bf \underline{Notes}}
\begin{enumerate}[label=${}^\ast$\arabic*]
    \item 
    Alice's disclosure on the values of the intensity, basis and bit selection of the transmitted pulses belonging to the $j$th 
    block must always be made after the measurement of the pulses belonging to the $j$th block is completed. 
    Here, from the time ordering assumption in Table~\ref{table:assumption}, it is guaranteed that after Alice receives the measurement outcome of the pulse belonging to the $j$th block from Bob, Bob will always have completed the measurement of the pulse belonging to the $j$th block. Thus, the flowchart in Fig.~\ref{fig:flowsift} guarantees that Alice will disclose the information after Bob's measurement with respect to the $j$th block is completed.
\item
If $\alpha_i=Z$, then $a_i$ must not be disclosed at any time. This means that not only $a_i$ itself must not be disclosed, but also any variables dependent on $a_i$ must not be disclosed. This includes, for instance, the output of a function whose input contains $a_i$.
\item
If $y_i=\U{No~click}$, $b_i$ and $\beta_i$ do not have to be disclosed. \\
If $\alpha_i=Z$ and $y_i=\U{click}$, $b_i$ must not be disclosed at any time. This means that not only $b_i$ itself must not be disclosed, but also any variables dependent on $b_i$ must not be disclosed. This includes, for instance, the output of a function whose input contains $b_i$.
\item
If $y_i=\U{No~click}$, $b_i$ and $\beta_i$ do not have to be disclosed. 
\end{enumerate}

\clearpage
\subsubsection{Key generation flowchart}
\label{sec:keydistil}
\begin{figure}[t]
    \centering
    \includegraphics[width=16.5cm]{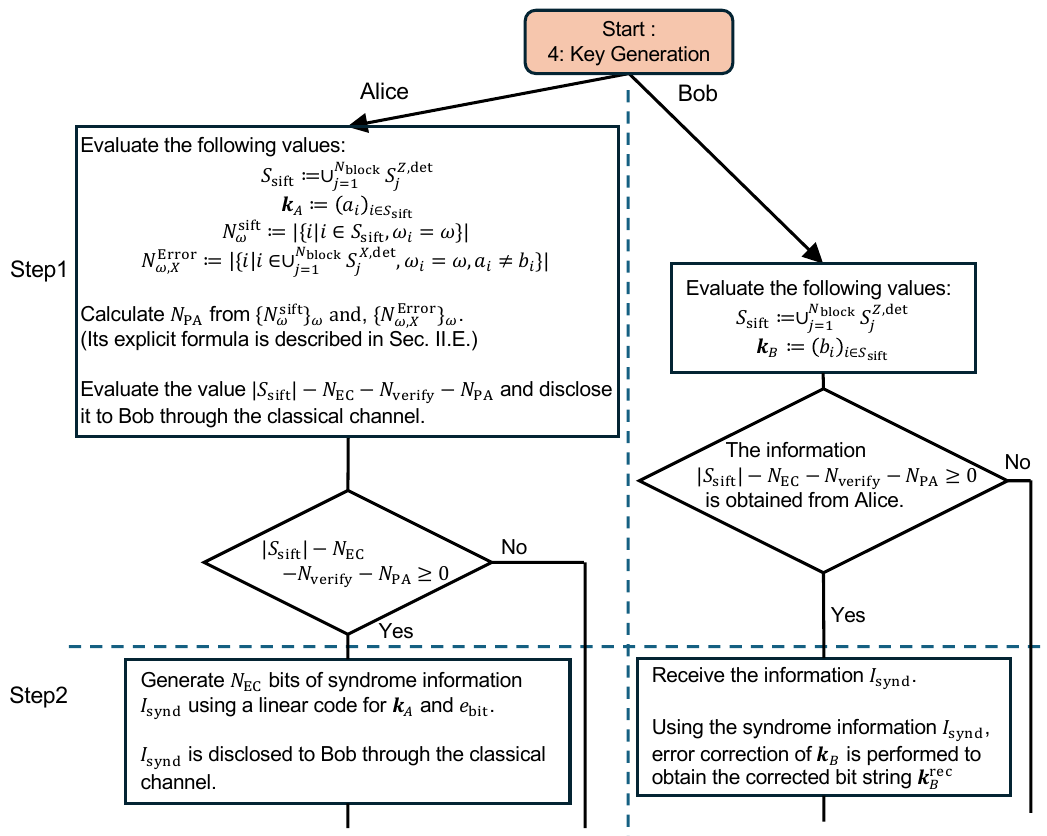}
    \caption{Key generation flowchart 1.}
    \label{fig:flowKD1}
\end{figure}

\begin{figure}[h]
    \centering
    \includegraphics[width=16.5cm]{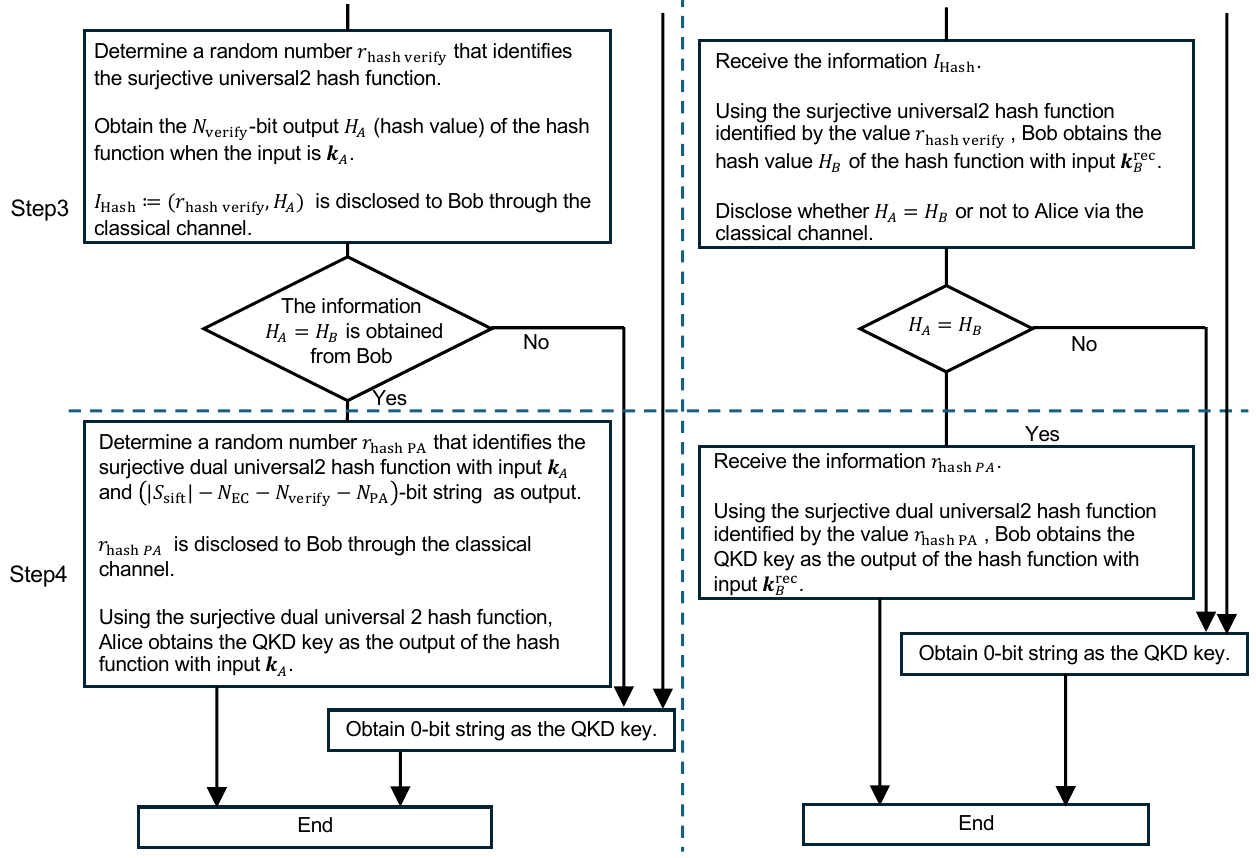}
    \caption{Key generation flowchart 2.}
    \label{fig:flowKD2}
\end{figure}
\clearpage
Figures~\ref{fig:flowKD1} and \ref{fig:flowKD2} show the QKD key generation procedure by the transmitter and the receiver 
(the key generation is a process that Alice and Bob perform on their classical computers to obtain the QKD key).

\subsection{Explicit expression for the amount of privacy amplification $N_{\text{PA}} + N_{\text{EC}} + N_{\text{verify}}$}
\label{sec:explicitPAform}

We define the following symbols: 
\begin{equation}
N_S^{\U{sift}}:=|\{i\in\{1,\cdots,N\}|\omega_i=S,\alpha_i=\beta_i=Z,b_i\in\{0,1\}\}|    
\end{equation}
\begin{equation}
N_D^{\U{sift}}:=|\{i\in\{1,\cdots,N\}|\omega_i=D,\alpha_i=\beta_i=Z,b_i\in\{0,1\}\}|    
\end{equation}
\begin{equation}
N_V^{\U{sift}}:=|\{i\in\{1,\cdots,N\}|\omega_i=V,\alpha_i=\beta_i=Z,b_i\in\{0,1\}\}|
\end{equation}
\begin{equation}
N_{D,X}^{\U{Error}}:=|\{i\in\{1,\cdots,N\}|\omega_i=D,\alpha_i=\beta_i=X,b_i\in\{0,1\},a_i\neq b_i\}|    
\end{equation}
\begin{equation}
N_{V,X}^{\U{Error}}:=|\{i\in\{1,\cdots,N\}|\omega_i=V,\alpha_i=\beta_i=X,b_i\in\{0,1\},a_i\neq b_i\}|.
\label{eq:NErrorVX}
\end{equation}
Here, $N_{\omega_i}^{\U{sift}}$ represents the random variable for the number of detected pulses in the $Z$ basis with the intensity label $\omega_i\in\{S,D,V\}$. Similarly, $N_{\omega_i,X}^{\U{Error}}$ represents the random variable for the number of bit errors in the $X$ basis with the intensity label $\omega_i\in\{D,V\}$.
\\\\
Alice and Bob estimate the expected values of $N_S^{\U{sift}},N_D^{\U{sift}},N_V^{\U{sift}},N_{D,X}^{\U{Error}},N_{V,X}^{\U{Error}}$ in the absence of eavesdroppers on the quantum channel used and denote them as $\tilde{N}_S^{\U{sift}},\tilde{N}_D^{\U{sift}},\tilde{N}_V^{\U{sift}},\tilde{N}_{D,X}^{\U{Error}},\tilde{N}_{V,X}^{\U{Error}}$. 
Note that these values must be determined before running the protocol. It may be determined by preliminary direct experiments or by the estimation based on the device and channel parameters. Our proof guarantees the security for any choice of the values, however, if these predetermined values deviate from the values actually obtained from Quantum Communication in Fig.~\ref{fig:flowQC}, 
the key rate will be reduced due to an increase of the abortion rate or an increase of the amount of privacy amplification.
\\\\
In Step~4 of Fig.~\ref{fig:flowKD2} (the final step of the QKD protocol), the length
\footnote{
This length is called the amount of privacy amplification.}
of the keys that Alice and Bob shorten is given by 
\begin{align}
 N_{\U{verify}}+N_{\U{EC}} + N_{\U{PA}}. 
\end{align}
The ratio of this value to the total number $N$ of double-pulses emitted by the transmitter is called privacy amplification rate. 
Here:
\begin{itemize}
    \item 
$N_{\U{verify}}$ is a constant defined in Table~\ref{table:talbe_constant}.
\item 
$N_{\U{EC}}$ is the bit length of the syndrome used for bit-error correction, which is determined in Step~2 of Fig.~\ref{fig:flowKD1}. 
\item 
$N_{\U{PA}}$ is given by 
\begin{equation}
 N_{\U{PA}} = N_{\U{sift}}-\lfloor \underline{N}_{1,Z} \rfloor+\lfloor \underline{N}_{1,Z} \rfloor h\left(\frac{
\lceil\overline{N}_{\U{ph}}\rceil
}{\lfloor \underline{N}_{1,Z} \rfloor}\right)+
\Bigg\lceil -\log_2\left(\frac{\epsilon_{\mathrm{secrecy}}^2}{4}\right) \Bigg\rceil.
\label{eq:NPAChapter2.5}
\end{equation}
\begin{itemize}
\item 
$N_{\U{sift}}$ is the length of the sifted key, which is determined in Step~1 of Fig.~\ref{fig:flowKD1}.
\item
$h(\bullet)$ is the binary entropy function. 
\item 
$\underline{N}_{1,Z}$ is the lower bound on the number $N_{1,Z}$ of events in which Alice emitted a single photon and Bob successfully detected the pulses, given that Alice and Bob used the $Z$ basis. The explicit expression of $\underline{N}_{1,Z}$ is given by Eq.~(\ref{eq:LowerN1zsection25}).
\item 
$\overline{N}_{\U{ph}}$ is the upper bound on the number $N_{\U{ph}}$ of events in which Alice emitted a single photon and Alice and Bob observed a bit error in the $X$ basis (this type of error is referred to as a phase error in Sec.~\ref{chap:security}), given that Alice and Bob used the $Z$ basis. The explicit expression of $\overline{N}_{\U{ph}}$ is given by Eq.~(\ref{eq:UpperN1zsection25}).
\item 
$\epsilon_{\U{secrecy}}$ is a constant value, which is defined in Table~\ref{table:talbe_constant}.
\end{itemize}
\end{itemize}
Then, the defined QKD protocol is guaranteed to be $(2^{-N_{\U{verify}}}+\epsilon_{\U{secrecy}})$-secure.
\\\\
Explicit expressions for $\underline{N}_{1,Z}$ and $\overline{N}_{\U{ph}}$, which appear in Eq.~(\ref{eq:NPAChapter2.5}), are provided in Secs.~\ref{subsec:N1Z} and \ref{subsec:M1Z}, respectively. 
When performing numerical computation of $N_{\U{PA}}$, numerical errors must be handled in such a way that they result 
in an increase in $N_{\U{PA}}$.
\\\\
First, the definitions appearing in these expressions are summarized below.
\begin{itemize}
    \item 
    Let
    \begin{align}
    p_{\mu_{\omega},n}^{\U{CS}}:=\frac{\mu_{\omega}^n}{n!}e^{-\mu_{\omega}}
    \label{eq:poisson2}
\end{align}
denote the probability that the two consecutive coherent light pulses (called a double pulse) sent by Alice contain $n\in\{0,1,2,...\}$ photons when she selects the intensity label $\omega\in\{S,D,V\}$. Recall that the intensity of the double-pulses specified by $\omega$ is $\mu_{\omega}$. Here, `CS' in Eq.~(\ref{eq:poisson2}) stands for a coherent signal.
\\
Let
\begin{align}
p_{\omega,n}^{\U{int}}:=p_{\omega}p_{\mu_\omega,n}^{\U{CS}}
\label{eq:def-p-omegajoint-n_2}
\end{align}
denote the joint probability that Alice selects the intensity label $\omega\in\{S,D,V\}$ and that the double pulse sent by Alice contains $n\in\{0,1,2,...\}$ photons.
\\
Let
\begin{equation}
    p^{\U{int}}_{\omega | n} := \frac{p_{\omega}p_{\mu_\omega,n}^{\U{CS}}}{\sum_{\omega\in\{S,D,V\}}p_{\omega}p_{\mu_\omega,n}^{\U{CS}}}
    \label{eq:def-p-omega-n_2_2}
\end{equation}
denote the conditional probability that the intensity label is $\omega\in\{S,D,V\}$, given that the double pulse sent by Alice contains $n\in\{0,1,2,...\}$ photons. 
\item 
When deriving $\underline{N}_{1,Z}$ and $\overline{N}_{\U{ph}}$, Kato's inequality~\cite{Katoinequality,tightnpj} is used. To represent the statistical deviation term in this inequality, the following functions are used. In these functions, positive real numbers are substituted into $s$ and $t$, while the probability is substituted into $\epsilon$.
\begin{align}
a_{\U{K}}(s,t,\epsilon):=&\frac{216\sqrt{s}t(s-t)\ln\epsilon-48s^{\frac{3}{2}}(\ln\epsilon)^2
+27\sqrt{2}(s-2t)\sqrt{-s^2(\ln\epsilon)[9t(s-t)-2s\ln\epsilon]}}
{4(9s-8\ln\epsilon)[9t(s-t)-2s\ln\epsilon]
}
\label{eq:katoafunction}
\end{align}
\begin{align}
b_{\U{K}}(s,t,\epsilon):=
\frac{\sqrt{18a_{\U{K}}(s,t,\epsilon)^2s-[16a_{\U{K}}(s,t,\epsilon)^2+24a_{\U{K}}(s,t,\epsilon)\sqrt{n}+9s]\ln\epsilon}}{3\sqrt{2s}}
\label{eq:katobfunction}
\end{align}
\begin{align}
a'_{\U{K}}(s,t,\epsilon):=
&\frac{-216\sqrt{s}t(s-t)\ln\epsilon+48s^{\frac{3}{2}}(\ln\epsilon)^2
+27\sqrt{2}(s-2t)\sqrt{-s^2(\ln\epsilon)[9t(s-t)-2s\ln\epsilon]}}
{4(9s-8\ln\epsilon)[9t(s-t)-2s\ln\epsilon]
}
\label{eq:katoafunction2}
\end{align}
\begin{align}
b'_{\U{K}}(s,t,\epsilon):=
\frac{\sqrt{18a_{\U{K}}(s,t,\epsilon)^2s-[16a_{\U{K}}(s,t,\epsilon)^2-24a_{\U{K}}(s,t,\epsilon)\sqrt{s}+9s]\ln\epsilon}}{3\sqrt{2s}}
\label{eq:katobfunction2}
\end{align}
\end{itemize}

\subsubsection{Explicit expression for $\underline{N}_{1,Z}$}
\label{subsec:N1Z}
The lower bound $\underline{N}_{1,Z}$ on $N_{1,Z}$ is given by
\begin{align}
&\underline{N}_{1,Z}\ge
\(1+a_{\U{K}}^{1,Z}\frac{2}{\sqrt{N}}\)^{-1}\notag\\
&\Bigg\{\lambda
\left[
N_S^{\U{sift}}(1+a_{\U{K}}^{S}\frac{2}{\sqrt{N}})+(b_{\U{K}}^{S}-a_{\U{K}}^{S})\sqrt{N}
\right]
+\gamma
\left[
N_V^{\U{sift}}(1+a_{\U{K}}^{V}\frac{2}{\sqrt{N}})+(b_{\U{K}}^{V}-a_{\U{K}}^{V})\sqrt{N}
\right]
\notag\\
&+\zeta
\left[
N_D^{\U{sift}}-
\left[b_{\U{K}}^{D}+a_{\U{K}}^{D}\left(\frac{2N_D^{\U{sift}}}{N}-1\right)
\right]\sqrt{N}
\right]
-(b_{\U{K}}^{1,Z}-a_{\U{K}}^{1,Z})\sqrt{N}
\Bigg\}
.
\label{eq:LowerN1zsection25}
\end{align}
The explanations for each symbol appearing in this equation are as follows.
\begin{itemize}
    \item 
$\lambda$, $\zeta$ and $\gamma$ are defined using Eqs.~(\ref{eq:def-p-omegajoint-n_2}) and (\ref{eq:def-p-omega-n_2_2}) as follows.
\begin{align}
 \lambda&:=-\left(
-\frac{\mu_D^2}{\mu_S^2}
\frac{p_{S|1}^{\U{int}}}{p_{S,0}^{\U{int}}}+\frac{p_{D|1}^{\U{int}}}{p_{D,0}^{\U{int}}}-\frac{p_{V|1}^{\U{int}}}{p_{V,0}^{\U{int}}}
\right)^{-1}\frac{1}{p_{S,0}^{\U{int}}}\frac{\mu_D^2}{\mu_S^2}
\le0
\\
\zeta&:=\left(-
\frac{\mu_D^2}{\mu_S^2}\frac{p_{S|1}^{\U{int}}}{p_{S,0}^{\U{int}}}+\frac{p_{D|1}^{\U{int}}}{p_{D,0}^{\U{int}}}-\frac{p_{V|1}^{\U{int}}}{p_{V,0}^{\U{int}}}
\right)^{-1}\frac{1}{p_{D,0}^{\U{int}}}\ge0
\\
\gamma&:=-\left(
-\frac{\mu_D^2}{\mu_S^2}\frac{p_{S|1}^{\U{int}}}{p_{S,0}^{\U{int}}}+\frac{p_{D|1}^{\U{int}}}{p_{D,0}^{\U{int}}}-\frac{p_{V|1}^{\U{int}}}{p_{V,0}^{\U{int}}}
\right)^{-1}\frac{1}{p_{V,0}^{\U{int}}}\le0
\end{align}
\item
$a_{\U{K}}^{1,Z}$ and $b_{\U{K}}^{1,Z}$ are defined using 
Eqs.~(\ref{eq:katoafunction}) and (\ref{eq:katobfunction}) as 
$a_{\U{K}}(N,\tilde{N}_{1,Z},\frac{\epsilon_{\U{secrecy}}^2}{32})$ 
and 
$b_{\U{K}}(N,\tilde{N}_{1,Z},\frac{\epsilon_{\U{secrecy}}^2}{32})$, 
respectively if $a_{\U{K}}(N,\tilde{N}_{1,Z},\frac{\epsilon_{\U{secrecy}}^2}{32})>-\frac{\sqrt{N}}{2}$. 
Similar to $\tilde{N}_{\omega}^{\U{sift}}$ introduced below Eq.~(\ref{eq:NErrorVX}), $\tilde{N}_{1,Z}$ represents the expected value of $N_{1,Z}$. 
Note that if $a_{\U{K}}(N,\tilde{N}_{1,Z},\frac{\epsilon_{\U{secrecy}}^2}{32})\le-\frac{\sqrt{N}}{2}$, $\underline{N}_{1,Z}=0$. 
\item
$a_{\U{K}}^{\omega}$ and $b_{\U{K}}^{\omega}$ for $\omega\in\{S,D,V\}$ are defined using 
Eqs.~(\ref{eq:katoafunction})-(\ref{eq:katobfunction2}) as follows.
\begin{align}
a_{\U{K}}^{\omega}:=a_{\U{K}}\(N,\tilde{N}_{\omega}^{\U{sift}},\frac{\epsilon_{\U{secrecy}}^2}{32}\),~~
b_{\U{K}}^{\omega}:=b_{\U{K}}\(N,\tilde{N}_{\omega}^{\U{sift}},\frac{\epsilon_{\U{secrecy}}^2}{32}\)
\end{align}
for $\omega\in\{S,V\}$, and 
\begin{align}
a_{\U{K}}^D:=a'_{\U{K}}\(N,\tilde{N}_{D}^{\U{sift}},\frac{\epsilon_{\U{secrecy}}^2}{32}\),~~
b_{\U{K}}^D:=b'_{\U{K}}\(N,\tilde{N}_D^{\U{sift}},\frac{\epsilon_{\U{secrecy}}^2}{32}\).
\end{align}
\end{itemize}

\subsubsection{Explicit expression for $\overline{N}_{\U{ph}}$
}
\label{subsec:M1Z}
The upper bound on $N_{\U{ph}}$ is given by
\begin{align}
&\overline{N}_{\U{ph}}
=
\left(1-\frac{2a_{\U{K}}^{\U{ph}}}{\sqrt{N}}\right)^{-1}
\Bigg\{
\frac{p^2_Z}{p^2_Xp_{D|1}^{\U{int}}}
\left[N_{D,X}^{\U{Error}}(1+a_{\U{K}}^{D,X}\frac{2}{\sqrt{N}})+(b_{\U{K}}^{D,X}-a_{\U{K}}^{D,X})\sqrt{N}
\right]
\notag\\
&-\frac{p^2_Zp_{D|0}^{\U{int}}}{p^2_Xp_{D|1}^{\U{int}}p_{V|0}^{\U{int}}}
\left[
N_{V,X}^{\U{Error}}-
\left[b_{\U{K}}^{V,X}+a_{\U{K}}^{V,X}\left(\frac{2N_{V,X}^{\U{Error}}}{N}-1\right)
\right]\sqrt{N}
\right]
+(b_{\U{K}}^{\U{ph}}-a_{\U{K}}^{\U{ph}})\sqrt{N}
\Bigg\}
.
\label{eq:UpperN1zsection25}
\end{align}
The explanations for each symbol appearing in this equation are as follows.
\begin{itemize}
        \item 
The probabilities $p_Z$ and $p_X$ are defined in Table~\ref{table:talbe_constant}.
    \item 
    The probabilities $p_{D|0}^{\U{int}}, p_{D|1}^{\U{int}}$ and $p_{V|0}^{\U{int}}$ are defined by Eq.~(\ref{eq:def-p-omega-n_2_2}). 
    \item 
$a_{\U{K}}^{\U{ph}}$ and $b_{\U{K}}^{\U{ph}}$ are defined using 
Eqs.~(\ref{eq:katoafunction2}) and (\ref{eq:katobfunction2}) as 
$a'_{\U{K}}(N,\tilde{N}_{\U{ph}},\frac{\epsilon_{\U{secrecy}}^2}{24})$
and 
$b'_{\U{K}}(N,\tilde{N}_{\U{ph}},\frac{\epsilon_{\U{secrecy}}^2}{24})$, respectively 
if $a'_{\U{K}}(N,\tilde{N}_{\U{ph}},\frac{\epsilon_{\U{secrecy}}^2}{24})<\frac{\sqrt{N}}{2}$. 
Similar to $\tilde{N}_{\omega}^{\U{sift}}$ introduced below Eq.~(\ref{eq:NErrorVX}), $\tilde{N}_{\U{ph}}$ represents the expected value of $N_{\U{ph}}$. Note that if $a'_{\U{K}}(N,\tilde{N}_{\U{ph}},\frac{\epsilon_{\U{secrecy}}^2}{24})\ge\frac{\sqrt{N}}{2}$, 
$\overline{N}_{\U{ph}}=N$.
\item  
$a_{\U{K}}^{\omega,X}$ and $b_{\U{K}}^{\omega,X}$ for $\omega\in\{D,V\}$ are defined using Eqs.~(\ref{eq:katoafunction})-(\ref{eq:katobfunction2}) as follows.
\begin{align}
a_{\U{K}}^{D,X}:=a_{\U{K}}\(N,\tilde{N}_{D,X}^{\U{Error}},\frac{\epsilon_{\U{secrecy}}^2}{24}\),~~
b_{\U{K}}^{D,X}:=b_{\U{K}}\(N,\tilde{N}_{D,X}^{\U{Error}},\frac{\epsilon_{\U{secrecy}}^2}{24}\),
\end{align}

\begin{align}
a_{\U{K}}^{V,X}:=a'_{\U{K}}\(N,\tilde{N}_{V,X}^{\U{Error}},\frac{\epsilon_{\U{secrecy}}^2}{24}\),~~
b_{\U{K}}^{V,X}:=b'_{\U{K}}\(N,\tilde{N}_{V,X}^{\U{Error}},\frac{\epsilon_{\U{secrecy}}^2}{24}\).
\end{align}
\end{itemize}

\section{Security proof}
\label{chap:security}
This section describes a rigorous security proof of the QKD protocol treated in the previous section. In the security proof, explanations by text are minimized. Instead, these are all included in mathematical formulation. Hence, it requires numerous symbols throughout the proof. These symbols are summarized in Sec.~\ref{sec:defsymbols}.

\subsection{Definitions of symbols}
\label{sec:defsymbols}

\begin{enumerate}
    \item 
Projective operator
\begin{align}
    \hat{P}[\ket{\cdot}]:=\ket{\cdot}\bra{\cdot}
\end{align}
\item
Pauli $Z$ operator 
\begin{equation}
 \hat{\sigma}_Z :=\hat{P}[\ket{0}]-\hat{P}[\ket{1}]
\end{equation}
\item
State
\begin{equation}
 \ket{+} := (\ket{0}+\ket{1})/\sqrt{2}
\end{equation}
\item
State
\begin{equation}
 \ket{-} := (\ket{0}-\ket{1})/\sqrt{2}
\end{equation}
\item
Pauli $X$ operator 
\begin{equation}
 \hat{\sigma}_X := \hat{P}[\ket{+}]-\hat{P}[\ket{-}]
\end{equation}
\item 
For an integer $n$ greater than or equal to 1, a set
\begin{equation}
    \left[n\right] := \{i\}_{i=1}^{n} 
\end{equation}
\item
For variables $X_1,...,X_i$,
\begin{align}
\vec{X}_{\le i}:=X_1X_2...X_i.
\end{align}
\item 
Schatten-1 norm for operator $\hat{A}$
\begin{align}
||\hat{A}||:=\U{tr}\sqrt{\hat{A}^\dagger \hat{A}}
\end{align}
\item 
The fidelity between two density operators (states) $\hat{\rho}$ and $\hat{\sigma}$ is defined as
\begin{align}
    F(\hat{\rho},\hat{\sigma})=\(\tr\sqrt{\sqrt{\hat{\rho}}\hat{\sigma}\sqrt{\hat{\rho}}}\)^2.
    \label{eq:fidelity}
\end{align}
\item 
Let
\begin{align}
\delta(x,y) =
\begin{cases}
1 & \text{if}~~x=y\\
0 & \U{otherwise}
\end{cases}
\end{align}
be the Kronecker delta function that outputs 1 if the variables are equal, and 0 otherwise.
\\
As a natural extension of $\delta(x,y)$, we define a function that outputs 1 if $x_i = y_i$ holds for all $i$ in two 
$n$-variables $x_1, \dots, x_n$ and $y_1, \dots, y_n$, and 0 otherwise:
\begin{align}
\delta(x_1,x_2,...,x_n ; y_1,y_2,...,y_n) =
\begin{cases}
1 & \text{if}~~x_i= y_i~\U{for~all~}i\\
0 & \U{otherwise}.
\end{cases}
\end{align}
\item
Let
\begin{align}
h(x) =
\begin{cases}
0 & x=0\\
-x\log_2x-(1-x)\log_2(1-x) & 0 < x\le1/2\\
1 & x>1/2
\end{cases}
\end{align}
be an increasing function of $x$ with $x\ge0$. $h(x)$ is identical to the binary entropy function for $0\le x\le1/2$. 
\item
\label{item:symbol23}

We assign the following symbols to refer to physical systems appearing in the quantum communication flowchart Fig.~\ref{fig:flowQC} and the information exchanging and processing flowchart Fig.~\ref{fig:flowsift}. 
These symbols are used in Secs.~\ref{sec:alicestate}-\ref{sec:zentai} to mathematically describe the procedures in these flowcharts.
\\\\
$A_i^{\U{sig}1}$: system of the first pulse in the double-pulse that Alice sends in the 
$i$th transmission

$A_i^{\U{sig}2}$: system of the second pulse in the double-pulse that Alice sends in the $i$th transmission

$A_i^{\U{sig}}$: composite system of systems $A_i^{\U{sig}1}$ and $A_i^{\U{sig}2}$

$A_{\rm sig}$: entire system of all the double-pulses emitted by Alice

$A_i^{\U{CR}}:$ system that stores the intensity, basis, and bit value information of the double-pulse that Alice sends in the $i$th transmission
\footnote{Note that ``CR" stands for ``classical register", which contains Alice's information about the emitted state.}

$A_{S_j}^{\U{CR}}$: system that stores the intensity, basis, and bit value information of the double-pulses that Alice sends in the $j$th block
\footnote{Note that $A_{S_j}^{\U{CR}}$ is a composite system of $A_i^{\U{CR}}$'s consisting only of the ones related to the $j$th block.}

${C_i^{A_B}}$: Alice's system that stores the information about Bob's $i$th measurement outcome

${C_{S_j}^{A_B}}$: Alice's system that stores the information about Bob's measurement outcomes regarding the $j$th block
\footnote{Note that ${C_{S_j}^{A_B}}$ is a composite system of ${C_i^{A_B}}$'s consisting only of the ones related to the 
$j$th block.}

$B_i^{\U{sig}}$: system of the $i$th optical pulse received by Bob

$B_{S_j}^{\U{sig}}$: system of the $j$th block received by Bob, which is a composite system of $B_{(j-1)M+1}^{\U{sig}},B_{(j-1)M+2}^{\U{sig}},...,B_{jM}^{\U{sig}}$. 

$B_i^{\U{CR}}$: Bob's system that stores the measurement basis and measurement outcome of the $i$th received optical pulse

$B_{S_j}^{\U{CR}}$: Bob's system that stores the measurement basis and measurement outcome of the optical pulses in the $j$th block
\footnote{Note that $B_{S_j}^{\U{CR}}$ is a composite system of $B_i^{\U{CR}}$'s consisting only of the ones related to the $j$th block.}

${C_i^{B_A}}$: Bob's system that stores the information about Alice's $i$th emitted state

${C_{S_j}^{B_A}}$: Bob's system that stores the information about Alice's 
emitted states for the $j$th block
\footnote{Note that ${C_{S_j}^{B_A}}$ is a composite system of ${C_i^{B_A}}$'s consisting only of the ones related to the 
$j$th block.}

$E$: Eve's system

$C^{E_A}_i$: Eve's system that stores the information about the $i$th emitted state that Alice has disclosed

$C^{E_A}_{S_j}$: Eve's system that stores the information about Alice's emitted states for the $j$th block

$C^{E_B}_i$: Eve's system that stores the information about the $i$th measurement outcome that Bob has disclosed

$C^{E_B}_{S_j}$: Eve's system that stores the information about the measurement outcomes of the $j$th block that Bob has disclosed

\item
We assign the following symbols to physical systems appearing in the key generation flowchart Figs.~\ref{fig:flowKD1} 
and \ref{fig:flowKD2}. 
These symbols are used in Sec.~\ref{subsec:kd} to mathematically describe the procedures in this flowchart.
\\\\
$A_{\U{CR}}$: composite system of $A_1^{\U{CR}},A_2^{\U{CR}},...,A_N^{\U{CR}}$

$C_{A_B}$: composite system of $C_1^{A_B},C_2^{A_B},...,C_N^{A_B}$

$B_{\U{CR}}$: composite system of $B_1^{\U{CR}},B_2^{\U{CR}},...,B_N^{\U{CR}}$

$C_{B_A}$: composite system of $C_1^{B_A},C_2^{B_A},...,C_N^{B_A}$

$A_{\U{sift}}$: system that stores Alice's sifted key

$B_{\U{sift}}$: system that stores Bob's sifted key

$C_{\U{PA}}$: System that stores the information of the random number that identifies the dual universal2 hash function~\cite{tsurumaru2010} in the key generation flowchart Figs.~\ref{fig:flowKD1} and \ref{fig:flowKD2}. 
This system is held by Alice, Bob and Eve.

$C_{\U{EC}}$: System that stores the public information exchanged between Alice and Bob for bit error correction. 
This system is held by Alice, Bob and Eve. 

$C^{\U{Hash}}_A$: System that stores the hash value of Alice's sifted key for verifying the success or failure of bit error correction. This system is held by Alice, Bob and Eve. 

$B_{\U{Hash}}$: Bob's system that stores the hash value of Bob's reconciled key (corrected bit string) for verifying the success or failure of bit error correction. This system is held by Alice, Bob and Eve. 

$C^{\U{Length}}_{\U{Key}}$: System that stores the information of the length of the secret key
\footnote{Note that this corresponds to the QKD key as described in previous sections.}. 
This system is held by Alice, Bob and Eve. 

$C^{\U{Length}}_{\U{Judge}}$: Immediately before the ``end'' in the key generation flowchart Fig.~\ref{fig:flowKD2}, there is a procedure called ``Obtain 0-bit string as the QKD key.". This system $C^{\U{Length}}_{\U{Judge}}$ holds the information on whether this procedure is executed or not. This system is held by Alice, Bob and Eve.

$C^{\U{HashResult}}_B$: System that stores information on whether the hash value of Alice's sifted key matches the hash value of Bob's reconciled key. This system is held by Alice, Bob and Eve. 

\item 
In defining a completely positive map $\mathcal{E}$, we use the following notations for simplicity.
Let $L(\mathcal{H}_A)$ denote a set of linear operators on the Hilbert space $\mathcal{H}_A$, which is another way of expressing system $A$.
Let a notation
\begin{equation}
    \mathcal{E}: A \to B
\end{equation}
denote a domain and a codomain of a map
\begin{equation}
    \mathcal{E}: L( \mathcal{H}_{A})
    \to L(\mathcal{H}_{B}).
\end{equation}
Let $\hat{\rho}$ be a positive operator in $L( \mathcal{H}_{A}\otimes\mathcal{H}_{C})$.
Let the notation
\begin{equation}
    \mathcal{E}(\hat{\rho})
\end{equation}
denote
\begin{equation}
    \mathcal{E}\otimes \mathbf{1}_{C}(\hat{\rho}),
\end{equation}
where $\mathbf{1}_{C}$ is the identity map on $L( \mathcal{H}_{C})$.
\end{enumerate}

\subsection{Mathematical description of the QKD protocol}
\label{sec:mathematicalQKDprotocol}
In this section, we mathematically describe the QKD protocol introduced in
Sec.~\ref{sec:flowchart}.

\subsubsection{Alice's transmission in the quantum communication flowchart}
\label{sec:alicestate}

From the perfect state-preparation and phase randomization assumptions in Sec.~\ref{sec:assumptions-of-actual-protocol}, 
the state of the double-pulse that Alice sends in the $i$th transmission is written as
\begin{align}
\hat{\rho}(\theta_{a_i,\alpha_i},\mu_{\omega_i})_{A_i^{\U{sig}}}
= \frac{1}{2\pi}\int_0^{2\pi}
    \hat{P}[\ket{e^{\U{i}(\delta+\theta_{a_i,\alpha_i})}\sqrt{\mu_{\omega_i}/2}}_{A_i^{\U{sig}1}}
    \ket{e^{\U{i}\delta}\sqrt{\mu_{\omega_i}/2}}_{A_i^{\U{sig}2}}]\U{d}\delta
    \label{eq:rhothetamuactual}
\end{align}
with
\begin{align}
\ket{e^{\U{i}\theta}\sqrt{\mu}}:=e^{-\mu/2}\sum_{k=0}^{\infty}
\frac{(e^{\U{i}\theta}\sqrt{\mu})^k}{\sqrt{k!}}\ket{k}.
\label{eq:coherent}
\end{align}
The former state is equivalent to the classical mixture of the photon number states in the double-pulse due to phase randomization, that is,
\begin{align}
\hat{\rho}(\theta_{a_i,\alpha_i},\mu_{\omega_i})_{A_i^{\U{sig}}}=
\sum_{n_i=0}^{\infty}
\hat{N}_{n_i}\hat{\rho}(\theta_{a_i,\alpha_i},\mu_{\omega_i})_{A_i^{\U{sig}}}\hat{N}_{n_i}
\label{eq:alicestate2}
\end{align}
with $\hat{N}_n$ being the projection to $n$-photon subspace as
\begin{align}
\hat{N}_n:=\sum_{k=0}^n \hat{P}[\ket{n-k}_{A_i^{\U{sig}1}}\ket{k}_{A_i^{\U{sig}2}}].
\label{eq:Nn}
\end{align}
Here, the ket vectors on the right-hand sides of Eqs.~(\ref{eq:coherent}) and (\ref{eq:Nn}) represent the photon number states of a single-mode light, as specified in the perfect state-preparation assumption in Sec.~\ref{sec:assumptions-of-actual-protocol}.

From the ideal random number assumption in Sec.~\ref{sec:assumptions-of-actual-protocol}, $\omega\in\{S,D,V\},\alpha\in\{Z,X\},$ and $a\in\{0,1\}$ are chosen with the prescribed probability distributions $p_\omega,p_\alpha$, and $p_a$, respectively. 
By taking the average over $\vec{\omega}:=\omega_1\omega_2...\omega_N,\vec{\alpha}:=\alpha_1\alpha_2...\alpha_N$, 
and $\vec{a}:=a_1a_2...a_N$, Alice's entire emitted state is described as 
\begin{align}
\hat{\rho}_{\U{in},A_{\U{CR}},A_{\U{sig}}}
&=\bigotimes_{i=1}^N\hat{\rho}_{\U{in},A_i^{\U{CR}},A_i^{\U{sig}}}
\notag\\
&=
\bigotimes_{i=1}^N\sum_{\omega_i\in\{S,D,V\}}
\sum_{\alpha_i\in\{Z,X\}}\sum_{a_i\in\{0,1\}}
p_{a_i}p_{\omega_i}p_{\alpha_i}
\hat{P}[\ket{\omega_i,\alpha_i,a_i}_{A_i^{\U{CR}}}]
\otimes
\sum_{n_i=0}^{\infty}
\hat{N}_{n_i}\hat{\rho}(\theta_{a_i,\alpha_i},\mu_{\omega_i})_{A_i^{\U{sig}}}\hat{N}_{n_i}.
\label{eq:alicestate}
\end{align}

\subsubsection{Eavesdropper's operation in the quantum communication flowchart}
\label{subsec:eveoperation}
From the no side channel assumption in Sec.~\ref{sec:assumptions-of-actual-protocol}, we can assume that Eve sends optical pulses to Bob without loss of generality.
As defined in Item~\ref{item:symbol23} in Sec.~\ref{sec:defsymbols}, system of the $i$th pulse ($j$th block) received by Bob is denoted by $B_i^{\U{sig}}$ ($B_{S_j}^{\U{sig}}$ with $j\in\{1,...,N_{\U{block}}\}$). 
From the no side channel assumption in Sec.~\ref{sec:assumptions-of-actual-protocol}, in constructing Eve's operation, we consider the most general scenario in which Bob receives pulses after Alice sends all the optical pulses (system $A_{\rm sig}$) in 
Eq.~(\ref{eq:alicestate}). We do not lose generality here because the order in which Alice's and Bob's operators are applied 
does not affect the result, with Alice's and Bob's composite system being a tensor product of  each system. Note that this scenario encompasses the actual situation where Eve has simultaneous access to only a subset of all the optical pulses. Consequently, our analysis never underestimates Eve's eavesdropping capability. Hence, Eve's quantum operation of outputting the state of Bob's system $B_{S_1}^{\U{sig}}$ (we denote this operation as $\mathcal{E}^E_{j=1}$)
can be written as the following CPTP (completely-positive trace preserving) map 
\begin{align}
    \mathcal{E}^E_{j=1}: A_{\U{sig}}E\to B_{S_1}^{\U{sig}} E
    \label{eq:eej1}
\end{align}
by use of the notation in Sec.~\ref{sec:defsymbols}.

Next, we consider Eve's operation of outputting the state of Bob's system $B_{S_j}^{\U{sig}}\; (j\in \{2,\cdots, N_{\U{block}}\})$. When Eve outputs this state, she can exploit the state of system $E$ and the information about the $(j-1)$-th block announced by Alice and Bob, which is specified by the information exchanging and processing flowchart Fig.~\ref{fig:flowsift}
\footnote{Note that when constructing Eve's operation to output the state of system $B_{S_j}^{\U{sig}}$, the information from up to the $(j-2)$-th block (which is stored in $C^{E_A}_1...C^{E_A}_{S_{j-2}}C^{E_B}_1...C^{E_B}_{S_{j-2}}$) is included in the input system $E$ of her operation. 
}.
\\
Using the definitions of
\begin{align}
C^{E_A}_{S_{j-1}}:=\bigotimes_{i=(j-2)M+1}^{(j-1)M}C^{E_A}_i,
\end{align}
and
\begin{align}
C^{E_B}_{S_{j-1}}:=
\bigotimes_{i=(j-2)M+1}^{(j-1)M}C^{E_B}_i
\label{eq:cebj}
\end{align}
given in Sec.~\ref{sec:defsymbols}, Eve's operation for outputting the state of system $B_{S_j}^{\U{sig}}$ with $j\in\{2,...,N_{\U{block}}\}$ is written as the following CPTP map
\begin{align}
\mathcal{E}^E_{j}: E C^{E_A}_{S_{j-1}} C^{E_B}_{S_{j-1}}
\to B_{S_j}^{\U{sig}} E.
    \label{eq:eej}
\end{align}

Finally, Eve can evolve her system using the information about the $N_{\U{block}}$th block (the final block) that Alice and Bob have disclosed. This CPTP map $\mathcal{E}^E_{N_{\U{block}}+1}$ is described as 
\begin{align}
\mathcal{E}^E_{N_{\U{block}}+1}: E C^{E_A}_{S_{N_{\U{block}}}} C^{E_B}_{S_{N_{\U{block}}}}
\to E.
    \label{eq:eejB}
\end{align}

We provide three remarks regarding the CPTP map $\mathcal{E}_j^E$ introduced above.

\begin{enumerate}
    \item 
Eq.~(\ref{eq:eej1}): 
When Eve sends the state of the first block (system $B_{S_1}^{\U{sig}}$) to Bob, 
Eve can utilize her system $E$ and $A_{\U{sig}}$ of Alice's emitted state. This is because, as shown in the flowchart Fig.~\ref{fig:flowsift}, no information is disclosed by Alice and Bob until Bob completes the measurement of the first block.
\item
Eq.~(\ref{eq:eej}): When Eve sends the state of the $j$th block (system $B_{S_j}^{\U{sig}}$) to Bob, Eve can not only utilize system $E$ but also the information about the blocks up to $(j-1)$ that Alice and Bob have disclosed. This is because, as shown in the information exchanging and processing flowchart Fig.~\ref{fig:flowsift}, 
once the $j$th measurement is completed, Bob and then Alice disclose the information about the $j$th block.
\item 
Eq.~(\ref{eq:eejB}): This equation describes the operation in which Eve evolves her system using the information disclosed by Bob and Alice about the $j$th block. Since Bob's measurement has already been completed, the output of $\mathcal{E}^E_{N_{\U{block}}+1}$ consists only of Eve's system.
\end{enumerate}

\subsubsection{Bob's measurement in the quantum communication flowchart}
\label{sec:bobmsn_infpro}
\begin{figure}
    \centering
    \includegraphics[width=8cm]{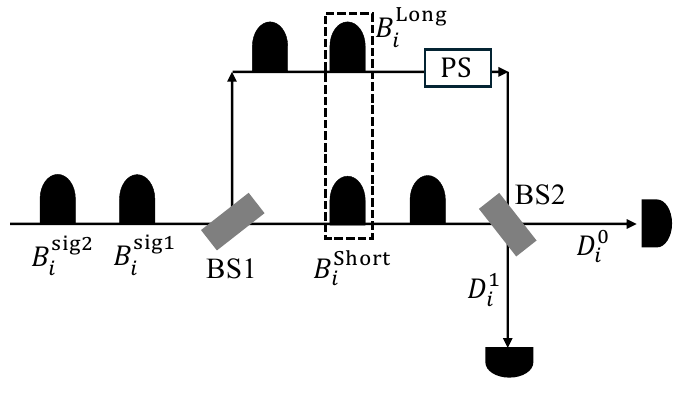}
    \caption{
    Bob's measurement setup. The input states of systems $B_i^{\U{sig}1}$ and $B_i^{\U{sig}2}$ are first split into long and short paths of the interferometer by the first beam splitter (BS1). When system $B_i^{\U{sig}1}$ (or $B_i^{\U{sig}2}$) passes through the long (or short) arm of the interferometer, it is denoted as $B_i^{\U{Long}}$ (or $B_i^{\U{Short}}$). After the state of system $B_i^{\U{Long}}$ undergoes phase modulation by the phase shifter (PS), the two pulses in systems $B_i^{\U{Long}}$ and $B_i^{\U{Short}}$ interfere at the second beam splitter (BS2). The output systems of BS2 are denoted by $D_i^0$ and $D_i^1$. 
    }
    \label{fig:bobactualmeasurement}
\end{figure}
In this section, we describe Bob's measurement operator (see Fig.~\ref{fig:bobactualmeasurement} for Bob's measurement setup). 

According to Bob's procedures in the quantum communication flowchart Fig.~\ref{fig:flowQC}, 
Bob's measurement for the $i$th received double pulse is described by the following CPTP map 
\begin{align}
    \mathcal{E}_i^B: B_i^{\U{sig}} (=B_i^{\U{sig1}}B_i^{\U{sig2}})\to B_i^{\U{CR}}:=
    B_i^{\U{basis}}B_i^{\U{bit}}.
    \label{eq:bobEB}
    \end{align}
Here, the input is the state of system $B_i^{\U{sig}}$ [$B_i^{\U{sig1}}$ ($B_i^{\U{sig2}}$) denotes system of the input first (second) pulse], which is generated by Eve through the CPTP map $\mathcal{E}_j^E$ given in Eqs.~(\ref{eq:eej1}) and (\ref{eq:eej}). Meanwhile, the output of $\mathcal{E}_i^B$ is system $B_i^{\U{CR}}$, which is held by Bob and stores the information of the $i$th measurement basis $\beta_i\in\{Z,X\}$ in system $B_i^{\U{basis}}$ and the $i$th measurement outcome $b_i\in\{0,1,\U{No~click}\}$ in system $B_i^{\U{bit}}$. 

To mathematically describe $\mathcal{E}_i^B$, the following CPTP maps and POVMs are introduced. 
\begin{enumerate}
\item
Let $\mathcal{E}^{\U{BS},JL}_{\theta}$ denote a CPTP map of the beam splitter with transmittance $\cos^2\theta$ acting on state $\hat{\rho}_{JL}$ of systems $J$ and $L$.
\begin{align}
    \mathcal{E}^{\U{BS},JL}_{\theta}(\hat{\rho}_{JL}):=
    e^{\U{i}\theta\sum_{k}(\hat{a}^\dagger_{J,k}\hat{a}_{L,k}+\hat{a}_{J,k}\hat{a}^\dagger_{L,k})}
    \hat{\rho}_{JL}
    e^{-\U{i}\theta\sum_{k}(\hat{a}^\dagger_{J,k}\hat{a}_{L,k}+\hat{a}_{J,k}\hat{a}^\dagger_{L,k})}.
    \end{align}
Here, $\hat{a}^\dagger_{J,k}$ and $\hat{a}_{L,k}^\dagger$ denote creation operators of the $k$th optical mode of systems $J$ and $L$, and the sum of $k$ is taken over all possible optical modes, including spatial modes, frequency modes, and others.
    \item 
Let $\mathcal{E}^{\U{pLoss}}_{\eta}$ denote a CPTP map corresponding to a quantum channel with 
transmittance $\eta$ acting on state $\hat{\rho}_{B}$ of system $B$:
\begin{equation}
    \mathcal{E}^{\U{pLoss},B}_{\eta}(\hat{\rho}_B):=
    \tr_{R} \mathcal{E}^{\U{BS},B,R}_{\arccos \sqrt{\eta}}
    \left(\hat{\rho}_B\otimes \hat{P}\left[\ket{\U{vac}}_{R}\right]\right).
    \end{equation}
 Here, $\ket{\U{vac}}_R$ denotes the vacuum state of system $R$.
    \item 
    Let $\mathcal{E}^{\U{PS},J}_{\theta}$ denote a CPTP map of a $\theta$-phase shifter acting on state $\hat{\rho}_J$ of system $J$:
\begin{align}
    \mathcal{E}^{\U{PS},J}_{\theta}(\hat{\rho}_J):=
    e^{\U{i}\theta\sum_{k}\hat{a}^\dagger_{J,k}\hat{a}_{J,k}}
    \hat{\rho}_J
    e^{-\U{i}\theta\sum_{k}\hat{a}^\dagger_{J,k}\hat{a}_{J,k}}.
\end{align}
Note that the value of phase modulation when the basis $\beta=Z~(X)$ is chosen is $\theta_Z=\pi/2~(\theta_X=\pi)$. 
\item 
Let $\{\hat{E}^{\U{detector1},B}_{\U{Click}}, \hat{E}^{\U{detector1},B}_{\U{Noclick}}\}$ denote a POVM of the threshold detector for system $B$ with dark count probability $p_{\U{dark}}$:
\begin{align}
    \hat{E}^{\U{detector1},B}_{\U{Noclick}}&=(1-p_{\U{dark}})\hat{P}\left[\ket{\U{vac}}_B\right],
    \label{eq:detector2E1noclick}
\\
    \hat{E}^{\U{detector1},B}_{\U{Click}}&=\hat{I}_B-(1-p_{\U{dark}})\hat{P}\left[\ket{\U{vac}}_B\right].
                \label{eq:detector2E1}
\end{align}
\item 
Let $\{\hat{E}^{\U{detector},B}_{\U{Click}}, \hat{E}^{\U{detector},B}_{\U{Noclick}} \}$ denote the POVM of the threshold detector for system $B$ with dark count probability $p_{\U{dark}}$ and the detection efficiency $\eta_{\U{det}}$:
\begin{equation}
   \hat{E}^{\U{detector},B}_{\U{Click}}:= \mathcal{E}^{\U{pLoss},B\dagger}_{\eta_{\U{det}}}
   (\hat{E}^{\U{detector1},B}_{\U{Click}}),
                   \label{eq:detectorE1click}
\end{equation}
\begin{equation}
   \hat{E}^{\U{detector},B}_{\U{Noclick}}
   := \mathcal{E}^{\U{pLoss},B\dagger}_{\eta_{\U{det}}}
   (\hat{E}^{\U{detector1},B}_{\U{Noclick}}).
                      \label{eq:detectorE1noclick}
\end{equation}
\item 
The POVM element corresponding to the click event on the threshold detector that outputs bit 0 when Bob chooses the  basis $\beta$ is given by
\begin{align}
 \hat{E}_{\beta,D^0_i}^{\U{Click}}:
 B^{\U{sig1}}_iB^{\U{sig2}}_i\to
 D_i^0D_i^1
\end{align}
with
\begin{align}
\hat{E}_{\beta,D^0_i}^{\U{Click}}
    :=
\left(\mathcal{E}^{\U{pLoss},B^{\U{sig1}}_i\dagger}_{1/2} 
    \otimes \mathcal{E}^{\U{pLoss},B^{\U{sig2}}_i\dagger}_{1/2}\right)\circ
\left(\mathcal{E}^{\U{PS},B^{\U{Long}\dagger}_i}_{\theta_\beta}
    \otimes \hat{I}^\dagger_{B^{\U{Short}}_i}
\right)\mathcal{E}^{\U{BS},B^{\U{Long}}_i,B^{\U{Short}}_i\dagger}_{\pi/4}
    \circ   (\hat{E}^{\U{detector},D^0_i}_{\U{Click}}\otimes \hat{I}_{D^1_i}).
\label{EclickD0}
\end{align}
From the assumption of identical performance of photon detectors in Sec.~\ref{table:assumption}, the POVM element corresponding to the click event on the threshold detector that outputs bit 1 when Bob chooses the  basis $\beta$ is given by
\begin{align}
 \hat{E}_{\beta,D^1_i}^{\U{Click}}:
 B^{\U{sig1}}_iB^{\U{sig2}}_i\to
 D_i^0D_i^1
\end{align}
with
\begin{align}
\hat{E}_{\beta,D^1_i}^{\U{Click}}
    :=
\left(\mathcal{E}^{\U{pLoss},B^{\U{sig1}}_i\dagger}_{1/2} 
    \otimes \mathcal{E}^{\U{pLoss},B^{\U{sig2}}_i\dagger}_{1/2}\right)\circ
\left(\mathcal{E}^{\U{PS},B^{\U{Long}\dagger}_i}_{\theta_\beta}
    \otimes \hat{I}^\dagger_{B^{\U{Short}}_i}
\right)\mathcal{E}^{\U{BS},B^{\U{Long}}_i,B^{\U{Short}}_i\dagger}_{\pi/4}
    \circ   (\hat{I}_{D^0_i}\otimes \hat{E}^{\U{detector},D^1_i}_{\U{Click}}).
\label{EclickD1}
\end{align}
From the ideal phase modulation assumption (receiver side)  in Table~\ref{table:assumption}, 
the value of phase modulation when the basis $\beta=Z$ ($X$) is chosen is $\theta_Z=\pi/2$ $(\theta_X=\pi)$. 
\\\\
Here, the action of these POVM elements on input state $\hat{\rho}_{B^{\U{sig1}}_iB^{\U{sig2}}_i}$ is explained as follows. First, the two input pulses experience a 50\% loss due to the first beam splitter (BS1), which is represented by the map $\mathcal{E}^{\U{pLoss},B^{\U{sig1}}_i}_{1/2}\otimes\mathcal{E}^{\U{pLoss},B^{\U{sig2}}_i}_{1/2}$. 
Next, for system that passes through the long arm of the interferometer originated from $B^{\U{sig1}}_i$, its phase is shifted by $\theta_\beta$ by the phase shifter, represented by $\mathcal{E}^{\U{PS},B^{\U{Long}}_i}_{\theta_\beta}$. Finally, the two pulses of systems $B^{\U{Short}}_iB^{\U{Long}}_i$ interfere at the second beam splitter (BS2), represented by the map $\mathcal{E}^{\U{BS},B^{\U{Long}}_i,B^{\U{Short}}_i\dagger}_{\pi/4}$, resulting in a click event at the detector that outputs the bit $b$, represented by the map $\hat{E}^{\U{detector},D^b_i}_{\U{Click}}$. 
\\\\
The `No click events' are the complements of Eqs.~(\ref{EclickD0}) and (\ref{EclickD1}), and their POVM elements are written as
\begin{align}
    \hat{E}_{\beta,D^b_i}^{\U{Noclick}}
    := \hat{I}_{B^{\U{sig1}}_i}\otimes\hat{I}_{B^{\U{sig2}}_i}-\hat{E}_{\beta,D^b_i}^{\U{Click}}.
\end{align}
\item 
The POVM $\{\hat{E}^{\U{meas},\beta,i}_{b}\}_{b\in\{0,1,\U{Noclick}\}}$, which corresponds to obtaining the measurement outcome $b$ with the measurement basis $\beta$, is written as
\begin{align}
     \hat{E}^{\U{meas},\beta,i}_{b}:=
    \begin{cases}
        \hat{E}_{\beta,D^b_i}^{\U{Click}}
        \hat{E}_{\beta,D^{b\oplus1}_i}^{\U{Noclick}}
        +\frac{1}{2}\hat{E}_{\beta,D^0_i}^{\U{Click}}
        \hat{E}_{\beta,D^1_i}^{\U{Click}} & b\in\{0,1\}\\\\
                \hat{E}_{\beta,D^0_i}^{\U{Noclick}}
        \hat{E}_{\beta,D^1_i}^{\U{Noclick}} & b=\U{Noclick}.
    \end{cases}
\label{eq:bobpovmlast}
\end{align}
\end{enumerate}
Using the POVM $\{\hat{E}^{\U{meas},\beta,i}_{b}\}_{b\in\{0,1,\U{Noclick}\}}$ in Eq.~(\ref{eq:bobpovmlast}), Bob's operation $\mathcal{E}^{B}_i$ in Eq.~(\ref{eq:bobEB}), which acts on the state of system $B_i^{\U{sig}}$, is given by
\begin{align}
    \mathcal{E}^B_i(\hat{\rho}_{B_i^{\U{sig}}})
    =
    \sum_{\beta\in\{Z,X\},b\in\{0,1,\U{Noclick}\}}p_{\beta}
     \hat{P}\left[\ket{\beta}_{B^{\U{basis}}_i}\ket{b}_{B^{\U{bit}}_i} \right]
    \tr_{B^{\U{sig1}}_iB^{\U{sig2}}_i} \left(\hat{E}^{\U{meas},\beta,i}_{b} 
    (\hat{\rho}_{B_i^{\U{sig}}})\right).
    \label{eq:EBformula}
\end{align}

\subsubsection{Bob's information disclosure in the information exchanging and processing flowchart}
\label{sec:bobinfpro}
According to the information exchanging and processing flowchart Fig.~\ref{fig:flowsift}, 
Bob announces the information about the measurement outcomes for the $j$th block after completing the measurement of this block through the classical channel. 
Bob's operation $\mathcal{E}^{B,\U{public}}_{S_j}$ 
is described by the following CPTP map
\begin{align}
\mathcal{E}^{B,\U{public}}_{S_j}: 
B_{S_j}^{\U{CR}}
\to
B_{S_j}^{\U{CR}}
C^{E_B}_{S_{j}} C_{S_j}^{A_B},
\label{eq:public}
\end{align}
where system $C^{E_B}_{S_{j}}$ is held by Eve, and system $C_{S_j}^{A_B}$ is held by Alice. From the 
authenticated classical channel assumption in Sec.~\ref{sec:assumptions-of-actual-protocol}, 
the information Bob announced through the classical channel reaches Alice without being tampered with. 

Hence, $\mathcal{E}^{B,\U{public}}_{S_j}$ can be written as
\begin{align}
    \mathcal{E}^{B,\U{public}}_{S_j}=
    \bigotimes_{i=(j-1)M+1}^{jM}\mathcal{E}^{B,\U{public}}_i
\label{eq:EsjBpublic}
\end{align}
with
\begin{align}
\mathcal{E}^{B,\U{public}}_i:
\begin{cases}
\ket{\U{Noclick}}_{B_i^{\U{bit}}}\ket{\beta}_{B_i^{\U{basis}}}
\\
  \mapsto 
  \ket{\U{Noclick}}_{B_i^{\U{bit}}}\ket{\beta}_{B_i^{\U{basis}}}
  \ket{\beta,\U{Noclick},\nul}_{C_i^{A_B}}\ket{\beta,\U{Noclick},\nul}_{C^{E_B}_i}\\
  ~~~~\U{for}~\beta\in\{Z,X\}
  \\\\
\ket{b}_{B_i^{\U{bit}}}\ket{\beta}_{B_i^{\U{basis}}}
\mapsto\ket{b}_{B_i^{\U{bit}}}\ket{\beta}_{B_i^{\U{basis}}}
\ket{\beta,\U{click},b}_{C_i^{A_B}}\ket{\beta,\U{click},b}_{C^{E_B}_i}
\\
  ~~~~\U{for}~\beta=X,b\in\{0,1\}
  \\\\
  \ket{b}_{B_i^{\U{bit}}}\ket{\beta}_{B_i^{\U{basis}}}
\mapsto\ket{b}_{B_i^{\U{bit}}}\ket{\beta}_{B_i^{\U{basis}}}
\ket{\beta,\U{click},\nul}_{C_i^{A_B}}\ket{\beta,\U{click},\nul}_{C^{E_B}_i}
\\
  ~~~~\U{for}~\beta=Z,b\in\{0,1\}.
  \end{cases}
\end{align}
Here, $\ket{\U{null}}$ represents the unique element of the trivial (zero-dimensional) Hilbert space. 

\subsubsection{Alice's information disclosure in the information exchanging and processing flowchart}
\label{sec:aliceinfpro}
According to the information exchanging and processing flowchart Fig.~\ref{fig:flowsift}, 
Alice announces the information about the emitted states for the $j$th block, which is stored in system $A_{S_j}^{\U{CR}}$,
after she receives the information about the measurement outcomes for the $j$th block ($1\le j\le N_{\U{block}}$) from Bob. This operation is described by the CPTP map
\begin{align}
    \mathcal{E}^{A,\U{public}}_{S_j}:=
    \bigotimes_{i=(j-1)M+1}^{jM}\mathcal{E}^{A,\U{public}}_i,
    \label{eq:publicA}
\end{align}
where
\begin{align}
    \mathcal{E}^{A,\U{public}}_i:
    A_i^{\U{CR}}{C_i^{A_B}}\to 
    A_i^{\U{CR}}{C_i^{A_B}}C_i^{B_A}C_i^{E_A}
    \label{eq:aliceINFPRO}
\end{align}
specifies the following CPTP map
\begin{align}
\label{EApublic}
\mathcal{E}^{A,\U{public}}_i:
\begin{cases}
\ket{\omega,Z,a}_{A_i^{\U{CR}}}
  \ket{Z,\U{click},\nul}_{{C_i^{A_B}}}\\
  \mapsto \ket{\omega,Z,a}_{A_i^{\U{CR}}}
  \ket{Z,\U{click},\nul}_{{C_i^{A_B}}}
  \ket{\omega,Z,\nul}_{C^{E_A}_i}
  \ket{\omega,Z,\nul}_{{C_i^{B_A}}}\\
  ~~~~\U{for}~\omega\in\{S,D,V\}, a\in\{0,1\}
  \\\\
  \ket{\omega,X,a}_{A_i^{\U{CR}}}
  \ket{X,\U{click},b}_{{C_i^{A_B}}}\\
  \mapsto \ket{\omega,X,a}_{A_i^{\U{CR}}}
  \ket{X,\U{click},b}_{{C_i^{A_B}}}\ket{\omega,X,a}_{C^{E_A}_i}
  \ket{\omega,X,a}_{{C_i^{B_A}}}
  \\
  ~~~~\U{for}~\omega\in\{S,D,V\}, a,b\in\{0,1\}
  \\\\
  \ket{\omega,\alpha,a}_{A_i^{\U{CR}}}
  \ket{\beta,\U{Noclick},\nul}_{{C_i^{A_B}}}\\
  \mapsto \ket{\omega,\alpha,a}_{A_i^{\U{CR}}}
  \ket{\beta,\U{Noclick},\nul}_{{C_i^{A_B}}}
  \ket{\nul,\nul,\nul}_{C^{E_A}_i}
  \ket{\nul,\nul,\nul}_{{C_i^{B_A}}}\\
  ~~~~\U{for}~\omega\in\{S,D,V\}, \alpha,\beta\in\{Z,X\},a\in\{0,1\}
   \\\\
  \ket{\omega,\alpha,a}_{A_i^{\U{CR}}}
  \ket{\beta,\U{click},b}_{{C_i^{A_B}}}\\
  \mapsto 
    \ket{\omega,\alpha,a}_{A_i^{\U{CR}}}
  \ket{\beta,\U{click},b}_{{C_i^{A_B}}}
  \ket{\nul,\nul,\nul}_{C^{E_A}_i}
  \ket{\nul,\nul,\nul}_{{C_i^{B_A}}}\\
  ~~~~\U{for}~\omega\in\{S,D,V\}, \alpha\neq\beta,a,b\in\{0,1\}.
  \end{cases}
\end{align}
Here, system ${C_i^{A_B}}$ is held by Alice, 
system $C_i^{B_A}$ is held by Bob, and  system $C_i^{E_A}$ is held by Eve.

\subsubsection{Entire operation of Alice, Bob and Eve}
\label{sec:zentai}
\begin{figure}
    \centering
    \includegraphics[width=14.5cm]{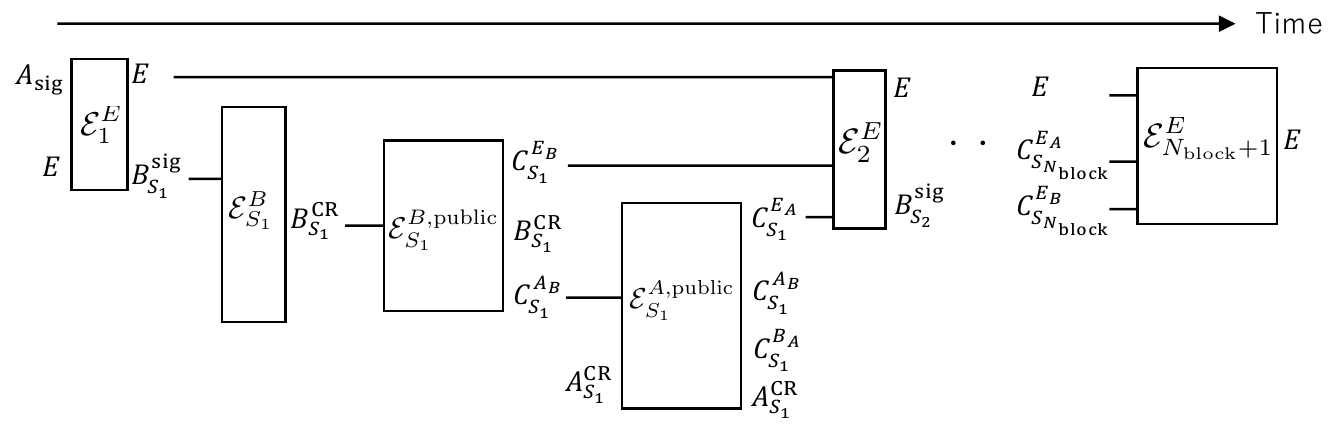}
    \caption{
    The time evolution of the entire system of Alice, Bob and Eve up to the quantum communication flowchart Fig.~\ref{fig:flowQC} and the information exchanging and processing flowchart Fig.~\ref{fig:flowsift}.
    }
    \label{fig:systemzentai}
\end{figure}

In Fig.~\ref{fig:systemzentai}, we depict the time evolution of the entire system held by Alice, Bob and Eve, as introduced in Secs.~\ref{sec:alicestate}-\ref{sec:aliceinfpro}. Below, we make three remarks on Fig.~\ref{fig:systemzentai}.
\begin{enumerate}
\item
Eve's operations $\mathcal{E}^E_{1}$, $\mathcal{E}^E_{j}$ ($j\in\{2,...,N_{\U{block}}\}$) and 
$\mathcal{E}^E_{N_{\U{block}}+1}$ are defined by Eqs.~(\ref{eq:eej1}), (\ref{eq:eej}) and (\ref{eq:eejB}), respectively. 
\item
Bob's measurement operation 
$\mathcal{E}_{S_j}^B$ is 
\begin{align}
    \mathcal{E}_{S_j}^B=\bigotimes _{i=(j-1)M+1}^{jM}\mathcal{E}_{i}^B, 
\end{align}
with $\mathcal{E}_i^B$ being defined by Eq.~(\ref{eq:bobEB}), and 
the output system of $\mathcal{E}_{S_j}^B$ is
\begin{align}
B_{S_j}^{\U{CR}}:=\bigotimes_{i=(j-1)M+1}^{jM}B_i^{\U{CR}}.
\end{align}
\item
$\mathcal{E}^{B,\U{public}}_{S_j}$ and 
$\mathcal{E}^{A,\U{public}}_{S_j}$ are defined by Eqs.~(\ref{eq:public}) and (\ref{eq:publicA}), respectively. 
The input and output system $A_{S_j}^{\U{CR}}$ of $\mathcal{E}^{A,\U{public}}_{S_j}$ is defined by
\begin{align}
A_{S_j}^{\U{CR}}:=\bigotimes_{i=(j-1)M+1}^{jM}A_i^{\U{CR}}, 
\end{align}
and output systems $C^{E_A}_{S_{j}}$ and $C_{S_j}^{B_A}$ of $\mathcal{E}^{A,\U{public}}_{S_j}$ are respectively defined by
\begin{align}
C^{E_A}_{S_{j}}&:=\bigotimes_{i=(j-1)M+1}^{jM}C^{E_A}_{i}, 
\\
C_{S_j}^{B_A}&:=\bigotimes_{i=(j-1)M+1}^{jM}C_i^{B_A}.
\end{align}
System ${C_{S_j}^{A_B}}$ is included as an input to Alice's operation $\mathcal{E}^{A,\U{public}}_{S_j}$. This is because Bob discloses the information of the $j$th block after the measurement of the $j$th block is completed. 
\end{enumerate}

Using the CPTP maps introduced so far and Eq.~(\ref{eq:alicestate}), state $\hat{\rho}_{\U{QC}}$ of Alice's system $A_{\U{CR}}C_{A_B}$, Bob's system $B_{\U{CR}}C_{B_A}$, and Eve's system $E$ immediately after completing the quantum communication flowchart Fig.~\ref{fig:flowQC} and the 
information exchanging and processing flowchart Fig.~\ref{fig:flowsift} is given by
\begin{align}
&\hat{\rho}_{\U{QC}}=
\mathcal{E}_{N_{\U{block}}+1}^E\circ 
(\mathcal{E}_{S_{N_{\U{block}}}}^{A,\U{public}}\circ 
    \mathcal{E}_{S_{N_{\U{block}}}}^{B,\U{public}}
    \circ\mathcal{E}_{S_{N_{\U{block}}}}^B\circ\mathcal{E}_{N_{\U{block}}}^E)\circ
     \cdots\circ \notag\\
&    
(\mathcal{E}_{S_2}^{A,\U{public}}\circ 
    \mathcal{E}_{S_2}^{B,\U{public}}\circ
    \mathcal{E}_{S_2}^B\circ\mathcal{E}_2^E)\circ
(\mathcal{E}_{S_1}^{A,\U{public}}\circ 
    \mathcal{E}_{S_1}^{B,\U{public}}\circ 
    \mathcal{E}_{S_1}^B\circ \mathcal{E}_1^E)
    (\hat{\rho}_{\U{in},A_{\U{CR}},A_{\U{sig}}}).
    \label{eq:rhoQC}
\end{align}

\subsubsection{Operation of key generation in the key generation flowchart}
\label{subsec:kd}
The key generation flowchart Figs.~\ref{fig:flowKD1} and \ref{fig:flowKD2} consists of four steps, and we will describe each step mathematically.

{\bf Alice's and Bob's Step 1}\\
The operation performed by Alice and Bob at Step 1 is described by the CPTP map
\begin{align}
\mathcal{E}^{\U{sift}}: A_{\U{CR}}C_{A_B}B_{\U{CR}} C_{B_A} \to A_{\U{sift}} B_{\U{sift}}
C^{\U{Length}}_{\U{Key}}
C^{\U{Length}}_{\U{Judge}}.
\end{align}
This CPTP map can be written by using the Kraus operators 
$$\{\hat{K}^{\U{sift}}_{\vec{\omega},\vec{\omega'},
\vec{\alpha},\vec{\alpha'},\vec{a},\vec{a'},
\vec{\beta},\vec{\beta'},\vec{b},\vec{y},\vec{b'}}\}_{\vec{\omega},\vec{\omega'},
\vec{\alpha},\vec{\alpha'},\vec{a},\vec{a'},
\vec{\beta},\vec{\beta'},\vec{b},\vec{y},\vec{b'}}$$ 
acting on state $\hat{\rho}_{A_{\U{CR}}C_{A_B}B_{\U{CR}} C_{B_A}}$ of systems $A_{\U{CR}}C_{A_B}B_{\U{CR}} C_{B_A}$ as
\begin{align}
&\mathcal{E}^{\U{sift}}(\hat{\rho}_{A_{\U{CR}}C_{A_B}B_{\U{CR}} C_{B_A}})=
\sum_{\vec{\omega}\in\{S,D,V\}^N,\vec{\omega'}\in\{S,D,V,\nul\}^N,
\vec{\alpha}\in\{Z,X\}^N,\vec{\alpha'}\in\{Z,X,\nul\}^N,\vec{a'}\in\{0,1,\nul\}^N}\notag\\
&\sum_{\vec{\beta},\vec{\beta'}\in\{Z,X\}^N,\vec{y}\in\{\U{Noclick,click}\}^N,\vec{b'}\in\{0,1,\nul\}^N}\notag\\
&
\left(
\sum_{\vec{a}\in\{0,1\}^N,\vec{b}\in\{0,1,\U{Noclick}\}^N}
\hat{K}^{\U{sift}}_{\vec{\omega},\vec{\omega'},
\vec{\alpha},\vec{\alpha'},\vec{a},\vec{a'},
\vec{\beta},\vec{\beta'},\vec{b},\vec{y},\vec{b'}}
\right)
\hat{\rho}_{A_{\U{CR}}C_{A_B}B_{\U{CR}}C_{B_A}}
\notag\\
&\left(
\sum_{\vec{a}\in\{0,1\}^N,\vec{b}\in\{0,1,\U{Noclick}\}^N}
\hat{K}^{\U{sift}\dagger}_{\vec{\omega},\vec{\omega'},
\vec{\alpha},\vec{\alpha'},\vec{a},\vec{a'},
\vec{\beta},\vec{\beta'},\vec{b},\vec{y},\vec{b'}}
\right)
.
\label{eq:esift}
\end{align}
In this CPTP map, we take the sum over all the information stored in systems $A_{\U{CR}},C_{A_B},B_{\U{CR}}$ and $C_{B_A}$. 
To write the Kraus operators explicitly, we introduce the following functions:
\begin{enumerate}
    \item 
Let $f(\vec{b})$ be the function that takes $\vec{b}$ as input and outputs $N$ elements, each of which can be either ``Noclick" or ``click". 
The $i$th element of the output is ``Noclick" if $b_i=$Noclick, and it takes ``click" if $b_i\in\{0,1\}$. 
\item     
Let $g(\vec{\beta},\vec{b})$ be the function that takes $\vec{\beta}$ and $\vec{b}$ as inputs and outputs $N$ elements, each of which can be either 0, 1 or $\nul$. 
The $i$th element of the output is $b_i$ if $\beta_i=X$ and $b_i\in\{0,1\}$, and it takes $\nul$ otherwise. 
\item 
Let $t(\vec{\omega},\vec{\alpha},\vec{a},\vec{\beta'},\vec{y},\vec{b'})$ be the function that takes $\vec{\omega},\vec{\alpha},\vec{a},\vec{\beta'},\vec{y}$ and $\vec{b'}$ as inputs and outputs $\vec{\omega'},\vec{\alpha'}$ and $\vec{a'}$, with the $i$th element defined by Eq.~(\ref{EApublic}).
\item 
Let $f^{\U{Length}}_{\U{Judge}}$ be the function that takes 1 if 
$N_{\U{sift}}-N_{\U{PA}}-N_{\U{EC}}-N_{\U{verify}}>0$ and 
takes 0 if $N_{\U{sift}}-N_{\U{PA}}-N_{\U{EC}}-N_{\U{verify}}\le0$.
\end{enumerate}
Using these functions, the Kraus operators can be expressed as follows:
\begin{align}
&
\hat{K}^{\U{sift}}_{\vec{\omega},\vec{\omega'},
\vec{\alpha},\vec{\alpha'},\vec{a},\vec{a'},
\vec{\beta},\vec{\beta'},\vec{b},\vec{y},\vec{b'}}
=\delta\(\vec{\beta'},\vec{\beta}\)
\delta\(\vec{y},f(\vec{b})\)\delta\(\vec{b'},g(\vec{\beta},\vec{b})\)
\delta\(\vec{\omega'},\vec{\alpha'},\vec{a'};
t(\vec{\omega},\vec{\alpha},\vec{a},\vec{\beta'},\vec{y},\vec{b'})\)
\notag\\
&\ket{N_{\U{sift}},\bm{k}_A}_{A_{\U{sift}}}
\bra{\vec{\omega},\vec{\alpha},\vec{a}}_{A_{\U{CR}}}
\bra{\vec{\beta'},\vec{y},\vec{b'}}_{C_{A_B}}
\otimes
\ket{N_{\U{sift}},\bm{k}_B}_{B_{\U{sift}}}
\bra{\vec{\beta},\vec{b}}_{B_{\U{CR}}}
\bra{\vec{\omega'},\vec{\alpha'},\vec{a'}}_{C_{B_A}}
\notag\\
&\otimes
\ket{N_{\U{sift}}-N_{\U{PA}}-N_{\U{EC}}-N_{\U{verify}}}_{C^{\U{Length}}_{\U{key}}}
\ket{f^{\U{Length}}_{\U{Judge}}(N_{\U{sift}}-N_{\U{PA}}-N_{\U{EC}}-N_{\U{verify}})}_{C^{\U{Length}}_{\U{Judge}}}\notag\\
&+
\left[1-\delta\(\vec{\beta'},\vec{\beta}\)
\delta\(\vec{y},f(\vec{b})\)\delta\(\vec{b'},g(\vec{\beta},\vec{b})\)
\delta\(\vec{\omega'},\vec{\alpha'},\vec{a'}
;t(\vec{\omega},\vec{\alpha},\vec{a},\vec{\beta'},\vec{y},\vec{b'})\)\right]
\notag\\
&\ket{0,\nul}_{A_{\U{sift}}}
\bra{\vec{\omega},\vec{\alpha},\vec{a}}_{A_{\U{CR}}}
\bra{\vec{\beta'},\vec{y},\vec{b'}}_{C_{A_B}}
\notag\\
&\otimes
\ket{0,\nul}_{B_{\U{sift}}}
\bra{\vec{\beta},\vec{b}}_{B_{\U{CR}}}
\bra{\vec{\omega'},\vec{\alpha'},\vec{a'}}_{C_{B_A}}
\otimes
\ket{0}_{C^{\U{Length}}_{\U{key}}}
\ket{0}_{C^{\U{Length}}_{\U{Judge}}}.
\label{eq:kraususiftoperation}
\end{align}
Here, $\bm{k}_A:=(a_i)_{i\in S_{\U{sift}}}$ and $\bm{k}_B:=(b_i)_{i\in S_{\U{sift}}}$ are Alice's and Bob's sifted keys, respectively, with $S_{\U{sift}}$ defined in the key generation flowchart Fig.~\ref{fig:flowKD1}. $N_{\U{PA}}$ is related to the amount of privacy amplification $N_{\U{EC}}+N_{\U{verify}}+N_{\U{PA}}$.
$N_{\U{PA}}$ is determined by the observables ($\{N_{\omega,X}^{\U{Error}}\}_{\omega}$) as shown in Sec.~\ref{subsec:M1Z}. 

{\bf Alice's and Bob's Step 2}\\
For state $\mathcal{E}^{\U{sift}}(\hat{\rho}_{\U{QC}})$ after Step~1, the operation performed by Alice and Bob at Step~2 is described by the CPTP map
\footnote{
Note that $\U{Im}(\mathcal{E}^{\U{sift}})$ represents the image of map $\mathcal{E}_{\U{sift}}$.
}
\begin{align}
\mathcal{E}^{\U{EC}}: \U{Im}(\mathcal{E}^{\U{sift}})
\to A_{\U{sift}} B_{\U{sift}} 
C_{\U{EC}}C^{\U{Length}}_{\U{Judge}}C^{\U{Length}}_{\U{Key}}.
\end{align}
This CPTP map can be written using the Kraus operators 
\[
\{\hat{K}^{\U{EC}}_{N_{\U{sift}},\bm{k}_A,\bm{k}_B}\}_{
N_{\U{sift}},\bm{k}_A,\bm{k}_B
}
\]
acting on state $\hat{\rho}_{A_{\U{sift}} B_{\U{sift}}C^{\U{Length}}_{\U{Judge}}}$ of systems $A_{\U{sift}} B_{\U{sift}}C^{\U{Length}}_{\U{Judge}}$ as
\begin{align}
\mathcal{E}^{\U{EC}}(\hat{\rho}_{A_{\U{sift}} B_{\U{sift}}C^{\U{Length}}_{\U{Judge}}})
=\sum_{N_{\U{sift}}=0}^N
\sum_{\bm{k}_A,\bm{k}_B\in\{0,1\}^{N_{\U{sift}}}} \hat{K}^{\U{EC}}_{N_{\U{sift}},\bm{k}_A,\bm{k}_B}
\hat{\rho}_{A_{\U{sift}} B_{\U{sift}}C^{\U{Length}}_{\U{Judge}}}
\hat{K}^{\U{EC}\dagger}_{N_{\U{sift}},\bm{k}_A,\bm{k}_B}
\label{eq:mathcalEC}
\end{align}
with
\begin{align}
&\hat{K}^{\U{EC}}_{N_{\U{sift}},\bm{k}_A,\bm{k}_B}=
\hat{P}[\ket{N_{\U{sift}},\bm{k}_A}_{A_{\U{sift}}}]\otimes
\ket{f_{\U{synd}}(N_{\U{sift}},N_{\U{EC}},\bm{k}_A)}_{C_{\U{EC}}}\otimes\notag\\
&\ket{N_{\U{sift}},\bm{k}_B\oplus f_{\U{EC}}\circ f_{\U{synd}}(N_{\U{sift}},N_{\U{EC}},\bm{k}_A)}
\bra{N_{\U{sift}},\bm{k}_B}_{B_{\U{sift}}}
\otimes\hat{P}[\ket{1}_{C^{\U{Length}}_{\U{Judge}}}]\notag\\
&+\ket{0,\nul}
\bra{N_{\U{sift}},\bm{k}_A}_{A_{\U{sift}}}\otimes
\ket{\nul}_{C_{\U{EC}}}\otimes
\ket{0,\nul}
\bra{N_{\U{sift}},\bm{k}_B}_{B_{\U{sift}}}
\otimes\hat{P}[\ket{0}_{C^{\U{Length}}_{\U{Judge}}}].
\label{eq:KECkraus}
\end{align}
Here, $f_{\U{synd}}$ represents the syndrome information of $N_{\U{EC}}$ bits based on a linear code, and $f_{\U{EC}}$ represents the function that inputs the syndrome information $f_{\U{synd}}$ sent from Alice and outputs $N_{\U{sift}}$ bits, where each bit is 1 if Bob finds a bit error in $\bm{k}_B$ and 0 if he finds no bit error in $\bm{k}_B$. 
Note that  as specified in the assumption regarding the syndrome for linear codes in Table~\ref{table:yougo}, we require that $f_{\U{EC}}$ always outputs an error vector with unit probability. 
\\\\
{\bf Alice's and Bob's Step 3}\\
For state $\mathcal{E}^{\U{EC}}\circ\mathcal{E}^{\U{sift}}(\hat{\rho}_{\U{QC}})$ after Step~2, the operation performed by Alice and Bob at Step~3 is described by the CPTP map
\begin{align}
\mathcal{E}^{\U{verify}}: \U{Im}(\mathcal{E}^{\U{EC}}\circ\mathcal{E}^{\U{sift}})\to
A_{\U{sift}}B_{\U{sift}} C^{\U{Hash}}_AC^{\U{HashResult}}_BC_{\U{EC}}C^{\U{Length}}_{\U{Judge}}C^{\U{Length}}_{\U{Key}}.
\end{align}
This CPTP map can be written by using the Kraus operators
\[
\{
\hat{K}^{\U{verify}}_{r_{\U{hash~verify}},
N_{\U{sift}},\bm{k}_A,\bm{k}_B}
\}_{r_{\U{hash~verify}},
N_{\U{sift}},\bm{k}_A,\bm{k}_B}
\] 
acting on state $\hat{\rho}_{A_{\U{sift}}B_{\U{sift}}C^{\U{Length}}_{\U{Judge}}}$ of systems $A_{\U{sift}}B_{\U{sift}}C^{\U{Length}}_{\U{Judge}}$ as
\begin{align}
&\mathcal{E}^{\U{verify}}(\hat{\rho}_{A_{\U{sift}}B_{\U{sift}}C^{\U{Length}}_{\U{Judge}}})=
\tr_{B_{\U{Hash}}}\frac{1}{2^{N_{\U{hash~verify}}}}
\sum_{r_{\U{hash~verify}}\in\{0,1\}^{N_{\U{hash~verify}}}}
\notag\\
&\sum_{N_{\U{sift}}=0}^N
\sum_{\bm{k}_A,\bm{k}_B\in\{0,1\}^{N_{\U{sift}}}}
\hat{K}^{\U{verify}}_{r_{\U{hash~verify}},
N_{\U{sift}},\bm{k}_A,\bm{k}_B}
\hat{\rho}_{A_{\U{sift}}B_{\U{sift}}C^{\U{Length}}_{\U{Judge}}}
\hat{K}^{\U{verify}\dagger}_{r_{\U{hash~verify}},
N_{\U{sift}},\bm{k}_A,\bm{k}_B}
\label{eq:mathcalverifyA}
\end{align}
with
\begin{align}
&\hat{K}^{\U{verify}}_{r_{\U{hash~verify}},N_{\U{sift}},\bm{k}_A,\bm{k}_B}:=
\notag\\
&\hat{P}\left[\ket{N_{\U{sift}},\bm{k}_A}_{A_{\U{sift}}}
\ket{N_{\U{sift}},\bm{k}_B\oplus f_{\U{EC}}\circ f_{\U{synd}}(N_{\U{sift}},N_{\U{EC}},\bm{k}_A)}_{B_{\U{sift}}}\right]\notag\\
\otimes
&
\ket{r_{\U{hash~verify}},
H(N_{\U{sift}},N_{\U{verify}},\bm{k}_A,r_{\U{hash~verify}})}_
{C^{\U{Hash}}_A}\notag\\
\otimes&
\ket{
H\left(N_{\U{sift}},N_{\U{verify}},\bm{k}_B\oplus f_{\U{EC}}\circ f_{\U{synd}}(N_{\U{sift}},N_{\U{EC}},\bm{k}_A),r_{\U{hash~verify}}\right)}_{B_{\U{Hash}}}
\notag\\
&
\Big|\delta\Bigg(
H(N_{\U{sift}},N_{\U{verify}},\bm{k}_A,r_{\U{hash~verify}})
,\notag\\
&H\left(N_{\U{sift}},N_{\U{verify}},\bm{k}_B\oplus f_{\U{EC}}\circ f_{\U{synd}} (N_{\U{sift}},N_{\U{EC}},\bm{k}_A),r_{\U{hash~verify}}\right)
\Bigg)\Big\rangle_{C^{\U{HashResult}}_B}
\notag\\
&
\Big|\delta\Bigg(
H(N_{\U{sift}},N_{\U{verify}},\bm{k}_A,r_{\U{hash~verify}})
,\notag\\
&H\left(N_{\U{sift}},N_{\U{verify}},\bm{k}_B\oplus f_{\U{EC}}\circ f_{\U{synd}}(N_{\U{sift}},N_{\U{EC}},\bm{k}_A),r_{\U{hash~verify}}\right)
\Bigg)\Big\rangle\bra{1}_{C^{\U{Length}}_{\U{Judge}}}
\notag\\
&+\ket{0,\nul}\bra{N_{\U{sift}},\bm{k}_A}_{A_{\U{sift}}}
\otimes
\notag\\
&
\ket{\nul}_{C_A^{\U{Hash}}}
\ket{\nul}_{B_{\U{Hash}}}
\ket{\nul}_{C_B^{\U{HashResult}}}\otimes
\ket{0,\nul}
\bra{N_{\U{sift}},\bm{k}_B}_{B_{\U{sift}}}
\otimes\hat{P}[\ket{0}_{C^{\U{Length}}_{\U{Judge}}}]
\end{align}
Here, $H$ is the universal2 hash function with an output of $N_{\U{verify}}$ bits.

{\bf Alice's and Bob's Step 4}\\
For state $\mathcal{E}^{\U{verify}}\circ\mathcal{E}^{\U{EC}}\circ\mathcal{E}^{\U{sift}}(\hat{\rho}_{\U{QC}})$ after Step~3, the operation performed by Alice and Bob at Step~4 is described by the CPTP map
\begin{align}
\mathcal{E}^{\U{PA}}:
\U{Im}(\mathcal{E}^{\U{verify}}\circ\mathcal{E}^{\U{EC}}\circ\mathcal{E}^{\U{sift}})
\to
A_{\U{sift}} B_{\U{sift}}C^{\U{Length}}_{\U{Key}}C^{\U{Length}}_{\U{Judge}}C^{\U{Hash}}_AC^{\U{HashResult}}_BC_{\U{EC}}C_{\U{PA}}
.
\end{align}
This CPTP map can be written by using the Kraus operators
\[
\{
\hat{K}^{\U{PA}}_{r_{\U{hashPA}},N_{\U{sift}},\bm{k}_A,\bm{k}_B,N_{\U{fin}}}
\}_{r_{\U{hashPA}},N_{\U{sift}},\bm{k}_A,\bm{k}_B,N_{\U{fin}}}
\]
acting on state $\hat{\rho}_{A_{\U{sift}}B_{\U{sift}}C^{\U{Length}}_{\U{Key}}C^{\U{Length}}_{\U{Judge}}}$ of systems $A_{\U{sift}}B_{\U{sift}}C^{\U{Length}}_{\U{Key}}C^{\U{Length}}_{\U{Judge}}$ as
\begin{align}
&\mathcal{E}^{\U{PA}}(\hat{\rho}_{A_{\U{sift}}B_{\U{sift}}C^{\U{Length}}_{\U{Key}}C^{\U{Length}}_{\U{Judge}}})
=
\frac{1}{2^{N_{\U{hashPA}}}}
\sum_{r_{\U{hashPA}}\in\{0,1\}^{N_{\U{hashPA}}}}
\sum^N_{N_{\U{sift}}=0}\sum_{\bm{k}_A,\bm{k}_B\in\{0,1\}^{N_{\U{sift}}}}
\sum_{N_{\U{fin}}=0}^N\notag\\
&
\hat{K}^{\U{PA}}_{r_{\U{hashPA}},N_{\U{sift}},\bm{k}_A,\bm{k}_B,N_{\U{fin}}}
\hat{\rho}_{A_{\U{sift}}B_{\U{sift}}C^{\U{Length}}_{\U{Key}}C^{\U{Length}}_{\U{Judge}}}
\hat{K}^{\U{PA}\dagger}_{r_{\U{hashPA}},N_{\U{sift}},\bm{k}_A,\bm{k}_B,N_{\U{fin}}}
\label{eq:epa}
\end{align}
with
\begin{align}
&\hat{K}^{\U{PA}}_{r_{\U{hashPA}},N_{\U{sift}},\bm{k}_A,\bm{k}_B,N_{\U{fin}}}:=
\notag\\
&
\ket{r_\U{hashPA}}_{C_{\U{PA}}}
\ket{N_{\U{fin}},f_{\U{PA}}(N_{\U{sift}},N_{\U{fin}},\bm{k}_A,r_{\U{hashPA}})}
\bra{N_{\U{sift}},\bm{k}_A}_{A_{\U{sift}}}\otimes\notag\\
&
\ket{N_{\U{fin}},f_{\U{PA}}\left(N_{\U{sift}},N_{\U{fin}},
\bm{k}_B\oplus f_{\U{EC}}\circ f_{\U{synd}}(N_{\U{sift}},N_{\U{EC}},\bm{k}_A)
,r_{\U{hashPA}}\right)}
\notag\\
&
\bra{N_{\U{sift}},\bm{k}_B\oplus f_{\U{EC}}\circ f_{\U{synd}}(N_{\U{sift}},N_{\U{EC}},\bm{k}_A)}_{B_{\U{sift}}}
\otimes
\hat{P}[\ket{1}_{C^{\U{Length}}_{\U{Judge}}}\ket{N_{\U{fin}}}_{C^{\U{Length}}_{\U{Key}}}]
\notag\\
&+
\ket{r_\U{hashPA}}_{C_{\U{PA}}}\otimes
\ket{0,\nul}\bra{N_{\U{sift}},\bm{k}_A}_{A_{\U{sift}}}\notag\\
&
\ket{0,\nul}\bra{N_{\U{sift}},\bm{k}_B\oplus f_{\U{EC}}\circ f_{\U{synd}}(N_{\U{sift}},N_{\U{EC}},\bm{k}_A)}
_{B_{\U{sift}}}
\hat{P}[\ket{0}_{C^{\U{Length}}_{\U{Judge}}}\ket{N_{\U{fin}}}_{C^{\U{Length}}_{\U{Key}}}]
.
\end{align}
Here, $f_{\U{PA}}$ is the surjective dual universal2 hash function with an output of $N_{\U{fin}}:=N_{\U{sift}}-N_{\U{PA}}-N_{\U{EC}}-N_{\U{verify}}$ bits.

Eve can evolve her system $E$ of $\hat{\rho}_{\U{QC}}$, which is given by Eq.~(\ref{eq:rhoQC}) and the rightmost $E$ in Fig.~\ref{fig:systemzentai}, using the information made public by Alice and Bob as described previously in this section. 
This CPTP map of Eve's is denoted by 
\begin{align}
\mathcal{E}^{\U{final}}: E
C_{\U{EC}}C_{\U{Key}}^{\U{Length}}C_{\U{Judge}}^{\U{Length}}C^{\U{Hash}}_{A}C^{\U{HashResult}}_BC_{\U{PA}}\to E .
\end{align}
The final state $\hat{\rho}_{\U{PA}}$ of systems $A_{\U{sift}} B_{\U{sift}} E$ after completing the QKD protocol can be represented by 
\begin{align}
\hat{\rho}_{\U{PA}}
    =\mathcal{E}^{\U{final}}\circ
    \mathcal{E}^{\U{PA}}\circ\mathcal{E}^{\U{verify}}\circ
    \mathcal{E}^{\U{EC}}\circ\mathcal{E}^{\U{sift}}(\hat{\rho}_{\U{QC}}).
    \label{eq:rhoPA}
\end{align}
Note that $\mathcal{E}^{\U{sift}}$, $\mathcal{E}^{\U{EC}}$, $\mathcal{E}^{\U{verify}}$ and $\mathcal{E}^{\U{PA}}$ are defined by Eqs.~(\ref{eq:esift}), (\ref{eq:mathcalEC}), (\ref{eq:mathcalverifyA}) and (\ref{eq:epa}), respectively.

\subsection{Security definition}
\label{sec:secdef}
In this section, we describe the definition of the security in the QKD protocol based on the universal composable security framework~\cite{ben2005,renner2009,renner2022}. 

For this, we first define the following CPTP map 
\begin{align}
\mathcal{E}^{\U{idealize}}: A_{\U{sift}} B_{\U{sift}}\to A_{\U{sift}} B_{\U{sift}},
\end{align}

\begin{align}
\mathcal{E}^{\U{idealize}}(\hat{\rho}_{A_{\U{sift}} B_{\U{sift}}}):=\sum_{N_{\U{fin}}=0}^N
\tr_{A_{\U{sift}}B_{\U{sift}}}\left(
\hat{E}_{N_{\U{fin}}}\hat{\rho}_{A_{\U{sift}} B_{\U{sift}}}
\right)\otimes\frac{1}{2^{N_{\U{fin}}}}\sum_{\vec{a}\in\{0,1\}^{N_{\U{fin}}}}
\hat{P}[\ket{N_{\U{fin}},\vec{a}}_{A_{\U{sift}}}\ket{N_{\U{fin}},\vec{a}}_{B_{\U{sift}}}]
\end{align}
that replaces Alice's and Bob's secret keys with ideal ones 
depending on the length $N_{\U{fin}}$ of the secret key
\footnote{
Note that $\ket{N_{\U{fin}},\vec{a}}$ with $N_{\U{fin}}=0$ represents the unique element of the trivial (zero-dimensional) Hilbert space. 
}
. 
Here, $\hat{\rho}_{A_{\U{sift}}B_{\U{sift}}}$ denotes the state of systems $A_{\U{sift}} B_{\U{sift}}$, and the projection operator corresponding to the secret key length of $N_{\U{fin}}$ is defined by
\begin{align}
\hat{E}_{N_{\U{fin}}}
:=
\sum_{\vec{a}\in\{0,1\}^{N_{\U{fin}}}}
\hat{P}\left[
\ket{N_{\U{fin}},
\vec{a}}_{A_{\U{sift}}}\right].
\label{eq:eqENfinPro}
\end{align}
Then, by using state $\hat{\rho}_{\U{PA}}$ given in Eq.~(\ref{eq:rhoPA}), the ideal state of systems $A_{\U{sift}}B_{\U{sift}}E$ is defined  as
\begin{align}
\hat{\rho}_{\U{ideal}}:=\mathcal{E}^{\U{idealize}}(\hat{\rho}_{\U{PA}})
\label{eq:rhoideal}.
\end{align}
For $\hat{\rho}_{\U{PA}}$ in Eq.~(\ref{eq:rhoPA}) and $\hat{\rho}_{\U{ideal}}$ in Eq.~(\ref{eq:rhoideal}), if there exists $\epsilon$ ($0<\epsilon<1$) such that
\begin{align}
\frac{1}{2}||\hat{\rho}_{\U{PA}}-\hat{\rho}_{\U{ideal}}||\le\epsilon
\label{eq:epsilon}
\end{align}
holds, 
then the QKD protocol is defined to be $\epsilon$-secure. 
We call $\epsilon$ the security parameter of the QKD protocol. 

For this security parameter, 
the following Proposition~\ref{prop_bunri} holds. To state this, we first explain the useful relations for $\hat{\rho}_{\U{PA}}$ and $\hat{\rho}_{\U{ideal}}$.
\\\\
\underline{Expression for $\hat{\rho}_{\U{PA}}$}
\\
Let
\begin{align}
\Pr_{\U{PA}}(N_{\U{fin}}):=
\tr\left(\hat{E}_{N_{\U{fin}}}\hat{\rho}_{\U{PA}}\right)
\end{align}
be the probability distribution of the secret key length for $\hat{\rho}_{\U{PA}}$ in Eq.~(\ref{eq:rhoPA}). 
Then, $\hat{\rho}_{\U{PA}}$ can be rewritten as
\begin{align}
\hat{\rho}_{\U{PA}}=\sum_{N_{\U{fin}}=0}^N\Pr(N_{\U{fin}})
\hat{\rho}_{\U{PA}|N_{\U{fin}}},~~~
\hat{\rho}_{\U{PA}|N_{\U{fin}}}:=
\frac{\hat{E}_{N_{\U{fin}}}
\hat{\rho}_{\U{PA}}
\hat{E}_{N_{\U{fin}}}}
{\Pr(N_{\U{fin}})}.
\label{eq:PA1}
\end{align}
Also, let 
\begin{align}
\Pr(k^{\U{fin}}_A,k^{\U{fin}}_B|N_{\U{fin}}):=
\tr(\hat{P}[\ket{N_{\U{fin}},k^{\U{fin}}_A}_{A_{\U{sift}}}
\ket{N_{\U{fin}},k^{\U{fin}}_B}_{B_{\U{sift}}}
]
\hat{\rho}_{\U{PA}|N_{\U{fin}}})
\end{align}
be the probability that Alice's and Bob's secret keys are 
$k^{\U{fin}}_A\in\{0,1\}^{N_{\U{fin}}}$ and $k^{\U{fin}}_B\in\{0,1\}^{N_{\U{fin}}}$,
respectively, when the secret key length is $N_{\U{fin}}$. 
Then, $\hat{\rho}_{\U{PA}|N_{\U{fin}}}$ can be written as
\begin{align}
\hat{\rho}_{\U{PA}|N_{\U{fin}}}
=\sum_{k^{\U{fin}}_A,k^{\U{fin}}_B\in\{0,1\}^{N_{\U{fin}}}}
\Pr(k^{\U{fin}}_A,k^{\U{fin}}_B|N_{\U{fin}})
\hat{P}[\ket{N_{\U{fin}},k^{\U{fin}}_A}_{A_{\U{sift}}}
\ket{N_{\U{fin}},k^{\U{fin}}_B}_{B_{\U{sift}}}]\otimes
\hat{\rho}^E_{\U{PA}|N_{\U{fin}},k^{\U{fin}}_A,k^{\U{fin}}_B},
\label{eq:PA2}
\end{align}

\begin{align}
\hat{\rho}^E_{\U{PA}|N_{\U{fin}},k^{\U{fin}}_A,k^{\U{fin}}_B}:=\frac{
\bra{N_{\U{fin}},k^{\U{fin}}_A}_{A_{\U{sift}}}
\bra{N_{\U{fin}},k^{\U{fin}}_B}_{B_{\U{sift}}}
\hat{\rho}_{\U{PA}|N_{\U{fin}}}
\ket{N_{\U{fin}},k^{\U{fin}}_A}_{A_{\U{sift}}}
\ket{N_{\U{fin}},k^{\U{fin}}_B}_{B_{\U{sift}}}
}
{\Pr(k^{\U{fin}}_A,k^{\U{fin}}_B|N_{\U{fin}})}.
\end{align}
Here, we used the fact that 
Alice and Bob's secret key lengths are always equal, 
which can be seen from Step~4 of the key generation flowchart Fig.~\ref{fig:flowKD2}. 

Finally, we define
\begin{align}
\hat{\rho}^{AE}_{\U{PA}|N_{\U{fin}}}&:=
\tr_{B_{\U{sift}}}\left(\hat{E}_{N_{\U{fin}}}\hat{\rho}_{\U{PA}}
\right)
/\Pr(N_{\U{fin}})
\label{eq:rhoPAAENfinactual}
\end{align}
as the actual state of Alice's and Eve's systems when the secret key length is $N_{\U{fin}}$.
\\\\
\underline{Expression for $\hat{\rho}_{\U{ideal}}$}
\\
From the definition of $\hat{\rho}_{\U{ideal}}$ in Eq.~(\ref{eq:rhoideal}), this ideal state 
can be described as
\begin{align}
\hat{\rho}_{\U{ideal}}=\sum_{N_{\U{fin}}=0}^N\Pr(N_{\U{fin}})
\underbrace{
\sum_{\vec{a}\in\{0,1\}^{N_{\U{fin}}}}\frac{1}{2^{N_{\U{fin}}}}
\hat{P}[\ket{N_{\U{fin}},\vec{a}}_{A_{\U{sift}}}\ket{N_{\U{fin}},\vec{a}}_{B_{\U{sift}}}]
\otimes
\frac{\tr_{A_{\U{sift}}B_{\U{sift}}}\left(
\hat{E}_{N_{\U{fin}}}\hat{\rho}_{\U{PA}}\right)}{\Pr(N_{\U{fin}})}
}
_{=:\hat{\rho}_{\U{ideal}|N_{\U{fin}}}}.
\label{eq:PA3}
\end{align}
We define 
\begin{align}
\hat{\rho}^{AE}_{\U{ideal}|N_{\U{fin}}}&:=
\frac{\tr_{B_{\U{sift}}}\left(
\hat{E}_{N_{\U{fin}}}\hat{\rho}_{\U{ideal}}
\right)}{\Pr(N_{\U{fin}})}
\label{eq:rhoAEidealNfin}
\end{align}
as the ideal state of Alice's and Eve's systems when the secret key length is $N_{\U{fin}}$.

\begin{proposition}
\label{prop_bunri}
{\bf Decomposition of the security parameter (Theorem in~\cite{koashi2009njp})}\\
For the states $\hat{\rho}_{\U{PA}}$ and 
$\hat{\rho}_{\U{ideal}}$ introduced above, 
if 
\begin{align}
\sum_{N_{\U{fin}}=0}^N\Pr(N_{\U{fin}})\Pr(k^{\U{fin}}_A\neq k^{\U{fin}}_B|N_{\U{fin}})\le\epsilon_{\U{correct}}
\label{eq:correctness}
\end{align}
and
\begin{align}
\frac{1}{2}\sum_{N_{\U{fin}}=0}^N\Pr(N_{\U{fin}})
||\hat{\rho}^{AE}_{\U{PA}|N_{\U{fin}}}-\hat{\rho}^{AE}_{\U{ideal}|N_{\U{fin}}}||\le\epsilon_{\U{secrecy}}
\label{eq:secrecy}
\end{align}
hold for $\epsilon_{\U{correct}}$ and $\epsilon_{\U{secrecy}}$ ($0<\epsilon_{\U{correct}},\epsilon_{\U{secrecy}}<1$), 
the security parameter $\epsilon$ 
in Eq.~(\ref{eq:epsilon}) satisfies  
\begin{align}
\epsilon=\epsilon_{\U{correct}}+\epsilon_{\U{secrecy}}.
\label{eq:epsilon_2}
\end{align}
Here, $\epsilon_{\U{correct}}$ is called the correctness parameter, and $\epsilon_{\U{secrecy}}$ is called the secrecy parameter.
\end{proposition}
{\bf Proof of Proposition~\ref{prop_bunri}}\\
We introduce the following state
\begin{align}
\hat{\sigma}_{ABE|N_{\U{fin}}}
:=\sum_{k^{\U{fin}}_A,k^{\U{fin}}_B\in\{0,1\}^{N_{\U{fin}}}}
\Pr(k^{\U{fin}}_A,k^{\U{fin}}_B|N_{\U{fin}})
\hat{P}[\ket{N_{\U{fin}},k^{\U{fin}}_A}_{A_{\U{sift}}}
\ket{N_{\U{fin}},k^{\U{fin}}_A}_{B_{\U{sift}}}]\otimes
\hat{\rho}^E_{\U{PA}|N_{\U{fin}},k^{\U{fin}}_A,k^{\U{fin}}_B}.
\label{eq:sigmaABENfin}
\end{align}
Employing Eqs.~(\ref{eq:PA1}), (\ref{eq:PA2}), (\ref{eq:PA3}), the strong convexity of the trace distance, and the triangle inequality lead to
\begin{align}
&\frac{1}{2}||\hat{\rho}_{\U{PA}}-\hat{\rho}_{\U{ideal}}||
\le
\frac{1}{2}\sum_{N_{\U{fin}}=0}^N\Pr(N_{\U{fin}})
||\hat{\rho}_{\U{PA}|N_{\U{fin}}}-\hat{\rho}_{\U{ideal}|N_{\U{fin}}}||
\notag\\
&\le
\frac{1}{2}\sum_{N_{\U{fin}}=0}^N\Pr(N_{\U{fin}})
(||\sigma_{ABE|N_{\U{fin}}}-\hat{\rho}_{\U{ideal}|N_{\U{fin}}}||+
||\hat{\rho}_{\U{PA}|N_{\U{fin}}}-\hat{\sigma}_{ABE|N_{\U{fin}}}||
).
\label{eq:bunkatu}
\end{align}
As for the first term of Eq.~(\ref{eq:bunkatu}), direct calculation leads to
\begin{align}
&\left|\left|\hat{\sigma}_{ABE|N_{\U{fin}}}-\hat{\rho}_{\U{ideal}|N_{\U{fin}}}\right|\right|
\notag\\
=&
\Bigg|\Bigg|
\sum_{k^{\U{fin}}_A,k^{\U{fin}}_B\in\{0,1\}^{N_{\U{fin}}}}
\Pr(k^{\U{fin}}_A,k^{\U{fin}}_B|N_{\U{fin}})
\hat{P}[\ket{N_{\U{fin}},k^{\U{fin}}_A}_{A_{\U{sift}}}
\ket{N_{\U{fin}},k^{\U{fin}}_A}_{B_{\U{sift}}}]\otimes
\hat{\rho}^E_{\U{PA}|N_{\U{fin}},k^{\U{fin}}_A,k^{\U{fin}}_B}
\notag\\
&-\sum_{\vec{a}\in\{0,1\}^{N_{\U{fin}}}}\frac{1}{2^{N_{\U{fin}}}}
\hat{P}[\ket{N_{\U{fin}},\vec{a}}_{A_{\U{sift}}}\ket{N_{\U{fin}},\vec{a}}_{B_{\U{sift}}}]
\otimes
\tr_{A_{\U{sift}}B_{\U{sift}}}[\hat{\rho}_{\U{PA}|N_{\U{fin}}}]
\Bigg|\Bigg|
\notag\\
=&\Bigg|\Bigg|
\hat{U}\Bigg[
\sum_{k^{\U{fin}}_A,k^{\U{fin}}_B\in\{0,1\}^{N_{\U{fin}}}}
\Pr(k^{\U{fin}}_A,k^{\U{fin}}_B|N_{\U{fin}})
\hat{P}[\ket{N_{\U{fin}},k^{\U{fin}}_A}_{A_{\U{sift}}}
\ket{N_{\U{fin}},k^{\U{fin}}_A}_{B_{\U{sift}}}]\otimes
\hat{\rho}^E_{\U{PA}|N_{\U{fin}},k^{\U{fin}}_A,k^{\U{fin}}_B}
\notag\\
&-\sum_{\vec{a}\in\{0,1\}^{N_{\U{fin}}}}\frac{1}{2^{N_{\U{fin}}}}
\hat{P}[\ket{N_{\U{fin}},\vec{a}}_{A_{\U{sift}}}\ket{N_{\U{fin}},\vec{a}}_{B_{\U{sift}}}]
\otimes
\tr_{A_{\U{sift}}B_{\U{sift}}}[\hat{\rho}_{\U{PA}|N_{\U{fin}}}]
\Bigg]\hat{U}^{\dagger}
\Bigg|\Bigg|
\notag\\
=&\Bigg|\Bigg|
\sum_{k^{\U{fin}}_A,k^{\U{fin}}_B\in\{0,1\}^{N_{\U{fin}}}}
\Pr(k^{\U{fin}}_A,k^{\U{fin}}_B|N_{\U{fin}})
\hat{P}[\ket{N_{\U{fin}},k^{\U{fin}}_A}_{A_{\U{sift}}}]\otimes
\hat{\rho}^E_{\U{PA}|N_{\U{fin}},k^{\U{fin}}_A,k^{\U{fin}}_B}
\notag\\
&-\sum_{\vec{a}\in\{0,1\}^{N_{\U{fin}}}}\frac{1}{2^{N_{\U{fin}}}}
\hat{P}[\ket{N_{\U{fin}},\vec{a}}_{A_{\U{sift}}}]
\otimes
\tr_{A_{\U{sift}}B_{\U{sift}}}[\hat{\rho}_{\U{PA}|N_{\U{fin}}}]
\Bigg|\Bigg|
\notag\\
=&
\Bigg|\Bigg|
\tr_{B_{\U{sift}}}[\hat{\rho}_{\U{PA}|N_{\U{fin}}}]-
\tr_{B_{\U{sift}}}[\hat{\rho}_{\U{ideal}|N_{\U{fin}}}]
\Bigg|\Bigg|
\notag\\
=&
||\hat{\rho}^{AE}_{\U{PA}|N_{\U{fin}}}-\hat{\rho}^{AE}_{\U{ideal}|N_{\U{fin}}}||.
\label{eq:triangle1}
\end{align}
The first equality is obtained by substituting the definitions in Eqs.~(\ref{eq:sigmaABENfin}) and (\ref{eq:PA3}). The second equality follows from the unitary-invariance property of the trace distance. The third equality follows by setting the unitary operation as
\[
\hat{U}=\sum_{k_A^{\U{fin}}\in\{0,1\}^{N_{\U{fin}}}}\hat{P}[\ket{N_{\U{fin}},k_A^{\U{fin}}}_{A_{\U{sift}}}]
\bigotimes_{i=1}^{N_{\U{fin}}}\hat{\sigma}^{(k_A^{\U{fin}})_i}_{X,B_{\U{sift}}}.
\]
The fourth equality follows by the definitions in Eqs.~(\ref{eq:PA2}) and (\ref{eq:PA3}). Finally, the fifth equality follows from Eqs.~(\ref{eq:PA1}) and (\ref{eq:rhoPAAENfinactual}).
\\\\
As for the second term of Eq.~(\ref{eq:bunkatu}), direct calculation leads to
\begin{align}
&||\hat{\rho}_{\U{PA}|N_{\U{fin}}}-\hat{\sigma}_{ABE|N_{\U{fin}}}||
\notag\\
=&
\Bigg|\Bigg|
\sum_{k^{\U{fin}}_A,k^{\U{fin}}_B\in\{0,1\}^{N_{\U{fin}}}}
\Pr(k^{\U{fin}}_A,k^{\U{fin}}_B|N_{\U{fin}})
\notag\\
&\left(
\hat{P}[\ket{N_{\U{fin}},k^{\U{fin}}_A}_{A_{\U{sift}}}
\ket{N_{\U{fin}},k^{\U{fin}}_B}_{B_{\U{sift}}}]
-
\hat{P}[\ket{N_{\U{fin}},k^{\U{fin}}_A}_{A_{\U{sift}}}
\ket{N_{\U{fin}},k^{\U{fin}}_A}_{B_{\U{sift}}}]
\right)
\otimes
\hat{\rho}^E_{\U{PA}|N_{\U{fin}},k^{\U{fin}}_A,k^{\U{fin}}_B}
\Bigg|\Bigg|
\notag\\
=&
\Bigg|\Bigg|
\sum_{k^{\U{fin}}_A,k^{\U{fin}}_B\in\{0,1\}^{N_{\U{fin}}}:k^{\U{fin}}_A\neq k^{\U{fin}}_B}
\Pr(k^{\U{fin}}_A,k^{\U{fin}}_B|N_{\U{fin}})
\notag\\
&\left(
\hat{P}[\ket{N_{\U{fin}},k^{\U{fin}}_A}_{A_{\U{sift}}}
\ket{N_{\U{fin}},k^{\U{fin}}_B}_{B_{\U{sift}}}]
-
\hat{P}[\ket{N_{\U{fin}},k^{\U{fin}}_A}_{A_{\U{sift}}}
\ket{N_{\U{fin}},k^{\U{fin}}_A}_{B_{\U{sift}}}]
\right)
\otimes
\hat{\rho}^E_{\U{PA}|N_{\U{fin}},k^{\U{fin}}_A,k^{\U{fin}}_B}
\Bigg|\Bigg|
\notag\\
=&
\Bigg|\Bigg|
\sum_{k^{\U{fin}}_A,k^{\U{fin}}_B\in\{0,1\}^{N_{\U{fin}}}:k^{\U{fin}}_A\neq k^{\U{fin}}_B}
\Pr(k^{\U{fin}}_A,k^{\U{fin}}_B|N_{\U{fin}})
\hat{P}[\ket{N_{\U{fin}},k^{\U{fin}}_A}_{A_{\U{sift}}}
\ket{N_{\U{fin}},k^{\U{fin}}_B}_{B_{\U{sift}}}]
\otimes
\hat{\rho}^E_{\U{PA}|N_{\U{fin}},k^{\U{fin}}_A,k^{\U{fin}}_B}
\Bigg|\Bigg|
\notag\\
+&
\Bigg|\Bigg|
\sum_{k^{\U{fin}}_A,k^{\U{fin}}_B\in\{0,1\}^{N_{\U{fin}}}:k^{\U{fin}}_A\neq k^{\U{fin}}_B}
\Pr(k^{\U{fin}}_A,k^{\U{fin}}_B|N_{\U{fin}})
\hat{P}[\ket{N_{\U{fin}},k^{\U{fin}}_A}_{A_{\U{sift}}}
\ket{N_{\U{fin}},k^{\U{fin}}_A}_{B_{\U{sift}}}]
\otimes
\hat{\rho}^E_{\U{PA}|N_{\U{fin}},k^{\U{fin}}_A,k^{\U{fin}}_B}
\Bigg|\Bigg|
\notag\\
=&
\sum_{k^{\U{fin}}_A,k^{\U{fin}}_B\in\{0,1\}^{N_{\U{fin}}}:k^{\U{fin}}_A\neq k^{\U{fin}}_B}
\Pr(k^{\U{fin}}_A,k^{\U{fin}}_B|N_{\U{fin}})
+\sum_{k^{\U{fin}}_A,k^{\U{fin}}_B\in\{0,1\}^{N_{\U{fin}}}:k^{\U{fin}}_A\neq k^{\U{fin}}_B}
\Pr(k^{\U{fin}}_A,k^{\U{fin}}_B|N_{\U{fin}})
\notag\\
=&2\Pr(k^{\U{fin}}_A\neq k^{\U{fin}}_B|N_{\U{fin}}).
\label{eq:triangle2}
\end{align}
The first equality follows from Eqs.~(\ref{eq:PA2}) and (\ref{eq:sigmaABENfin}). The third equality follows from 
\[
(\ket{N_{\U{fin}},k^{\U{fin}}_A}_{A_{\U{sift}}}
\ket{N_{\U{fin}},k^{\U{fin}}_B}_{B_{\U{sift}}})^{\dagger}
(\ket{N_{\U{fin}},k^{\U{fin}}_A}_{A_{\U{sift}}}
\ket{N_{\U{fin}},k^{\U{fin}}_A}_{B_{\U{sift}}})
=0
\]
for $k^{\U{fin}}_A\neq k^{\U{fin}}_B$. The third equality follows by $||\hat{X}||=\tr(\hat{X})$ for $\hat{X}\ge0$.
\\\\
Substituting Eqs.~(\ref{eq:triangle1}) and (\ref{eq:triangle2}) to 
Eq.~(\ref{eq:bunkatu}) results in
\begin{align}
\frac{1}{2}||\hat{\rho}_{\U{PA}}-\hat{\rho}_{\U{ideal}}||
\le
\frac{1}{2}\sum_{N_{\U{fin}}=0}^N\Pr(N_{\U{fin}})
||\hat{\rho}^{AE}_{\U{ideal}|N_{\U{fin}}}-\hat{\rho}^{AE}_{\U{PA}|N_{\U{fin}}}||
+\sum_{N_{\U{fin}}=0}^N\Pr(N_{\U{fin}})
\Pr(k^{\U{fin}}_A\neq k^{\U{fin}}_B|N_{\U{fin}}).
\end{align}
From this inequality, we can see that the proposition holds.
\\\\
Next, we state the following proposition regarding the correctness parameter $\epsilon_c$ defined in 
Eq.~(\ref{eq:correctness}).

\begin{proposition}
\label{prop_correct}
{\bf Derivation of the correctness parameter}\\
When $\epsilon_{\U{correct}}$ is defined as 
\begin{align}
\epsilon_{\U{correct}}:=2^{-N_{\U{verify}}},
\end{align}
Eq.~(\ref{eq:correctness}) holds. 
Recall that $N_{\U{verify}}$ represents the output bit 
length of the universal2 hash function used to verify the identity of the reconciled keys appearing in Step~3 of the key generation flowchart Fig.~\ref{fig:flowKD2}.
\end{proposition}

{\bf Proof of Proposition~\ref{prop_correct}}
This proof is based on the proof of Theorem~2 in~\cite{tomamichel2017}.
\begin{align}
\sum_{N_{\U{fin}}=0}^N\Pr(N_{\U{fin}})\Pr(k^{\U{fin}}_A\neq k^{\U{fin}}_B|N_{\U{fin}})
&=\sum_{N_{\U{fin}}=1}^N\Pr(N_{\U{fin}})\Pr(k^{\U{fin}}_A\neq k^{\U{fin}}_B|N_{\U{fin}})\\
&=\sum_{N_{\U{fin}}=1}^N\Pr(N_{\U{fin}})
\Pr(k^{\U{fin}}_A\neq k^{\U{fin}}_B,H_A=H_B|N_{\U{fin}})\\
&\le\sum_{N_{\U{fin}}=1}^N\Pr(N_{\U{fin}})
\Pr(\bm{k}_A\neq \bm{k}^{rec}_B,H_A=H_B|N_{\U{fin}})\\
&=\sum_{N_{\U{fin}}=1}^N\Pr(N_{\U{fin}})
\Pr(\bm{k}_A\neq \bm{k}^{rec}_B|N_{\U{fin}})
\Pr(H_A=H_B|N_{\U{fin}},\bm{k}_A\neq \bm{k}^{rec}_B)\\
&\le\sum_{N_{\U{fin}}=1}^N\Pr(N_{\U{fin}})
\Pr(H_A=H_B|N_{\U{fin}},\bm{k}_A\neq \bm{k}^{rec}_B)\\
&\le\sum_{N_{\U{fin}}=1}^N\Pr(N_{\U{fin}})2^{-N_{\U{verify}}}\\
&\le 2^{-N_{\U{verify}}}. 
\end{align}

The reason why each equation holds is explained below.

\begin{itemize}
\item
The first equality follows by 
$\Pr(k^{\U{fin}}_A\neq k^{\U{fin}}_B,N_{\U{fin}}=0)=0$.
\item
The second equality comes from the fact that 
the length of the secret key is zero when 
Alice's and Bob's hash values are different 
(namely, $H_A\neq H_B$), which can be seen from Alice's Step~3 
in the key generation flowchart Fig.~\ref{fig:flowKD2}. 
By contraposition, we can conclude that if $\forall N_{\U{fin}}\ge1$, $H_A=H_B$ holds. Therefore, 
we have 
$\Pr(k^{\U{fin}}_A\neq k^{\U{fin}}_B,H_A\neq H_B|N_{\U{fin}})=0$ 
for $\forall N_{\U{fin}}\ge1$. 
    \item  
    Let $\bm{k}^{\U{rec}}_B$ be the reconciled key after bit error correction in the sifted key that appears in Bob's Step~2 of the key generation flowchart Fig.~\ref{fig:flowKD1}. Then, if  
    $\bm{k}_A=\bm{k}^{\U{rec}}_B$, 
    $k^{\U{fin}}_A=k^{\U{fin}}_B$ holds. 
    By contraposition, we obtain 
    $\Pr(k^{\U{fin}}_A\neq k^{\U{fin}}_B,H_A=H_B|N_{\U{fin}})
    \le
    \Pr(\bm{k}_A\neq \bm{k}^{\U{rec}}_B,H_A=H_B|N_{\U{fin}})$, 
    which implies the first inequality. 
    \item  
        The third equality follows by Bayes' theorem. 
    \item  
    The second inequality is from 
    $\Pr(\bm{k}_A\neq \bm{k}^{\U{rec}}_B|N_{\U{fin}})\le1$. 
    \item  
    The third equality is satisfied by the definition of 
    the universal2 hash function. 
    \item  
    The fourth inequality follows because $\sum_{N_{\U{fin}}=1}^N\Pr(N_{\U{fin}})\le1$.
\end{itemize}

\subsection{Derivation of secrecy parameter}
\label{sec:dervationsecrecy}
In this section, we derive the secrecy parameter $\epsilon_{\U{secrecy}}$ in Eq.~(\ref{eq:secrecy}). 
In Sec.~\ref{sec:squash}, we rewrite Bob's measurement operator $\mathcal{E}_i^B$ in Eq.~(\ref{eq:bobEB}) using the squashing map to describe Bob's measurement in the virtual QKD protocol. In Sec.~\ref{subsec:virtualprotocol}, we introduce the virtual QKD protocol corresponding to the actual QKD protocol described in Sec.~\ref{sec:mathematicalQKDprotocol}. This virtual protocol is constructed such that Eve cannot discriminate the virtual protocol from the actual one and for any strategy Eve adopts, the probability distribution of Alice's secret key is equivalent to that of the actual protocol, as explained in Sec.~\ref{subsec:equivalence_actual_virtual}. 
Based on the virtual protocol, we calculate the trace distance in Eq.~(\ref{eq:secrecy}) in Sec.~\ref{subsec:mainprops}.

\subsubsection{Representation of Bob's measurement with squashing map}
\label{sec:squash}
The CPTP map $\mathcal{E}^B_i$ in Eq.~(\ref{eq:bobEB}) can be rewritten using the squashing map $\mathcal{E}^{\U{squash}}$, as shown in proposition~\ref{prop:squash}. 
Here, the squashing map~\cite{Beaudry2008,tsurumaru2008,Tsurumaru2010,fung2011,Gittsovich2014,Zhang2021,Upadhyaya2021,Nahar2025} 
is a CPTP map with the following property. $\mathcal{E}^{\U{squash}}$ maps the state of system $B_i^{\U{sig}}$ to a qubit state if Bob's measurement outcome is not ``No click", and after applying $\mathcal{E}^{\U{squash}}$, performing a measurement in the Pauli $Z$- or $X$-basis yields outcomes equivalent to those obtained by a measurement with $\mathcal{E}_i^B$. 
\begin{proposition}
\label{prop:squash}
There exists a CPTP map
\begin{align}
 \mathcal{E}^{\U{squash}}: L(\mathcal{H}_{B_i^{\U{sig}}}) \to 
 L(\U{span}\{\ket{0}_{B_i^{\U{bit}}},\ket{1}_{B_i^{\U{bit}}}\})\oplus L(
 \U{span}\{\ket{\U{Noclick}}_{B_i^{\U{bit}}}\}),
 \label{eq:Esquash}
\end{align}
and for any state $\hat{\rho}_{B_i^{\U{sig}}}$ of system $B_i^{\U{sig}}$, 
\begin{align}
\mathcal{E}_i^B(\hat{\rho}_{B_i^{\U{sig}}})
=&\sum_{\beta\in\{Z,X\}}p_{\beta}    
	\hat{P}[\ket{\U{Noclick}}_{B_i^{\U{bit}}}]
        \hat{P}[\ket{\beta}_{B_i^{\U{basis}}}]
        \tr\(
        \hat{P}[\ket{\U{Noclick}}_{B_i^{\U{bit}}}]
        \mathcal{E}^{\U{squash}}(\hat{\rho}_{B_i^{\U{sig}}})\)
        \notag\\
	&+\sum_{b\in\{0,1\},\beta\in\{Z,X\}}p_{\beta}
        \hat{P}[\ket{b}_{B_i^{\U{bit}}}]
        \hat{P}[\ket{\beta}_{B_i^{\U{basis}}}]
        \tr
        \(
        \frac{\hat{I}_2+(-1)^b\hat{\sigma}_{\beta}}{2}
        \mathcal{E}^{\U{squash}}(\hat{\rho}_{B_i^{\U{sig}}})\)
        \label{eq:MB}
\end{align}
holds. Here, 
$\hat{I}_2:=\hat{P}[\ket{0}]+\hat{P}[\ket{1}]+0\hat{P}[\ket{\U{Noclick}}]$, 
$\hat{\sigma}_Z:=\hat{P}[\ket{0}]-\hat{P}[\ket{1}]+0\hat{P}[\ket{\U{Noclick}}]$
and $\hat{\sigma}_X:=\ket{0}\bra{1}+\ket{1}\bra{1}+0\hat{P}[\ket{\U{Noclick}}]$. 
The explicit form of the squashing map $\mathcal{E}^{\U{squash}}$ is given in Eq.~(\ref{eq:squashnew}).
\end{proposition}
{\bf Proof of Proposition \ref{prop:squash}}\\
The squashing map for detectors with dark counts and detection efficiency is constructed in Sec.~III A of~\cite{Nahar2025}, which is essentially equivalent to $\mathcal{E}^{\U{squash}}$ we construct here. 
However, for completeness of this paper, we explicitly present the construction of the squashing map below. 
\\
{\bf Step~1. Squashing map for the POVM with ideal threshold detectors}
\\
We first construct the POVM $\{\hat{E}^{\U{meas1},\beta,i}_{b}\}_{b\in\{0,1,\U{Noclick}\}}$, which corresponds to obtaining measurement outcome $b\in\{0,1,\U{Noclick}\}$ with basis $\beta\in\{Z,X\}$ using ideal threshold detectors. Here, an ``ideal threshold detector" refers to a threshold detector with unit detection efficiency and zero dark count probability. Then, we show that there exists a squashing map for this POVM, derived from the known argument~\cite{Tsurumaru2010}.

In the next step~2, we relate $\hat{E}^{\U{meas1},\beta,i}_b$ to the actual POVM $\hat{E}^{\U{meas},\beta,i}_b$ defined in Eq.~(\ref{eq:bobpovmlast}) and construct the squashing map for $\hat{E}^{\U{meas},\beta,i}_b$.
\\\\
Let $\{\hat{E}^{\U{IdealDetector},A}_{\U{Click}}, \hat{E}^{\U{IdealDetector},A}_{\U{Noclick}} \}$ denote the POVM of the ideal threshold detector for system $A$:
\begin{align}
    \hat{E}^{\U{IdealDetector},A}_{\U{Noclick}} = \hat{P}\left[\ket{\U{vac}}_A\right],~~ 
    \hat{E}^{\U{IdealDetector},A}_{\U{Click}} = \hat{I}_A-\hat{P}\left[\ket{\U{vac}}_A\right].
                \label{eq:detector1E1}
\end{align}
Let us consider a setup where a BS (beam splitter) and a PS (phase shifter) are placed before two ideal threshold detectors, with systems $B^{\U{sig1}}_i$ and $B^{\U{sig1}}_i$ as the input to this setup. The POVM elements associated with detection by the detector corresponding to bits 0 and 1 are given by
\begin{align}
    \hat{E}^{\U{detect1},\beta,B^{\U{sig1}}_i}_{\U{Click}}:=& 
    \left(\mathcal{E}^{\U{PS},B^{\U{sig1}}_i\dagger}_{\theta_\beta}
    \otimes \hat{I}_{B^{\U{sig2}}_i}
    \right)
    \mathcal{E}^{\U{BS},B^{\U{sig1}}_iB^{\U{sig2}}_i\dagger}_{\pi/4}
   (\hat{E}^{\U{IdealDetector},B^{\U{sig1}}_i}_{\U{Click}}\otimes \hat{I}_{B^{\U{sig2}}_i})
       \label{eq:proofideal1}
\end{align}
and
\begin{align}
    \hat{E}^{\U{detect1},\beta,B^{\U{sig2}}_i}_{\U{Click}}
    :=& 
    \left(\mathcal{E}^{\U{PS},B^{\U{sig1}}_i\dagger}_{\theta_\beta}
    \otimes \hat{I}_{B^{\U{sig2}}_i}
    \right)
   \mathcal{E}^{\U{BS},B^{\U{sig1}}_iB^{\U{sig2}}_i\dagger}_{\pi/4}(\hat{I}_{B^{\U{sig1}}_i}\otimes\hat{E}^{\U{IdealDetector},B^{\U{sig2}}_i}_{\U{Click}} ),
\end{align}
respectively. Also, the POVM elements associated with no detection by the detector, corresponding to bit $b\in\{0,1\}$ are denoted by
\begin{align}
    \hat{E}^{\U{detect1},\beta,B^{\U{sig}b}_i}_{\U{Noclick}}
    := \hat{I}_{B^{\U{sig1}}_i}\otimes\hat{I}_{B^{\U{sig2}}_i}-\hat{E}^{\U{detect1},\beta,B^{\U{sig}b}_i}_{\U{Click}}.
        \label{eq:proofideal2}
\end{align}
Using Eqs.~(\ref{eq:proofideal1})-(\ref{eq:proofideal2}), the POVM element $\hat{E}^{\U{meas1},\beta,i}_{b}$ associated with obtaining measurement outcome $b\in\{0,1,\U{Noclick}\}$ is given by
\begin{equation}
     \hat{E}^{\U{meas1},\beta,i}_{b}:=
    \begin{cases}
        \hat{E}^{\U{detect1},\beta,B^{\U{sig1}}_i}_{\U{Click}}
        \hat{E}^{\U{detect1},\beta,B^{\U{sig2}}_i}_{\U{Noclick}}
        +\frac{1}{2}\hat{E}^{\U{detect1},\beta,B^{\U{sig1}}_i}_{\U{Click}}
        \hat{E}^{\U{detect1},\beta,B^{\U{sig2}}_i}_{\U{Click}} & b=0\\
        \hat{E}^{\U{detect1},\beta,B^{\U{sig1}}_i}_{\U{Noclick}}
        \hat{E}^{\U{detect1},\beta,B^{\U{sig2}}_i}_{\U{Click}}
        +\frac{1}{2}\hat{E}^{\U{detect1},\beta,B^{\U{sig1}}_i}_{\U{Click}}
        \hat{E}^{\U{detect1},\beta,B^{\U{sig2}}_i}_{\U{Click}} & b=1\\
        \hat{E}^{\U{detect1},\beta,B^{\U{sig1}}_i}_{\U{Noclick}}
        \hat{E}^{\U{detect1},\beta,B^{\U{sig2}}_i}_{\U{Noclick}} & b=\U{Noclick}.
    \end{cases}
\label{eq:Emeas1}
\end{equation}
From this definition, $\{\hat{E}^{\U{meas1},\beta,i}_{b}\}_{\beta,b}$ satisfies
\begin{equation}
   \mathcal{E}^{\U{change}\dagger}
    (\hat{E}^{\U{meas1},Z,i}_{b})
     =\hat{E}^{\U{meas1},X,i}_{b}
     \label{eq:squash-cond1}
\end{equation}
and
\begin{equation}
   \mathcal{E}^{\U{change}\dagger}\circ
   \mathcal{E}^{\U{change}\dagger}
    (\hat{E}^{\U{meas1},\beta,i}_{b})
     =\hat{E}^{\U{meas1},\beta,i}_{b\oplus1}
\label{eq:squash-cond2}
\end{equation}
for $\beta\in\{Z,X\}$ and $b\in\{0,1\}$, where
\begin{equation}
    \mathcal{E}^{\U{change}}
    :=\mathcal{E}^{\U{PS},B^{\U{sig1}}_i}_{\pi/2}
    \otimes \hat{I}_{B^{\U{sig2}}_i}.
\end{equation}
Let $\hat{\Pi}^{\U{photon},i}_{\vec{n}}$ denote a projection onto the subspace where the $k$th optical mode has $n_k$ photons:
\begin{align}
    \hat{\Pi}^{\U{photon},i}_{\vec{n}} := \sum_{\vec{m}, 0\leq m_k\leq n_k} \hat{P}\left[\ket{\vec{m}}_{B^{\U{sig1}}_i}\ket{\vec{n}-\vec{m}}_{B^{\U{sig2}}_i}\right],
   \label{eq:defPin}
\end{align}
where $\ket{\vec{m}}$ is a state that the $k$th mode has $m_k$ photons. 

Then, consider that
\begin{align}
    &\mathcal{E}^{\U{BS},B^{\U{sig1}}_iB^{\U{sig2}}_i\dagger}_{-\frac{\pi}{4}}\circ\left(\mathcal{E}^{\U{PS},B^{\U{sig1}\dagger}_i}_{-\pi/2}
    \otimes \hat{I}_{B^{\U{sig2}}_i}
    \right)
    \(\hat{\Pi}^{\U{photon},i}_{\vec{n}} (\hat{E}^{\U{meas1},Z,i}_{0} - \hat{E}^{\U{meas1},Z,i}_{1})
     \hat{\Pi}^{\U{photon},i}_{\vec{n}}\)\\
     =&
     \mathcal{E}^{\U{BS},B^{\U{sig1}}_iB^{\U{sig2}}_i\dagger}_{-\frac{\pi}{4}}\circ\left(\mathcal{E}^{\U{PS},B^{\U{sig1}\dagger}_i}_{-\pi/2}
    \otimes \hat{I}_{B^{\U{sig2}}_i}
    \right)\notag\\
    &\left( \hat{\Pi}^{\U{photon},i}_{\vec{n}}\left(
     \hat{E}^{\U{detect1},\beta,B^{\U{sig1}}_i}_{\U{Click}}
        \hat{E}^{\U{detect1},\beta,B^{\U{sig2}}_i}_{\U{Noclick}}
        -\hat{E}^{\U{detect1},\beta,B^{\U{sig1}}_i}_{\U{Noclick}}
        \hat{E}^{\U{detect1},\beta,B^{\U{sig2}}_i}_{\U{Click}}
        \right) \hat{\Pi}^{\U{photon},i}_{\vec{n}}\right)
        \\
         =&
         \hat{\Pi}^{\U{photon},i}_{\vec{n}}\notag\\
&\Bigg\{
\mathcal{E}^{\U{BS},B^{\U{sig1}}_iB^{\U{sig2}}_i\dagger}_{-\frac{\pi}{4}}\circ\left(\mathcal{E}^{\U{PS},B^{\U{sig1}\dagger}_i}_{-\pi/2}\otimes \hat{I}_{B^{\U{sig2}}_i}\right)(\hat{E}^{\U{detect1},\beta,B^{\U{sig1}}_i}_{\U{Click}})   
      \mathcal{E}^{\U{BS},B^{\U{sig1}}_iB^{\U{sig2}}_i\dagger}_{-\frac{\pi}{4}}\circ\left(\mathcal{E}^{\U{PS},B^{\U{sig1}\dagger}_i}_{-\pi/2}\otimes \hat{I}_{B^{\U{sig2}}_i}\right)(\hat{E}^{\U{detect1},\beta,B^{\U{sig2}}_i}_{\U{Noclick}})     
     \notag\\
    -&
    \mathcal{E}^{\U{BS},B^{\U{sig1}}_iB^{\U{sig2}}_i\dagger}_{-\frac{\pi}{4}}\circ\left(\mathcal{E}^{\U{PS},B^{\U{sig1}\dagger}_i}_{-\pi/2}\otimes \hat{I}_{B^{\U{sig2}}_i}\right)(\hat{E}^{\U{detect1},\beta,B^{\U{sig1}}_i}_{\U{Noclick}})   
      \mathcal{E}^{\U{BS},B^{\U{sig1}}_iB^{\U{sig2}}_i\dagger}_{-\frac{\pi}{4}}\circ\left(\mathcal{E}^{\U{PS},B^{\U{sig1}\dagger}_i}_{-\pi/2}\otimes \hat{I}_{B^{\U{sig2}}_i}\right)(\hat{E}^{\U{detect1},\beta,B^{\U{sig2}}_i}_{\U{Click}})
    \notag\\
&    \Bigg\}\hat{\Pi}^{\U{photon},i}_{\vec{n}}
\\
=&        
       \hat{\Pi}^{\U{photon},i}_{\vec{n}}\left(
       \hat{E}^{\U{IdealDetector},B^{\U{sig1}}_i}_{\U{Click}}\hat{E}^{\U{IdealDetector},B^{\U{sig2}}_i}_{\U{Noclick}}-
        \hat{E}^{\U{IdealDetector},B^{\U{sig1}}_i}_{\U{Noclick}}\hat{E}^{\U{IdealDetector},B^{\U{sig2}}_i}_{\U{Click}}
       \right)\hat{\Pi}^{\U{photon},i}_{\vec{n}}
\\        
     =& 
     \hat{P}\left[\ket{\vec{n}}_{B^{\U{sig1}}_i}\ket{\U{vac}}_{B^{\U{sig2}}_i}\right] 
     -\hat{P}\left[\ket{\U{vac}}_{B^{\U{sig1}}_i}\ket{\vec{n}}_{B^{\U{sig2}}_i}\right].
     \label{eq:squash-cond3}
\end{align}
The first equation follows from  Eq.~(\ref{eq:Emeas1}). The second equation follows from the commutativity of 
$[\hat{\Pi}^{\U{photon},i}_{\vec{n}}, \mathcal{E}^{\U{BS},B^{\U{sig1}}_{i}B^{\U{sig2}}_i\dagger}_{-\frac{\pi}{4}}]=
[\hat{\Pi}^{\U{photon},i}_{\vec{n}}, \mathcal{E}^{\U{PS},B^{\U{sig1}\dagger}_i}_{-\pi/2}]=0$ and the fact that 
$\mathcal{E}^{\U{BS},B^{\U{sig1}}_{i}B^{\U{sig2}}_i\dagger}_{-\frac{\pi}{4}}$ and $\mathcal{E}^{\U{PS},B^{\U{sig1}}_i}_{-\pi/2}$ are unitary. The third equation follows by Eqs.~(\ref{eq:proofideal1})-(\ref{eq:proofideal2}). 
The final equation comes from Eqs.~(\ref{eq:detector1E1}) and (\ref{eq:defPin}).

Equation~(\ref{eq:squash-cond3}) and the reversibility of $\mathcal{E}^{\U{PS},B^{\U{sig1}\dagger}_i}_{\theta}$ and $\mathcal{E}^{\U{BS},B^{\U{sig1}}_i,B^{\U{sig2}}_i\dagger}_{-\pi/4}$ 
\footnote{
The map $\mathcal{E}$ is reversible if an inverse map exists. Since 
$\mathcal{E}^{\U{PS},B^{\U{sig1}}_i}_{\theta}$ and $\mathcal{E}^{\U{BS},B^{\U{sig1}}_i,B^{\U{sig2}}_i}_{-\frac{\pi}{4}}$ 
are unitary maps, an inverse map always exists.}
imply that the rank of the operator
\[
\hat{\Pi}^{\U{photon},i}_{\vec{n}} (\hat{E}^{\U{meas1},Z,i}_{0} - \hat{E}^{\U{meas1},Z,i}_{1})
     \hat{\Pi}^{\U{photon},i}_{\vec{n}}
     \]
     is two. Furthermore, Eqs.~(\ref{eq:squash-cond1}) and (\ref{eq:squash-cond2}) imply that 
     \[
     \{\hat{\Pi}^{\U{photon},i}_{\vec{n}} \hat{E}^{\U{meas1},\beta,i}_{b}\hat{\Pi}^{\U{photon},i}_{\vec{n}}\}_{b\in\{0,1\},\beta\in\{Z,X\}}
     \]
     is $C_4$ symmetric (a cyclic group of order 4). Therefore, $\{\hat{\Pi}^{\U{photon},i}_{\vec{n}} \hat{E}^{\U{meas1},\beta,i}_{b}\hat{\Pi}^{\U{photon},i}_{\vec{n}}\}_{b,\beta}$ satisfies the precondition of Theorem 1 in \cite{Tsurumaru2010}. It means that there exists a squashing map $\mathcal{F}_{\vec{n}}$ for any $\vec{n} \in \{\vec{n} \mid  n_k\in \mathbb{N}^{\ge 0}, \vec{n}\neq \vec{0}\}$\footnote{Note that $\mathbb{N}^{\ge 0}$ represents the set of non-negative integers.} satisfying
\begin{equation}
    \mathcal{F}_{\vec{n}}^\dagger (\hat{\sigma}_{\beta})
    = \hat{\Pi}^{\U{photon},i}_{\vec{n}} (\hat{E}^{\U{meas1},\beta,i}_{0} - \hat{E}^{\U{meas1},\beta,i}_{1})
     \hat{\Pi}^{\U{photon},i}_{\vec{n}}.
\end{equation}
Here, $\mathbb{N}^{\ge 0}$ denotes the set of non-negative integers, and 
$\vec{0}$ is a vector that all elements are zero, which satisfies $\ket{\vec{0}} = \ket{\U{vac}}$.
By defining the operation
\begin{align}
\mathcal{F}^{\U{squash}}: L(\mathcal{H}_{B_i^{\U{sig}}}) \to 
 L(\U{span}\{\ket{0}_{B_i^{\U{bit}}},\ket{1}_{B_i^{\U{bit}}}\})\oplus L(
 \U{span}\{\ket{\U{Noclick}}_{B_i^{\U{bit}}}\}),
\end{align}
\begin{align}
        \mathcal{F}^{\U{squash}}(\hat{\rho})
        :=\tr\left(\hat{\Pi}^{\U{photon},i}_{\vec{0}}\hat{\rho}\right)\hat{P}\left[\ket{\U{Noclick}}_{B^{\U{bit}}_i}\right]+
        \sum_{\vec{n}\neq \vec{0},n_k\in \mathbb{N}^{\ge0}} \mathcal{F}_{\vec{n}}(\hat{\Pi}^{\U{photon},i}_{n}\hat{\rho}\hat{\Pi}^{\U{photon},i}_{\vec{n}}),
\end{align}
it satisfies
\begin{equation}
\begin{split}
   & \sum_{\beta\in\{Z,X\},b\in\{0,1,\U{Noclick}\}}p_{\beta}
   \hat{P}\left[\ket{\beta}_{B^{\U{basis}}_i}\ket{b}_{B^{\U{bit}}_i} \right]
    \tr_{B^{\U{sig1}}_iB^{\U{sig2}}_i} \left(\hat{E}^{\U{meas1},\beta,i}_{b} 
    \hat{\rho}\right) \\
    =&\sum_{\beta\in\{Z,X\}}p_{\beta}    
	\hat{P}[\ket{\U{Noclick}}_{B_i^{\U{bit}}}]
        \hat{P}[\ket{\beta}_{B_i^{\U{basis}}}]
        \tr\(
        \hat{P}[\ket{\U{Noclick}}_{B_i^{\U{bit}}}]
        \mathcal{F}^{\U{squash}}(\hat{\rho}_{B_i^{\U{sig}}})\)\\
	&+\sum_{b\in\{0,1\},\beta\in\{Z,X\}}p_{\beta}
        \hat{P}[\ket{b}_{B_i^{\U{bit}}}]
        \hat{P}[\ket{\beta}_{B_i^{\U{basis}}}]
        \tr
        \(
        \frac{\hat{I}_2+(-1)^b\hat{\sigma}_{\beta}}{2}
        \mathcal{F}^{\U{squash}}(\hat{\rho}_{B_i^{\U{sig}}})\) .  
\end{split}
\label{eq:orig-squash}
\end{equation}

{\bf Step~2. Squashing map for the POVM with actual threshold detectors}
\\
We relate $\hat{E}^{\U{meas1},\beta,i}_{b}$ defined in Eq.~(\ref{eq:Emeas1}) to the actual POVM $\hat{E}^{\U{meas},\beta,i}_{b}$ defined in Eq.~(\ref{eq:bobpovmlast}). For this, 
let $\{\hat{E}^{\U{actual},i}_{b}\}_{b\in\{0,1,\U{Noclick}\}}$ denote a POVM:
\begin{align}
    \hat{E}^{\U{actual},i}_{0} &:= \hat{E}^{\U{detector1},B^{\U{sig1}}_i}_{\U{Click}}\otimes  \hat{E}^{\U{detector1},B^{\U{sig2}}_i}_{\U{Noclick}}
    +\frac{1}{2} \hat{E}^{\U{detector1},B^{\U{sig1}}_i}_{\U{Click}}\otimes \hat{E}^{\U{detector1},B^{\U{sig2}}_i}_{\U{Click}}
    \label{Etmp20}
    \\
    \hat{E}^{\U{actual},i}_{1} &:= \hat{E}^{\U{detector1},B^{\U{sig1}}_i}_{\U{Noclick}}\otimes  \hat{E}^{\U{detector1},B^{\U{sig2}}_i}_{\U{Click}}
    +\frac{1}{2} \hat{E}^{\U{detector1},B^{\U{sig1}}_i}_{\U{Click}}\otimes \hat{E}^{\U{detector1},B^{\U{sig2}}_i}_{\U{Click}}\\
    \hat{E}^{\U{actual},i}_{\U{Noclick}} &:= \hat{E}^{\U{detector1},B^{\U{sig1}}_i}_{\U{Noclick}}\otimes  \hat{E}^{\U{detector1},B^{\U{sig2}}_i}_{\U{Noclick}},
    \label{Etmp2noclick}
\end{align}
where $\hat{E}^{\U{detector1},B}_{\U{Noclick}}$ and $\hat{E}^{\U{detector1},B}_{\U{Click}}$ are defined in Eqs.~(\ref{eq:detector2E1noclick}) and (\ref{eq:detector2E1}), respectively, and 
$\hat{E}^{\U{meas},\beta,i}_{b}$ can be written as
\begin{align}
    \hat{E}^{\U{meas},\beta,i}_{b}
    =&\left(\mathcal{E}^{\U{pLoss},B^{\U{sig1}}_i\dagger}_{1/2}
    \otimes \mathcal{E}^{\U{pLoss},B^{\U{sig2}}_i\dagger}_{1/2}\right)\circ\notag\\
    &\left(\mathcal{E}^{\U{PS},B^{\U{sig1}}_i\dagger}_{\theta_\beta}
    \otimes \hat{I}_{B^{\U{sig2}}_i}
    \right)\circ
    \mathcal{E}^{\U{BS},B^{\U{sig1}}_i,B^{\U{sig2}}_i\dagger}_{\pi/4} \circ \left(\mathcal{E}^{\U{pLoss},B^{\U{sig1}}_i\dagger}_{\eta_{\U{det}}}\otimes \mathcal{E}^{\U{pLoss},B^{\U{sig2}}_i\dagger}_{\eta_{\U{det}}}\right)
    (\hat{E}^{\U{actual},i}_{b}).
\label{Ebmeasrewritten}
\end{align}
Also, the POVM element $\hat{E}^{\U{meas1},\beta,i}_{b}$, defined in Eq.~(\ref{eq:Emeas1}), is rewritten as
\begin{align}
    \hat{E}^{\U{meas1},\beta,i}_{b}=\left(\mathcal{E}^{\U{PS},B^{\U{sig1}}_i\dagger}_{\theta_\beta}
    \otimes \hat{I}_{B^{\U{sig2}}_i}
    \right) \circ
        \mathcal{E}^{\U{BS},B^{\U{sig1}}_i,B^{\U{sig2}}_i\dagger}_{\pi/4}(E^{\U{ideal},i}_{b})
    \label{meas1tmp1}
\end{align}
with a POVM $\{\hat{E}^{\U{ideal},i}_{b}\}_{b\in\{0,1,\U{Noclick}\}}$: 
\begin{align}
    \hat{E}^{\U{ideal},i}_{0} &:= \hat{E}^{\U{IdealDetector},B^{\U{sig1}}_i}_{\U{Click}}\otimes  \hat{E}^{\U{IdealDetector},B^{\U{sig2}}_i}_{\U{Noclick}}
    +\frac{1}{2} \hat{E}^{\U{IdealDetector},B^{\U{sig1}}_i}_{\U{Click}}\otimes \hat{E}^{\U{IdealDetector},B^{\U{sig2}}_i}_{\U{Click}},
        \label{Etmp10}
    \\
    \hat{E}^{\U{ideal},i}_{1} &:= \hat{E}^{\U{IdealDetector},B^{\U{sig1}}_i}_{\U{Noclick}}\otimes  \hat{E}^{\U{IdealDetector},B^{\U{sig2}}_i}_{\U{Click}}
    +\frac{1}{2} \hat{E}^{\U{IdealDetector},B^{\U{sig1}}_i}_{\U{Click}}\otimes \hat{E}^{\U{IdealDetector},B^{\U{sig2}}_i}_{\U{Click}},
            \label{Etmp11}
    \\
    \hat{E}^{\U{ideal},i}_{\U{Noclick}} &:= \hat{E}^{\U{IdealDetector},B^{\U{sig1}}_i}_{\U{Noclick}}\otimes  \hat{E}^{\U{IdealDetector},B^{\U{sig2}}_i}_{\U{Noclick}}.
            \label{Etmp1noclick}
\end{align}

From Eqs.~(\ref{Etmp20})-(\ref{Etmp2noclick}) and Eqs.~(\ref{Etmp10})-(\ref{Etmp1noclick}), 
$\hat{E}^{\U{ideal},i}_b$ and $\hat{E}^{\U{actual},i}_b$ are related as
\begin{align}
    \hat{E}^{\U{actual},i}_b
    =& \left(1-\frac{p_{\U{dark}}}{2}\right) \hat{E}^{\U{ideal},i}_b
    +\frac{p_{\U{dark}}}{2} \hat{E}^{\U{ideal},i}_{b\oplus1} 
    + \left(p_{\U{dark}}-\frac{p_{\U{dark}}^2}{2}\right) \hat{E}^{\U{ideal},i}_{\U{Noclick}}~~~b\in\{0,1\}
\label{eq:Etmp2b}
\\
    \hat{E}^{\U{actual},i}_{\U{Noclick}}
    =& (1-p_{\U{dark}})^2 \hat{E}^{\U{ideal},i}_{\U{Noclick}}.
    \label{eq:Etmp2noclick}
\end{align}
Substituting Eqs.~(\ref{eq:Etmp2b}) and (\ref{eq:Etmp2noclick}) into the right-hand side of Eq.~(\ref{Ebmeasrewritten}), and then using Eq.~(\ref{meas1tmp1}), along with the fact that the loss map $\mathcal{E}^{\U{pLoss},B^{\U{sig1}}_i}_{\eta}\circ\mathcal{E}^{\U{pLoss},B^{\U{sig2}}_i}_{\eta}$ commutes with the beam splitter and the phase shifter, gives the relationship between $\hat{E}^{\U{meas},\beta,i}_b$ and $\hat{E}^{\U{meas1},\beta,i}_b$ as
\begin{align}
       \hat{E}^{\U{meas},\beta,i}_b
    =\left(1-\frac{p_{\U{dark}}}{2}\right) \hat{E}^{\U{meas1'},\beta,i}_b
    +\frac{p_{\U{dark}}}{2} \hat{E}^{\U{meas1'},\beta,i}_{b\oplus1} 
    + \left(p_{\U{dark}}-\frac{p_{\U{dark}}^2}{2}\right) \hat{E}^{\U{meas1'},\beta,i}_{\U{Noclick}},\;\; b\in\{0,1\}
\label{eq:E-dark-meas1}
\end{align}
and
\begin{align}
    \hat{E}^{\U{meas},\beta,i}_{\U{Noclick}}
    = (1-p_{\U{dark}})^2 \hat{E}^{\U{meas1'},\beta,i}_{\U{Noclick}},
    \label{eq:E-dark-meas2}
\end{align}
where
\begin{align}
    \hat{E}^{\U{meas1'},\beta,i}_{b} := \mathcal{E}^{\U{pLoss},B^{\U{sig1}}_i\dagger}_{\frac{\eta_{\U{det}}}{2}}\otimes \mathcal{E}^{\U{pLoss},B^{\U{sig2}}_i\dagger}_{\frac{\eta_{\U{det}}}{2}} \left(\hat{E}^{\U{meas1},\beta,i}_{b} \right).
    \label{eq:def-meas1dash}
\end{align}
For instance, Eq.~(\ref{eq:E-dark-meas1}) means that the POVM element $\hat{E}^{\U{meas},\beta,i}_0$ that gives the outcome 
$b=0$ in the actual measurement is associated with the following process in a measurement 
$\{\hat{E}^{\U{meas1'},\beta,i}_{b'}\}_{b'\in\{0,1,\U{Noclick}\}}$ using ideal detectors:

- With probability $1-\frac{p_{\U{dark}}}{2}$, set $b=0$ when the outcome $b'=0$ is obtained.  

- With probability $\frac{p_{\U{dark}}}{2}$, set $b=0$ when the outcome $b'=1$ is obtained.  

- With probability $p_{\U{dark}}$, set $b=0$ when the outcome $b'=\U{Noclick}$ is obtained.  
\\

The effect of the dark count probability $p_{\U{dark}}$ can be described by the following CPTP map on the state $\hat{\rho}$ after applying the squashing map to $\hat{E}^{\U{meas1'},\beta,i}_b$:
\begin{align}
    \mathcal{E}^{\U{dark},i}(\hat{\rho}) :=\sum_{k=1}^5 \hat{K}^{\U{dark},i}_k\hat{\rho}\hat{K}^{\U{dark},i\dagger}_k
\end{align}
with the Kraus operators being
\begin{align}
    \hat{K}^{\U{dark},i}_1
    = \sqrt{1-\frac{p_{\U{dark}}}{2}}\left(\hat{P}\left[\ket{0}_{B^{\U{bit}}_i}\right]+\hat{P}\left[\ket{1}_{B^{\U{bit}}_i}\right]  \right)
    \label{eq:E-dark-1}
\end{align}
\begin{align}
    \hat{K}^{\U{dark},i}_2
    = \sqrt{\frac{p_{\U{dark}}}{2}}\left(\U{i}\ket{0}\bra{1}_{B^{\U{bit}}_i}-\U{i}\ket{1}\bra{0}_{B^{\U{bit}}_i}  \right)
    \label{eq:E-dark-2}
\end{align}
\begin{align}
    \hat{K}^{\U{dark},i}_3
    = \sqrt{p_{\U{dark}}-\frac{p_{\U{dark}}^2}{2}}\ket{0}\bra{\U{Noclick}}_{B^{\U{bit}}_i} 
    \label{eq:E-dark-3}
\end{align}
\begin{align}
    \hat{K}^{\U{dark},i}_4
    = \sqrt{p_{\U{dark}}-\frac{p_{\U{dark}}^2}{2}}\ket{1}\bra{\U{Noclick}}_{B^{\U{bit}}_i}
    \label{eq:E-dark-4}
\end{align}
\begin{align}
    \hat{K}^{\U{dark},i}_5
    = (1-p_{\U{dark}})\hat{P}\left[\ket{\U{Noclick}}_{B^{\U{bit}}_i}\right].
    \label{eq:E-dark-5}
\end{align}
These Kraus operators are constructed from Eqs.~(\ref{eq:E-dark-meas1}) and (\ref{eq:E-dark-meas2}) and represent how the outcome $b\in\{0,1,\U{Noclick}\}$ obtained using the actual detector is generated after mapping to the qubit using the ideal detector. For example, when the measurement outcome after mapping to the qubit with the ideal detector is $b'=0$, 1, or Noclick, 
it is converted to $b=0$ with probabilities $1-p_{\U{dark}}/2$, $p_{\U{dark}}/2$, and $p_{\U{dark}}-p^2_{\U{dark}}/2$, respectively. 
These transitions correspond to Eq.~(\ref{eq:E-dark-1}), the first term of Eq.~(\ref{eq:E-dark-2}), and Eq.~(\ref{eq:E-dark-3}), 
respectively. 

From the definitions of these Kraus operators, we have
\begin{align}
   \mathcal{E}^{\U{dark},i\dagger} \left(\frac{\hat{I}_{B_i^{\U{bit}}}+(-1)^b\hat{\sigma}_{\beta}}{2}\right)
    =&\left(1-\frac{p_{\U{dark}}}{2}\right)\left(\frac{\hat{I}_{B_i^{\U{bit}}}+(-1)^b\hat{\sigma}_{\beta}}{2}\right)
    +\frac{p_{\U{dark}}}{2}\left(\frac{\hat{I}_{B_i^{\U{bit}}}+(-1)^{b\oplus1}\hat{\sigma}_{\beta}}{2}\right)\notag\\
   +&\left(p_{\U{dark}}-\frac{p_{\U{dark}}^2}{2}\right)\hat{P}\left[\ket{\U{Noclick}}_{B^{\U{bit}}_i}\right]
\label{eq:dark-qubit1}
\end{align}
and
\begin{align}
    \mathcal{E}^{\U{dark},i\dagger} \left(\hat{P}\left[\ket{\U{Noclick}}_{B^{\U{bit}}_i}\right]\right)
    = (1-p_{\U{dark}})^2 \hat{P}\left[\ket{\U{Noclick}}_{B^{\U{bit}}_i}\right].
    \label{eq:dark-qubit2}
\end{align}
By setting $\mathcal{E}^{\U{squash}}$ as
\begin{align}
    \mathcal{E}^{\U{squash}} = \mathcal{E}^{\U{dark},i}\circ \mathcal{F}^{\U{squash}}\circ\left(\mathcal{E}^{\U{pLoss},B^{\U{sig1}}_i\dagger}_{\frac{\eta_{\U{det}}}{2}}\otimes \mathcal{E}^{\U{pLoss},B^{\U{sig2}}_i\dagger}_{\frac{\eta_{\U{det}}}{2}}\right),
    \label{eq:squashnew}
\end{align}
we have
\begin{align}
    &\sum_{\beta\in\{Z,X\}}p_{\beta}    
	\hat{P}[\ket{\U{Noclick}}_{B_i^{\U{bit}}}]
        \hat{P}[\ket{\beta}_{B_i^{\U{basis}}}]
        \tr\(
        \hat{P}[\ket{\U{Noclick}}_{B_i^{\U{bit}}}]
        \mathcal{E}^{\U{squash}}(\hat{\rho}_{B_i^{\U{sig}}})\)
        \notag\\
	&+\sum_{b\in\{0,1\},\beta\in\{Z,X\}}p_{\beta}
        \hat{P}[\ket{b}_{B_i^{\U{bit}}}]
        \hat{P}[\ket{\beta}_{B_i^{\U{basis}}}]
        \tr
        \(
        \frac{\hat{I}_2+(-1)^b\hat{\sigma}_{\beta}}{2}
        \mathcal{E}^{\U{squash}}(\hat{\rho}_{B_i^{\U{sig}}})\)
                \\
    =& \sum_{\beta\in\{Z,X\}}p_{\beta}    
	\hat{P}[\ket{\U{Noclick}}_{B_i^{\U{bit}}}]
        \hat{P}[\ket{\beta}_{B_i^{\U{basis}}}]
        (1-p_{\U{dark}})^2
        \tr\(
        \hat{E}^{\U{meas1'},\beta,i}_{\U{Noclick}}
        \hat{\rho}_{B_i^{\U{sig}}}\)
                \notag\\
	&+\sum_{b\in\{0,1\},\beta\in\{Z,X\}}p_{\beta}
        \hat{P}[\ket{b}_{B_i^{\U{bit}}}]
        \hat{P}[\ket{\beta}_{B_i^{\U{basis}}}]
        \left(1-\frac{p_{\U{dark}}}{2}\right)
        \tr
        \(
        \hat{E}^{\U{meas1'},\beta,i}_b
        \hat{\rho}_{B_i^{\U{sig}}}\)
                \notag\\
        &+\sum_{b\in\{0,1\},\beta\in\{Z,X\}}p_{\beta}
        \hat{P}[\ket{b}_{B_i^{\U{bit}}}]
        \hat{P}[\ket{\beta}_{B_i^{\U{basis}}}]
        \frac{p_{\U{dark}}}{2}
        \tr
        \(
        \hat{E}^{\U{meas1'},\beta,i}_{b\oplus1}
        \hat{\rho}_{B_i^{\U{sig}}}\)
                \notag\\
        &+\sum_{b\in\{0,1\},\beta\in\{Z,X\}}p_{\beta}
        \hat{P}[\ket{b}_{B_i^{\U{bit}}}]
        \hat{P}[\ket{\beta}_{B_i^{\U{basis}}}]
        \left(p_{\U{dark}}-\frac{p_{\U{dark}}^2}{2}\right)
        \tr
        \(
        \hat{E}^{\U{meas1'},\beta,i}_{\U{Noclick}}
        \hat{\rho}_{B_i^{\U{sig}}}\)
\\
    =&\sum_{\beta\in\{Z,X\},b\in\{0,1,\U{Noclick}\}}p_{\beta}
    \tr_{B^{\U{sig1}}_iB^{\U{sig2}}_i} \left(\hat{E}^{\U{meas},\beta,i}_{b} 
    (\hat{\rho}_{B_i^{\U{sig}}})\right) \hat{P}\left[\ket{\beta}_{B^{\U{basis}}_i}\ket{b}_{B^{\U{bit}}_i} \right].
    \label{eq:finalsquash}
\end{align}
The first equality comes from Eqs.~(\ref{eq:orig-squash}), (\ref{eq:def-meas1dash}), (\ref{eq:dark-qubit1}) and (\ref{eq:dark-qubit2}). The second equality follows from Eqs.~(\ref{eq:E-dark-meas1}) and (\ref{eq:E-dark-meas2}). 

The top equation shows the probability distribution obtained from the Pauli-$\beta$ measurement on a single-qubit when 
$b\neq\U{Noclick}$ after applying the squashing map $\mathcal{E}^{\U{squash}}$. 
On the other hand, the bottom equation is equal to $\mathcal{E}^B_i(\hat{\rho}_{B_i^{\U{sig}}})$ 
from Eq.~(\ref{eq:EBformula}). Equation~(\ref{eq:finalsquash}) implies the existence of a squashing map for the actual measurement POVM $\{\hat{E}^{\U{meas},\beta,i}_{b}\}_{b}$ incorporating the effect of the dark count, and therefore demonstrates Eq.~(\ref{eq:MB}) of Theorem~\ref{prop:squash}. 
This ends the proof of Proposition \ref{prop:squash}.

\clearpage
\subsubsection{Virtual protocol}
\label{subsec:virtualprotocol}
Alice performs Step~\ref{step:viaa}, and Bob performs Step~\ref{step:viab} for $i=1,2,...,N$. 
\begin{enumerate}
\item
\begin{enumerate}
\item
\label{step:viaa}
Alice prepares systems $A_i^{\U{CR}}$, $R_i$ and $A_i^{\U{sig}}$ in the following state 
\begin{align}
\ket{\Psi_{\mathrm{in,vir}}}_{A_i^{\U{CR}},R_i,A_i^{\U{sig}}}=
&\sum_{\omega_i\in\{S,D,V\}}
\sum_{\alpha_i\in\{Z,X\}}\sum_{a_i\in\{0,1\}}
\sqrt{
p_{a_i}p_{\omega_i}p_{\alpha_i}}
\ket{\omega_i,\alpha_i,a_i}_{A_i^{\U{CR}}}\nonumber\\
&\otimes
\sum_{n_i=0}^{\infty}\sqrt{p_{\mu_{\omega_i},n_i}^{\U{CS}}}
\ket{n_i}_{R_i}
\ket{\psi_{n_i,\theta_{a_i,\alpha_i}}}_{A_i^{\U{sig}}}
\label{eq:virtualstateAlice}
\end{align}
and sends state of system $A_i^{\U{sig}}$ to Bob. Here, $\tr_{R_i}\hat{P}[\ket{\Psi_{\mathrm{in,vir}}}]$ is equal to $\hat{\rho}_{\U{in},A_i^{\U{CR}}A_i^{\U{sig}}}$ in Eq.~(\ref{eq:alicestate}),
\begin{align}
    p_{\mu_{\omega},n}^{\U{CS}}:=\frac{\mu_{\omega}^n}{n!}e^{-\mu_{\omega}}
    \label{eq:poisson}
\end{align}
denotes the probability that a double pulse sent by Alice contains $n\in\{0,1,2,...\}$ photons when she selects the intensity label $\omega\in\{S,D,V\}$, and $\ket{\psi_{n_i,\theta_{a_i,\alpha_i}}}$ denotes the superposition state of the $i$th double pulse with a total photon number of $n_i$, namely, 
\begin{align}
\ket{\psi_{n_i,\theta_{a_i,\alpha_i}}}_{A_i^{\U{sig}}}
=\sum_{k=0}^{n_i}
\sqrt{\frac{1}{2^n}\binom{n}{k}}
e^{\U{i}\theta_{a_i,\alpha_i}(n_i-k)}\ket{n_i-k}_{A_i^{\U{sig1}}}\ket{k}_{A_i^{\U{sig2}}}.
\label{eq:psinthetaalpha}
\end{align}
\item
\label{step:viab}
Bob chooses the measurement basis $\beta_i$ with probability $p_{\beta_i}$ ($\beta_i\in\{Z,X\}$) and stores this information in system $B_i^{\U{basis}}$. Bob applies the squashing map $\mathcal{E}^{\U{squash}}$ in Eq.~(\ref{eq:Esquash}) to system $B_i^{\U{sig}}$ of the $i$th received pulse, and then performs the projective measurement 
\begin{align}
\{\hat{I}_{B_i^{\U{bit}}}-\hat{P}[\ket{\U{Noclick}}_{B_i^{\U{bit}} }],\hat{P}[\ket{\U{Noclick}}_{B_i^{\U{bit}} }]\}    
\end{align}
on system $B_i^{\U{bit}}$. Additionally, if $\beta_i=X$, Bob measures the post-projected state in the $X$ basis and stores the measurement outcome in system $B_i^{\U{bit}}$.\\
Mathematically, Bob's measurement is described by the following CPTP map acting on state $\hat{\rho}_{B_i^{\U{sig}}}$ of system $B_i^{\U{sig}}$: 
\begin{align}
    \mathcal{E}_i^{B,\U{vir}}: B_i^{\U{sig}}\to 
    B_i^{\U{CR}}=B_i^{\U{basis}}B_i^{\U{bit}},
    \label{eq:bobEBvir}
\end{align}
with
\begin{equation}
\begin{split}
&\mathcal{E}_i^{B,\rm vir}(\hat{\rho}_{B_i^{\U{sig}}})
=\sum_{\beta_i\in\{Z,X\}}p_{\beta_i}
\hat{P}[\ket{\beta_i}_{B_i^{\U{basis}}}]
\hat{P}[\ket{\U{Noclick}}_{B_i^{\U{bit}}}]\mathcal{E}^{\U{squash}}(\hat{\rho}_{B_i^{\U{sig}}})\hat{P}[\ket{\U{Noclick}}_{B_i^{\U{bit}}}]\\
		&+\sum_{b\in\{0,1\}}p_{\beta_i=X}\hat{P}[\ket{\beta_i=X}_{B_i^{\U{basis}}}]
        \(\frac{\hat{I}_{B_i^{\U{bit}}}+(-1)^b\hat{\sigma}_{X}}{2}
        \mathcal{E}^{\U{squash}}(\hat{\rho}_{B_i^{\U{sig}}})\frac{\hat{I}_{B_i^{\U{bit}}}+(-1)^b\hat{\sigma}_{X}}{2}\)\\
&+p_{\beta_i=Z}\hat{P}[\ket{\beta_i=Z}_{B_i^{\U{basis}}}]
        (\hat{I}_{B_i^{\U{bit}}}-\hat{P}[\ket{\U{Noclick}}_{B_i^{\U{bit}}}])\mathcal{E}^{\U{squash}}(\hat{\rho}_{B_i^{\U{sig}}})(\hat{I}_{B_i^{\U{bit}}}-\hat{P}[\ket{\U{Noclick}}_{B_i^{\U{bit}}}]).
        \label{eq:MBvir}
\end{split}
\end{equation}
\end{enumerate}
\item
\label{step:virtualInfShare}
Alice and Bob execute the procedures specified in the information exchanging and processing flowchart. Specifically, Alice and Bob each perform the CPTP maps $\mathcal{E}_{S_j}^{A,\U{public}}$ in Eq.~(\ref{eq:publicA}) and $\mathcal{E}_{S_j}^{B,\U{public}}$ in Eq.~(\ref{eq:EsjBpublic}) for each of the $j$th blocks.
\item[]
Alice and Bob perform Steps~\ref{step:3fig} and \ref{step:3figbob}, respectively. Their procedures in Steps~\ref{step:3fig} and \ref{step:3figbob} are illustrated in Figs.~\ref{fig:virtual} and \ref{figbobv}, respectively.
\item
\label{step:3fig}
\begin{figure}[t]
\centering
    \includegraphics[width=14cm]{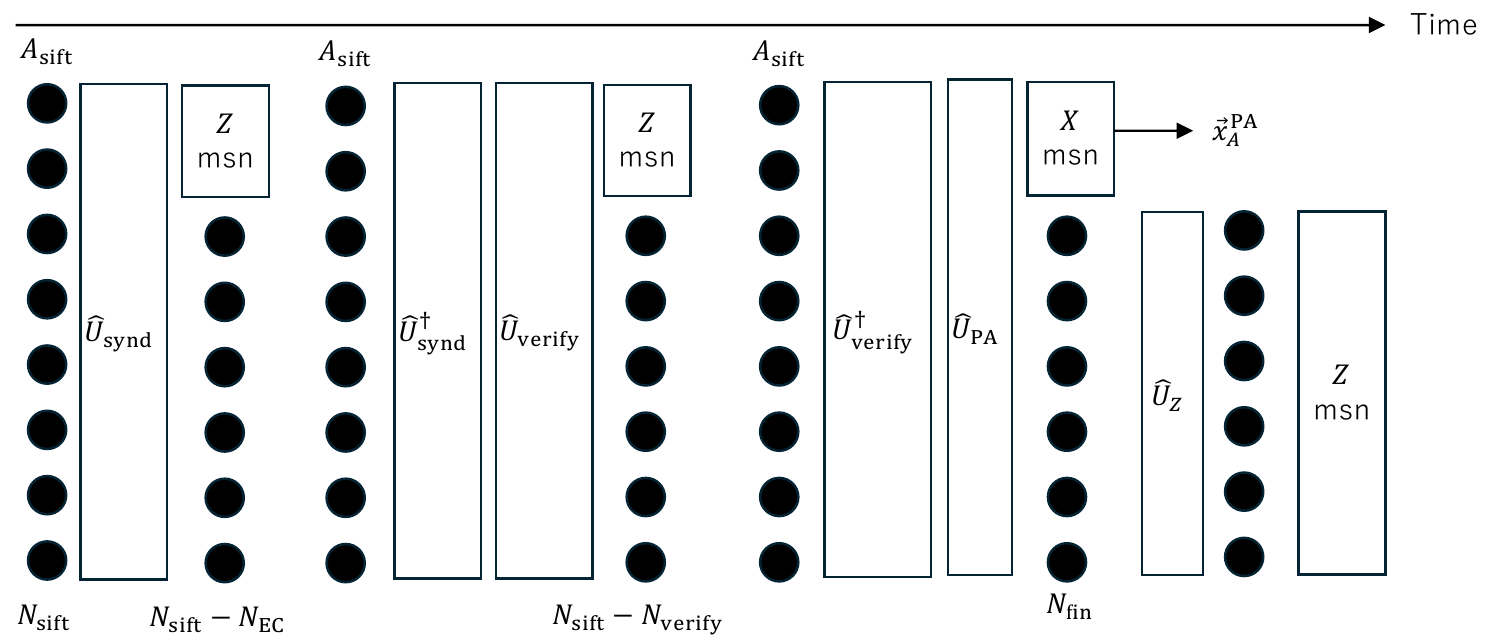}
    \caption{
    Alice's operations of Step~\ref{step:3fig} in the virtual protocol to generate the secret key.
    }
    \label{fig:virtual}
\end{figure}
\begin{figure}[t]
    \centering
    \includegraphics[width=11cm]{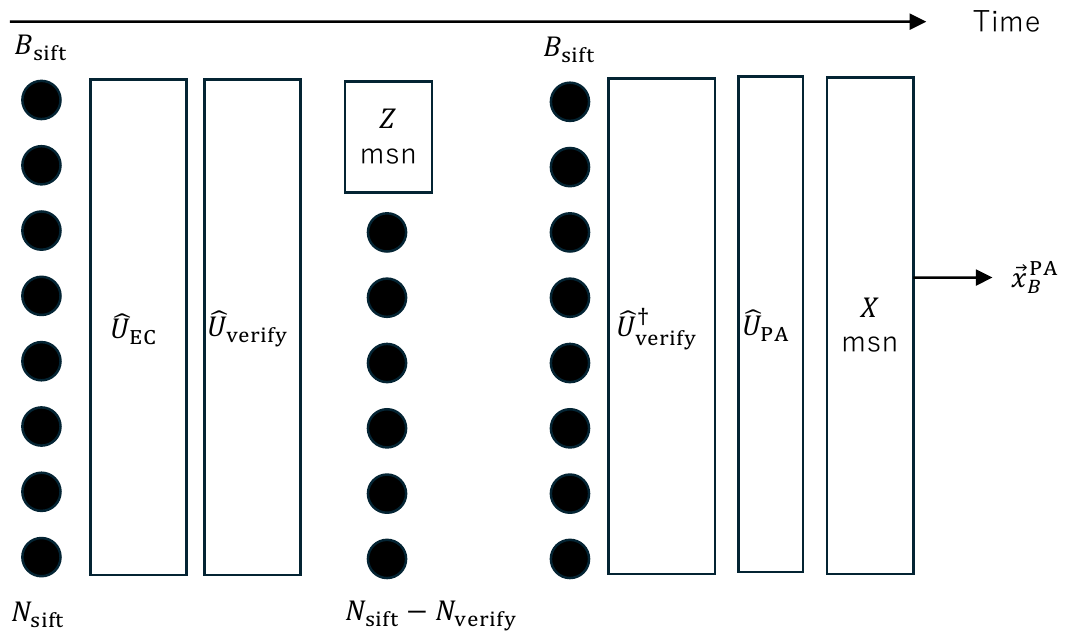}
    \caption{
    Bob's operations of Step~\ref{step:3figbob} in the virtual protocol.
    }
\label{figbobv}
\end{figure}
\begin{enumerate}
\item
\label{step3avirtual}
Alice performs Step~1 of the key generation flowchart and obtains her sifted key in system $A_{\U{sift}}$ consisting of $N_{\U{sift}}$ qubits. 
\\
At this point, the information regarding whether the secret key length is positive or not is stored in system $C_{\U{Judge}}^{\U{Length}}$, and this information is sent to Bob via a classical channel. Since Alice aborts the protocol if the secret key length is not positive, below we will only describe the steps when the secret key length is positive.
\item
\label{step:ec}
Alice applies the unitary operation $\hat{U}_{\U{synd}}(\tilde{\mathcal{C}}_{\U{synd}})$ to system $A_{\U{sift}}$; this unitary operation is constructed from the $N_{\U{sift}}\times N_{\U{sift}}$ invertible binary matrix $\tilde{\mathcal{C}}_{\U{synd}}$, defined by the parity check matrix $\mathcal{C}_{\U{synd}}$ employed for bit error correction in Step~2 of the key generation flowchart. Alice then measures the first $N_{\U{EC}}$ qubits of $A_{\U{sift}}$ in the $Z$ basis and sends the measurement outcome to Bob via a classical channel. The unitary operation $\hat{U}_{\U{synd}}(\tilde{\mathcal{C}}_{\U{synd}})$, which will be defined in Eq.~(\ref{eq:unitarysyndtildeCec}), is constructed such that the $Z$ basis measurement outcome is equal to the syndrome information $I_{\U{synd}}$ in Step~2 of the key generation flowchart. Alice applies the unitary operation $\hat{U}^{\dagger}_{\U{synd}}(\tilde{\mathcal{C}}_{\U{synd}})$ to system $A_{\U{sift}}$. 
\item 
\label{step:verify}
Alice chooses a random number $r_{\U{hash~verify}}$ that determines the surjective universal2 hash function. Alice applies the unitary operation $\hat{U}_{\U{verify}}(\tilde{\mathcal{C}}_{\U{verify}})$ to system $A_{\U{sift}}$; this unitary operation is constructed from the $N_{\U{sift}}\times N_{\U{sift}}$ invertible binary matrix $\tilde{\mathcal{C}}_{\U{verify}}$, defined by the surjective universal2 hash function $\mathcal{C}_{\U{verify}}$ used in Step~3 of the key generation flowchart. Alice then measures the first $N_{\U{verify}}$ qubits of $A_{\U{sift}}$ in the $Z$ basis. The unitary operation $\hat{U}_{\U{verify}}(\tilde{\mathcal{C}}_{\U{verify}})$, which will be defined in Eq.~(\ref{eq:unitaryverify}), is constructed such that the $Z$ basis measurement outcome is equal to the hash value $H_A$ in Step~3 of the key generation flowchart. Alice applies the unitary operation $\hat{U}^{\dagger}_{\U{verify}}(\tilde{\mathcal{C}}_{\U{verify}})$ to system $A_{\U{sift}}$.
\\
Alice sends the information $I_{\U{Hash}}:=(r_{\U{hash~verify}},H_A)$ to Bob via a classical channel. According to the information sent by Bob in Step~\ref{stepBobvirverify}, Alice either aborts the protocol or continues, depending on whether $H_A=H_B$. Below, we will only describe the steps when Alice does not abort the protocol.
\item
\label{step:pa}
Alice chooses a random number $r_{\U{hashPA}}$ that determines the surjective dual universal2 hash function. Alice sends $r_{\U{hashPA}}$ to Bob via a classical channel. Alice applies the unitary operation $\hat{U}_{\U{PA}}(\tilde{\mathcal{C}}_{\U{PA}})$ to system $A_{\U{sift}}$; this unitary operation is constructed from the $N_{\U{sift}}\times N_{\U{sift}}$ invertible binary matrix $\tilde{\mathcal{C}}_{\U{PA}}$, defined by the surjective dual universal2 hash function $\mathcal{C}_{\U{PA}}$ used in Step~4 of the key generation flowchart. Alice then measures the first $N_{\U{sift}}-N_{\U{fin}}$ qubits of system $A_{\U{sift}}$ in the $X$ basis and obtains the measurement outcome $\vec{x}^{\U{PA}}_A\in\{+,-\}^{N_{\U{sift}}-N_{\U{fin}}}$.
\item 
\label{step:pabitflip}
Alice performs the bit-flip operation $\hat{U}_Z$ in the $X$ basis to $A_{\U{sift}}$, which is defined by $\vec{x}_B^{\U{PA}}$ 
(to be sent from Bob at Step~\ref{virbob_xB}) and $\vec{x}^{\U{PA}}_A$.
\item
\label{step:virAliceZkey}
Alice measures $A_{\U{sift}}$ in the $Z$ basis to obtain the QKD key. 
\end{enumerate}
\item 
\label{step:3figbob}
\begin{enumerate}
\item
\label{step3avirtualBob}
Bob performs Step~1 of the key generation flowchart and obtains his sifted key in system $B_{\U{sift}}$ consisting of $N_{\U{sift}}$ qubits. Bob receives the information of whether the secret key length is positive or not, which is determined by Alice in Step~\ref{step3avirtual}. Since Bob aborts the protocol if the secret key length is not positive, below we will only describe the steps when the secret key length is positive.
\item
\label{step3bvirtualbob}
After receiving the syndrome information $I_{\U{synd}}$ sent by Alice at Step~\ref{step:ec}, Bob applies the unitary operation $\hat{U}_{\U{EC}}$ to system $B_{\U{sift}}$, which corresponds to the bit error correction operation defined by $I_{\U{synd}}$, and obtains the corrected bit string. 
\item
\label{stepBobvirverify}
Using the information $I_{\U{Hash}}$ sent by Alice at Step~\ref{step:verify}, Bob applies the unitary operation $\hat{U}_{\U{verify}}(\tilde{\mathcal{C}}_{\U{verify}})$ to system $B_{\U{sift}}$; this unitary operation is constructed from the $N_{\U{sift}}\times N_{\U{sift}}$ invertible binary matrix $\tilde{\mathcal{C}}_{\U{verify}}$, defined by the surjective universal2 hash function $\mathcal{C}_{\U{verify}}$ used in Step~3 of the key generation flowchart. Bob then measures the first $N_{\U{verify}}$ qubits of system $B_{\U{sift}}$ in the $Z$ basis and obtains the measurement outcome. The unitary operation $\hat{U}_{\U{verify}}(\tilde{\mathcal{C}}_{\U{verify}})$, which will be defined in Eq.~(\ref{eq:unitaryverify}), is constructed such that the $Z$-basis measurement outcome is equal to the hash value $H_B$ in Step~3. Bob applies the unitary operation $\hat{U}^{\dagger}_{\U{verify}}(\tilde{\mathcal{C}}_{\U{verify}})$ to system $B_{\U{sift}}$. Bob compares the hash value $H_A$ sent from Alice at Step~\ref{step:verify} and $H_B$, and sends the one-bit information of whether $H_A=H_B$ or not to Alice via a classical channel. Since Bob aborts the protocol if $H_A\neq H_B$, we will only describe the steps when Bob does not abort the protocol.
\item
\label{virbob_xB}
After Bob receives the information of $r_{\U{hashPA}}$, Bob applies the unitary operation $\hat{U}_{\U{PA}}(\tilde{\mathcal{C}}_{\U{PA}})$ to system $B_{\U{sift}}$; this unitary operation is constructed from the $N_{\U{sift}}\times N_{\U{sift}}$ invertible binary matrix $\tilde{\mathcal{C}}_{\U{PA}}$, defined by the surjective dual universal2 hash function $\mathcal{C}_{\U{PA}}$ used in Step~4 of the key generation flowchart. Bob then measures $B_{\U{sift}}$ in the $X$ basis and obtains the measurement outcome $\vec{x}_B^{\U{PA}}\in\{+,-\}^{N_{\U{sift}}}$ and sends this information to Alice. 
\end{enumerate}
\end{enumerate}

Below, we discuss each step of the above virtual protocol in detail.
\\
{\bf Step~\ref{step:viaa}}
\\
If Alice measures system $A_i^{\U{CR}}$ of the state given in Eq.~(\ref{eq:virtualstateAlice}), she obtains the outcomes $\omega_i,\alpha_i,a_i$ with probability $p_{\omega_i}p_{\alpha_i}p_{a_i}$. The $i$th emitted state corresponding to the outcomes $\omega_i,\alpha_i,a_i$ is given by
\begin{align}
\tr_{R_i}\hat{P}\left[\sum_{n_i=0}^{\infty}\sqrt{p_{\mu_{\omega_i},n_i}^{\U{CS}}}\ket{n_i}_{R_i}
\ket{\psi_{n_i,\theta_{a_i,\alpha_i}}}_{A_i^{\U{sig}}}
\right]
=&
\sum_{n_i=0}^{\infty}p^{\U{CS}}_{\mu_{\omega_i},n_i}\hat{P}[\ket{\psi_{n_i,\theta_{a_i,\alpha_i}}}_{A^{\U{sig}}_i}]
\notag\\
=&
\sum_{n_i=0}^{\infty}\hat{N}_{n_i}\hat{\rho}(\theta_{a_i,\alpha_i},\mu_{\omega_i})_{A^{\U{sig}}_i}
\hat{N}_{n_i}
\notag\\
=&
\hat{\rho}(\theta_{a_i,\alpha_i},\mu_{\omega_i})_{A^{\U{sig}}_i}.
\end{align}
The second equality follows by Eqs.~(\ref{eq:rhothetamuactual}), (\ref{eq:Nn}) and (\ref{eq:psinthetaalpha}). The third equality comes from Eq.~(\ref{eq:alicestate2}). This equation implies that the actual emitted state, given the choice of $\omega_i,\alpha_i,a_i$, can be obtained by Alice measuring system $A_i^{\U{CR}}$ and obtaining $\omega_i,\alpha_i,a_i$ in the virtual protocol.
\\\\
{\bf Step~\ref{step:viab}}
\\
The difference of Bob's measurements between the actual and virtual protocols is that in the virtual protocol, the measurement outcome is not determined when a detection event occurs with $\beta_i=Z$. This can be understood from Eqs.~(\ref{eq:MB}) and (\ref{eq:MBvir}).
\\\\
{\bf Step~\ref{step:virtualInfShare}}
\\
    In the virtual protocol, the bit value detected in the $Z$ basis is not determined, which differs from the actual protocol. However, in the information exchanging and processing flowchart, the $Z$-basis bit values are never made public, so the operations in this flowchart remain the same for both the virtual and actual protocols, even though the $Z$-basis values are not determined in the virtual protocol. Therefore, the operations of Alice's and Bob's information disclosure are the same as those in the actual protocol. 
\\\\
From the discussions so far, the state in the virtual protocol immediately before Steps~\ref{step:3fig} and \ref{step:3figbob} is written as
    \begin{align}
&\hat{\rho}_{\U{QC,vir}}:=
\mathcal{E}_{N_{\U{block}}+1}^E\circ
(\mathcal{E}_{S_{N_{\U{block}}}}^{A,\U{public}}\circ
    \mathcal{E}_{S_{N_{\U{block}}}}^{B,\U{public}}\circ
    \mathcal{E}_{S_{N_{\U{block}}}}^{B,\U{vir}}\circ\mathcal{E}_{N_{\U{block}}}^E)\circ
    \cdots\circ\notag\\
&    
(\mathcal{E}_{S_2}^{A,\U{public}}\circ
    \mathcal{E}_{S_2}^{B,\U{public}}\circ
    \mathcal{E}_{S_2}^{B,\U{vir}}\circ\mathcal{E}_2^E)\circ
(\mathcal{E}_{S_1}^{A,\U{public}}\circ
    \mathcal{E}_{S_1}^{B,\U{public}}\circ
    \mathcal{E}_{S_1}^{B,\U{vir}}\circ\mathcal{E}_1^E)\circ
    \left(
    \bigotimes_{i=1}^N 
\hat{P}[\ket{\Psi_{\mathrm{in,vir}}}_{A_i^{\U{CR}},R_i,A_i^{\U{sig}}}]
    \right)
    \label{eq:rhoQCvir}
\end{align}
with
\begin{align}
\mathcal{E}_{S_j}^{B,\U{vir}}:=\bigotimes_{i=(j-1)M+1}^{jM}\mathcal{E}_i^{B,\U{vir}}
\end{align}
for $j\in\{1,2,...,N_{\U{block}}\}$. 
\\\\
{\bf Steps~\ref{step3avirtual} and \ref{step3avirtualBob}}
\\
    The difference between the actual and virtual protocols is that in the virtual protocol, the measurement outcome is not determined when a detection event occurs with $\beta_i=Z$. 
\\\\
{\bf Steps~\ref{step:ec} and \ref{step3bvirtualbob}}
\\
To mathematically describe Steps~\ref{step:ec}-\ref{step:pa} and Steps~\ref{step3bvirtualbob}-\ref{virbob_xB}, let
\begin{equation}
    \ket{N_{\U{sift}},(\vec{x})_X} := \sum_{\vec{z}\in\{0,1\}^{N_{\U{sift}}}}2^{-N_{\U{sift}}/2}(-1)^{-\vec{z}\cdot\vec{x}}\ket{N_{\U{sift}},\vec{z}} 
\end{equation}
represent the eigenstate in the $X$-basis.
\\\\
The parity check matrices employed for bit error correction in this step is denoted by $\{\mathcal{C}_{\U{synd}}\}_{N_{\U{sift}}=1}^{N}$, and the $N_{\U{EC}}$-bit syndrome information (where $N_{\U{EC}}$ is determined by $N_{\U{sift}}$), as defined immediately below Eq.~(\ref{eq:KECkraus}), can be expressed as
\begin{align}
    f_{\U{synd}}(N_{\U{sift}}, N_{\U{EC}}, \bm{k}_A) =
    \vec{k}_A\mathcal{C}_{\U{synd}}.
\end{align}
Here, classical information consisting of multiple elements, such as Alice's sifted key $\bm{k}_A$ of $N_{\U{sift}}$ bits, is regarded as a row vector, and the matrix $\mathcal{C}_{\U{synd}}$ is a binary matrix of size $N_{\U{sift}}\times N_{\U{EC}}$. Note that for a linear code, the column vectors of $\mathcal{C}_{\U{synd}}$ are linearly independent, which implies that the $N_{\U{EC}}$-bit syndrome information is mutually independent.
\\
To make an invertible $N_{\U{sift}}\times N_{\U{sift}}$ matrix $\tilde{\mathcal{C}}_{\U{synd}}$ from $\mathcal{C}_{\U{synd}}$, 
we add $(N_{\U{sift}}-N_{\U{EC}})$ linearly independent columns to $\mathcal{C}_{\U{synd}}$. 
Here, $\mathcal{C}_{\U{synd}}$ is described by 
$\tilde{\mathcal{C}}_{\U{synd}}$ as
\begin{align}
\mathcal{C}_{\U{synd}}=\tilde{\mathcal{C}}_{\U{synd}}
    \overbrace{\left(
\begin{array}{c}
I_{N_{\U{EC}}} \\
0_{(N_{\U{sift}}-N_{\U{EC}})\times N_{\U{EC}}}
\end{array}
\right)}^{N_{\U{EC}}}
\Big\} N_{\U{sift}},
\end{align}
where $I_{N_{\U{EC}}}$ is the $N_{\U{EC}}\times N_{\U{EC}}$ identity matrix, and $0_{(N_{\U{sift}}-N_{\U{EC}})\times N_{\U{EC}}}$ is the $(N_{\U{sift}}-N_{\U{EC}})\times N_{\U{EC}}$ zero matrix. Although $N_{\U{EC}}$ depends on $N_{\U{sift}}$, it is not written explicitly for simplicity of notation. Then, the unitary operation that Alice applies to system $A_{\U{sift}}$ at Step~\ref{step:ec} is written as
\footnote{
The proof of Eq.~(\ref{eq:unitarysyndtildeCec}) is as follows. 
\begin{align}
\left(
\sum_{\vec{z}\in\{0,1\}^{N_{\U{sift}}}}\ket{N_{\U{sift}},\vec{z}\tilde{\mathcal{C}}_{\U{synd}}}\bra{N_{\U{sift}},\vec{z}}
\right)\ket{N_{\U{sift}},(\vec{x})_X}=\sum_{\vec{z}\in\{0,1\}^{N_{\U{sift}}}}\frac{1}{\sqrt{2^{N_{\U{sift}}}}}
(-1)^{\vec{x}\cdot\vec{z}}\ket{N_{\U{sift}},\vec{z}\tilde{\mathcal{C}}_{\U{synd}}}
\label{eq:XZtransform}
\end{align}
Since
\[
\vec{x}\cdot\vec{z}
=\left(\vec{x}(\tilde{\mathcal{C}}^{-1}_{\U{synd}})^{\U{T}}\right)(\vec{z}\tilde{\mathcal{C}}_{\U{synd}})^{\U{T}}
=\left(\vec{x}(\tilde{\mathcal{C}}^{-1}_{\U{synd}})^{\U{T}}\right)\cdot(\vec{z}\tilde{\mathcal{C}}_{\U{synd}}),
\]
holds, by setting $\vec{z}':=\vec{z}\tilde{\mathcal{C}}_{\U{synd}}$, Eq.~(\ref{eq:XZtransform}) is equal to
\begin{align}
\sum_{\vec{z}\in\{0,1\}^{N_{\U{sift}}}}\frac{1}{\sqrt{2^{N_{\U{sift}}}}}
(-1)^{\vec{x}(\tilde{\mathcal{C}}^{-1}_{\U{synd}})^{\U{T}}\cdot\vec{z}'}\ket{N_{\U{sift}},\vec{z}'}
=\ket{N_{\U{sift}},(\vec{x}(\tilde{\mathcal{C}}^{-1}_{\U{synd}})^{\U{T}})_X}.
\end{align}
}
\begin{align}
    \hat{U}_{\U{synd}}(\tilde{\mathcal{C}}_{\U{synd}})
    &:=\sum_{N_{\U{sift}}=1}^N
    \sum_{\vec{z}\in\{0,1\}^{N_{\U{sift}}}}\ket{N_{\U{sift}},\vec{z}\tilde{\mathcal{C}}_{\U{synd}}}\bra{N_{\U{sift}},\vec{z}}_{A_{\U{sift}}} \notag\\
    &=\sum_{N_{\U{sift}}=1}^N
    \sum_{\vec{x}\in\{0,1\}^{N_{\U{sift}}}}\ket{N_{\U{sift}},(\vec{x}(\tilde{\mathcal{C}}^{-1}_{\U{synd}})^{\U{T}})_X}
    \bra{N_{\U{sift}},(\vec{x})_X}_{A_{\U{sift}}}.
    \label{eq:unitarysyndtildeCec}
\end{align}
Here, as the matrix $\tilde{\mathcal{C}}_{\U{synd}}$ is full rank, its inverse matrix $\tilde{\mathcal{C}}^{-1}_{\U{synd}}$ exists, and `T' represents the transpose of a matrix.
\\\\
Alice's $Z$-basis measurement on the first $N_{\U{EC}}$ qubits of system $A_{\U{sift}}$, after performing the unitary operation $\hat{U}_{\U{synd}}$, is characterized by the Kraus operators
\begin{equation}
    \hat{K}^{\U{EC,vir}}_{N_{\U{sift}},\vec{a}_{N_{\U{EC}}}}:= 
    \sum_{\vec{a}\in 
    \{\vec{a}\in\{0,1\}^{N_{\U{sift}}}\mid 
    \vec{a}_{\leq N_{\U{EC}}} = \vec{a}_{N_\U{EC}}\}}
    \ket{\vec{a}_{N_{\U{EC}}}}_{C_{\U{EC}}}
    \otimes \hat{P}\left[ 
    \ket{N_{\U{sift}},\vec{a}}_{A_{\U{sift}}}
    \right].
    \label{eq:krausECvir}
\end{equation}
Here, $\vec{a}_{N_{\U{EC}}}\in\{0,1\}^{N_{\U{EC}}}$ represents the resulting $Z$-basis measurement outcome, which corresponds to the syndrome information $I_{\U{synd}}$.
\\
For state $\mathcal{E}^{\U{sift}}(\hat{\rho}_{\U{QC,vir}})$ immediately after Steps~\ref{step3avirtual} and \ref{step3avirtualBob}, the operation performed by Alice at Step~\ref{step:ec} is described by the CPTP map
\begin{align}
\mathcal{E}^{\U{EC}A,\U{vir}}: \U{Im}(\mathcal{E}^{\U{sift}})
\to A_{\U{sift}} B_{\U{sift}} 
C_{\U{EC}}C^{\U{Length}}_{\U{Judge}}C^{\U{Length}}_{\U{Key}}.
\end{align}
This CPTP map acting on state $\hat{\rho}_{A_{\U{sift}}C^{\U{Length}}_{\U{Judge}}}$ of systems $A_{\U{sift}}C^{\U{Length}}_{\U{Judge}}$ can be written as
\begin{equation}
\begin{split}
    &\mathcal{E}^{\U{EC},A,\U{vir}}(\hat{\rho}_{A_{\U{sift}}C^{\U{Length}}_{\U{Judge}}})
    \notag\\
    :=&
    \sum_{N_{\U{sift}}=1}^N 
    \sum_{\vec{a}_{N_{\U{EC}}}
    \in \{0,1\}^{N_{\U{EC}}}}
    \hat{U}_{\U{synd}}^{\dagger}(\tilde{\mathcal{C}}_{\U{synd}})
    \hat{K}^{\U{EC,vir}}_{N_{\U{sift}},\vec{a}_{N_{\U{EC}}}}
    \hat{U}_{\U{synd}}(\tilde{\mathcal{C}}_{\U{synd}})\hat{P}[\ket{1}_{C^{\U{Length}}_{\U{Judge}}}]
    \hat{\rho}_{A_{\U{sift}}C^{\U{Length}}_{\U{Judge}}}\hat{P}[\ket{1}_{C^{\U{Length}}_{\U{Judge}}}] \\
    & \hat{U}_{\U{synd}}^{\dagger}(\tilde{\mathcal{C}}_{\U{synd}})
    \hat{K}^{\U{EC,vir}{\dagger}}_{N_{\U{sift}},\vec{a}_{N_{\U{EC}}}}
    \hat{U}_{\U{synd}}(\tilde{\mathcal{C}}_{\U{synd}})
   +\hat{P}[\ket{0}_{C_{\U{Judge}}^{\U{Length}}}]\hat{\rho}_{A_{\U{sift}}C^{\U{Length}}_{\U{Judge}}} \hat{P}[\ket{0}_{C_{\U{Judge}}^{\U{Length}}}]\otimes
    \hat{P}[\ket{\nul}_{C_{\U{EC}}}]
    .
    \label{eq:EECAvir}
\end{split}
\end{equation}
Depending on the syndrome information $\vec{a}_{N_{\U{EC}}}$ sent by Alice at Step~\ref{step:ec}, Bob applies the following unitary operation to system $B_{\U{sift}}$ that corresponds to the bit error correction operation:
\begin{equation}
    \begin{split}
        \hat{U}_{N_{\U{sift}},N_{\U{EC}},\vec{a}_{N_{\U{EC}}}}^{\U{EC}}
        :=&
        \hat{P}\left[\ket{\vec{a}_{N_{\U{EC}}}}_{C_{\U{EC}}}\right]
        \otimes
        \prod_{i\in\{i\in [N_{\U{sift}}] \mid (f_{\U{EC}}(\vec{a}_{N_{\U{EC}}}))_i = 1\}} \hat{X}_{B,N_{\U{sift}},i}.
    \end{split}
\end{equation}
Here, 
\begin{equation}
    \hat{X}_{B,N_{\U{sift}},i}:= \sum_{\vec{x}\in\{0,1\}^{N_{\U{sift}}}} (-1)^{x_i}\hat{P}\left[\ket{N_{\U{sift}},(\vec{x})_X}_{B_{\U{sift}}}\right]
    \label{eq:BobsXoeratorEC}
\end{equation}
represents the bit-flip operation in the $Z$-basis for the $i$th qubit of system $B_{\U{sift}}$. 
\\
For state $\mathcal{E}^{\U{EC}A,\U{vir}}\circ\mathcal{E}^{\U{sift}}(\hat{\rho}_{\U{QC,vir}})$, the operation performed by Bob at Step~\ref{step3bvirtualbob} is described by the CPTP map
\begin{align}
\mathcal{E}^{\U{EC}B,\U{vir}}: \U{Im}(\mathcal{E}^{\U{EC}A,\U{vir}}\circ\mathcal{E}^{\U{sift}})
\to A_{\U{sift}} B_{\U{sift}} 
C_{\U{EC}}C^{\U{Length}}_{\U{Judge}}C^{\U{Length}}_{\U{Key}}.
\end{align}
This CPTP map acting on state $\hat{\rho}_{B_{\U{sift}}C_{\U{EC}}C^{\U{Length}}_{\U{Judge}}}$ of systems $B_{\U{sift}}C_{\U{EC}}C^{\U{Length}}_{\U{Judge}}$ can be written as
\begin{align}
         &\mathcal{E}^{\U{EC},B,\U{vir}}(\hat{\rho}_{B_{\U{sift}}C_{\U{EC}}C^{\U{Length}}_{\U{Judge}}})
         \notag\\
         :=&
    \sum_{N_{\U{sift}}=0}^N \sum_{\vec{a}_{N_{\U{EC}}}\in\{0,1\}^{N_{\U{EC}}}}\hat{U}_{N_{\U{sift}},N_{\U{EC}},\vec{a}_{N_{\U{EC}}}}^{\U{EC}}
    \hat{P}[\ket{1}_{C_{\U{Judge}}^{\U{Length}}}]
\hat{\rho}_{B_{\U{sift}}C_{\U{EC}}C^{\U{Length}}_{\U{Judge}}}
\hat{P}[\ket{1}_{C_{\U{Judge}}^{\U{Length}}}]
\hat{U}_{N_{\U{sift}},N_{\U{EC}},\vec{a}_{N_{\U{EC}}}}^{\U{EC}\dagger}
\notag\\
    &+\hat{P}[\ket{0}_{C_{\U{Judge}}^{\U{Length}}}\ket{\nul}_{C_{\U{EC}}}]
    \hat{\rho}_{B_{\U{sift}}C_{\U{EC}}C^{\U{Length}}_{\U{Judge}}} 
    \hat{P}[\ket{0}_{C_{\U{Judge}}^{\U{Length}}}\ket{\nul}_{C_{\U{EC}}}].
        \label{eq:EECBvir}
\end{align}
Combining Eqs.~(\ref{eq:EECAvir}) and (\ref{eq:EECBvir}), Steps~\ref{step:ec} and \ref{step3bvirtualbob} are described by the following CPTP map
\begin{align}
    \mathcal{E}^{\U{EC,vir}}:= \mathcal{E}^{\U{EC},B,\U{vir}}\circ \mathcal{E}^{\U{EC},A,\U{vir}}.
     \label{eq:EecvirAB}
\end{align}
\\
{\bf Steps \ref{step:verify} and \ref{stepBobvirverify}}
\\
For a surjective universal2 hash function $\mathcal{C}_{\U{verify}}$, we choose an $N_{\U{sift}}\times N_{\U{sift}}$ invertible binary matrix $\tilde{\mathcal{C}}_{\U{verify}}$ satisfying 
\begin{align}
    \mathcal{C}_{\U{verify}}=\tilde{\mathcal{C}}_{\U{verify}}
    \overbrace{\left(
\begin{array}{c}
I_{N_{\U{verify}}} \\
0_{(N_{\U{sift}}-N_{\U{verify}})\times N_{\U{verify}}}
\end{array}
\right)}^{N_{\U{verify}}}
\Big\} N_{\U{sift}},
\end{align}
where $I_{N_{\U{verify}}}$ is the $N_{\U{verify}}\times N_{\U{verify}}$ identity matrix, and $0_{(N_{\U{sift}}-N_{\U{verify}})\times N_{\U{verify}}}$ is the $(N_{\U{sift}}-N_{\U{verify}})\times N_{\U{verify}}$ zero matrix. Although $\tilde{\mathcal{C}}_{\mathrm{verify}}$ depends on $N_{\U{sift}}$, $N_{\U{verify}}$ and $r_{\U{hash~verify}}$, it is not written explicitly for simplicity of notation. Then, the unitary operation that Alice and Bob apply to their respective systems $A_{\U{sift}}$ and $B_{\U{sift}}$ at Steps~\ref{step:verify} and \ref{stepBobvirverify} is written as
\begin{align}
    \hat{U}_{\U{verify}}(\tilde{\mathcal{C}}_{\U{verify}})
    =&\sum_{\vec{z}\in\{0,1\}^{N_{\U{sift}}}}\ket{N_{\U{sift}},\vec{z}\tilde{\mathcal{C}}_{\U{verify}}}\bra{N_{\U{sift}},\vec{z}}_{A_{\U{sift}}} \notag \\
    &\otimes 
    \sum_{\vec{z}'\in\{0,1\}^{N_{\U{sift}}}}\ket{N_{\U{sift}},\vec{z}'\tilde{\mathcal{C}}_{\U{verify}}}\bra{N_{\U{sift}},\vec{z}'}_{B_{\U{sift}}}
         \notag\\
    =&\sum_{\vec{x}\in\{0,1\}^{N_{\U{sift}}}}
    \ket{N_{\U{sift}}, (\vec{x}(\tilde{\mathcal{C}}^{-1}_{\U{verify}})^{\U{T}})_X}
    \bra{N_{\U{sift}},(\vec{x})_X}_{A_{\U{sift}}}\notag\\
    &\otimes
    \sum_{\vec{x}'\in\{0,1\}^{N_{\U{sift}}}}
    \ket{N_{\U{sift}}, (\vec{x}'(\tilde{\mathcal{C}}^{-1}_{\U{verify}})^{\U{T}})_X}
    \bra{N_{\U{sift}},(\vec{x}')_X}_{B_{\U{sift}}}.
\label{eq:unitaryverify}    
\end{align}
The Kraus operators corresponding to the steps in Steps~\ref{step:verify} and \ref{stepBobvirverify}, where Alice and Bob each measure $N_{\U{verify}}$ qubits to obtain their hash values, and Bob obtains information about whether his hash value matches Alice's, are as follows:
\begin{align}
    &\hat{K}^{\U{verify,vir}}_{N_{\U{sift}},\vec{a}_{N_{\U{verify}}},\vec{b}_{N_{\U{verify}}}}\notag\\
    :=& 
    \sum_{\vec{a}\in 
    \{\vec{a}\in\{0,1\}^{N_{\U{sift}}}\mid 
    \vec{a}_{\leq N_{\U{verify}}} = \vec{a}_{N_{\U{verify}}}\}}
    \sum_{\vec{b}\in 
    \{\vec{b}\in\{0,1\}^{N_{\U{sift}}}\mid 
    \vec{b}_{\leq N_{\U{verify}}} = \vec{b}_{N_{\U{verify}}}\}}
    \notag\\
    &\Biggl(\delta(\vec{a}_{N_{\U{verify}}},\vec{b}_{N_{\U{verify}}})\hat{P}\left[ 
    \ket{N_{\U{sift}},\vec{a}}_{A_{\U{sift}}}
    \right]
    \hat{P}\left[ 
    \ket{N_{\U{sift}},\vec{b}}_{B_{\U{sift}}}
    \right]
    \notag\\
    &\otimes \ket{r_{\text{hash verify}},\vec{a}_{N_{\U{verify}}}}_{C_A^{\U{Hash}}} 
    \ket{\vec{b}_{N_{\U{verify}}}}_{B_{\U{hash}}}
\ket{1}_{C_B^{\U{HashResult}}}
    \otimes\ket{1}\bra{1}_{C_{\U{Judge}}^{\U{Length}}}
    \notag\\
    &+(1-\delta(\vec{a}_{N_{\U{verify}}},\vec{b}_{N_{\U{verify}}}))
    \ket{0,\nul}
    \bra{N_{\U{sift}},\vec{a}}_{A_{\U{sift}}}
    \otimes\ket{0,\nul}
    \bra{N_{\U{sift}},\vec{b}}_{B_{\U{sift}}}
    \notag\\
    &\otimes \ket{r_{\text{hash verify}},\vec{a}_{N_{\U{verify}}}}_{C_A^{\U{Hash}}} 
    \ket{\vec{b}_{N_{\U{verify}}}}_{B_{\U{hash}}}
\ket{0}_{C_B^{\U{HashResult}}}
    \otimes\ket{0}\bra{1}_{C_{\U{Judge}}^{\U{Length}}} \Biggr)
    \notag\\
    &+\hat{P}\left[\ket{0,\nul}_{A_{\U{sift}}}\right]
    \otimes \hat{P}\left[\ket{0,\nul}_{B_{\U{sift}}}\right]
    \otimes \ket{0,\nul}_{C_{A}^{\U{Hash}}}
    \ket{\nul}_{B_{\U{hash}}}
    \ket{\nul}_{C_{B}^{\U{HashResult}}}
\otimes\hat{P}\left[\ket{0}_{C_{\U{Judge}}^{\U{Length}}}\right].
\label{eq:krausverify} 
\end{align}
Here, $\vec{a}_{N_{\U{verify}}}\in\{0,1\}^{N_{\U{verify}}}$ ($\vec{b}_{N_{\U{verify}}}\in\{0,1\}^{N_{\U{verify}}}$) represents Alice's (Bob's) resulting $Z$-basis measurement outcome, which corresponds to the hash value $H_A$ ($H_B$).\\
For state $\mathcal{E}^{\U{EC},\U{vir}}\circ\mathcal{E}^{\U{sift}}(\hat{\rho}_{\U{QC,vir}})$, the operation performed by Bob at Steps~\ref{step:verify} and \ref{stepBobvirverify} is described by the CPTP map
\begin{align}
\mathcal{E}^{\U{verify},\U{vir}}: \U{Im}(\mathcal{E}^{\U{EC},\U{vir}}\circ\mathcal{E}^{\U{sift}})
\to A_{\U{sift}}B_{\U{sift}} C^{\U{Hash}}_AC^{\U{HashResult}}_BC_{\U{EC}}C^{\U{Length}}_{\U{Judge}}C^{\U{Length}}_{\U{Key}}.
\end{align}
This CPTP map acting on state $\hat{\rho}_{A_{\U{sift}}B_{\U{sift}}C^{\U{Length}}_{\U{Judge}}}$ of systems $A_{\U{sift}}B_{\U{sift}}C^{\U{Length}}_{\U{Judge}}$ can be written employing Eqs.~(\ref{eq:unitaryverify}) and (\ref{eq:krausverify}) as
\begin{equation}
\begin{split}
    &\mathcal{E}^{\U{verify,vir}}(\hat{\rho}_{A_{\U{sift}}B_{\U{sift}}C^{\U{Length}}_{\U{Judge}}})
    :=\frac{1}{2^{N_{\U{verify}}}}  
     \sum_{N_{\U{sift}}=0}^N 
     \sum_{r_{\text{hash verify}}\in \{0,1\}^{N_{\U{verify}}}}
     \sum_{\vec{a}_{N_{\U{verify}}}\in \{0,1\}^{N_{\U{verify}}}}
     \sum_{\vec{b}_{N_{\U{verify}}}\in \{0,1\}^{N_{\U{verify}}}}\\
     &\hat{U}^\dagger_{\U{verify}}(\tilde{\mathcal{C}}_{\U{verify}})
     \hat{K}^{\U{verify,vir}}_{N_{\U{sift}},\vec{a}_{N_{\U{verify}}},\vec{b}_{N_\U{verify}}}
     \hat{U}_{\U{verify}}(\tilde{\mathcal{C}}_{\U{verify}})\\
     &\hat{\rho}_{A_{\U{sift}}B_{\U{sift}}C^{\U{Length}}_{\U{Judge}}}\hat{U}^\dagger_{\U{verify}}(\tilde{\mathcal{C}}_{\U{verify}})
     \hat{K}^{\U{verify,vir}\dagger}_{N_{\U{sift}},\vec{a}_{N_{\U{verify}}},\vec{b}_{N_{\U{verify}}}}
     \hat{U}_{\U{verify}}(\tilde{\mathcal{C}}_{\U{verify}}).
     \label{eq:mapverifyvir}
\end{split}
\end{equation}
The state of Alice's, Bob's, and Eve's systems, immediately after completing error verification, namely Steps~\ref{step:verify} and \ref{stepBobvirverify}, is given by
\begin{align}
\hat{\rho}_{\U{verify,vir}}
:=
\mathcal{E}^{\U{verify,vir}}\circ\mathcal{E}^{\U{EC,vir}}\circ\mathcal{E}^{\U{sift}}(\hat{\rho}_{\U{QC,vir}}).
\label{eq:rhoverifyvir}
\end{align}

{\bf Steps~\ref{step:pa}, \ref{step:pabitflip} and \ref{virbob_xB}}
\\
For a surjective dual universal2 hash function $\mathcal{C}_{\U{PA}}$, we choose an $N_{\U{sift}}\times N_{\U{sift}}$ invertible binary matrix $\tilde{\mathcal{C}}_{\U{PA}}$ satisfying
\begin{align}
    \mathcal{C}_{\U{PA}}=\tilde{\mathcal{C}}_{\U{PA}}
    \overbrace{\left(
\begin{array}{c}
I_{N_{\U{fin}}} \\
0_{(N_{\U{sift}}-N_{\U{fin}})\times N_{\U{fin}}}
\end{array}
\right)}^{N_{\U{fin}}}
\Big\} N_{\U{sift}},
\end{align}
where $I_{N_{\U{fin}}}$ is the $N_{\U{fin}}\times N_{\U{fin}}$ identity matrix, and $0_{(N_{\U{sift}}-N_{\U{fin}})\times N_{\U{fin}}}$ is the $(N_{\U{sift}}-N_{\U{fin}})\times N_{\U{fin}}$ zero matrix. Although $\tilde{\mathcal{C}}_{\mathrm{PA}}$ depends on $N_{\U{sift}}$, $N_{\U{fin}}$ and $r_{\U{hashPA}}$, it is not written explicitly for simplicity of notation. Then, the unitary operation that Alice (Bob) applies to system $A_{\U{sift}}$ ($B_{\U{sift}}$) at Step~\ref{step:pa} (Step~\ref{virbob_xB}) is written as
\begin{align}
    \hat{U}_{\U{PA}}(\tilde{\mathcal{C}}_{\U{PA}})
    &=\sum_{\vec{z}\in\{0,1\}^{N_{\U{sift}}}}\ket{N_{\U{sift}},\vec{z}\tilde{\mathcal{C}}_{\U{PA}}}\bra{N_{\U{sift}},\vec{z}}_{A_{\U{sift}}(B_{\U{sift}})}
         \notag\\
    &=\sum_{\vec{x}\in\{0,1\}^{N_{\U{sift}}}}\ket{N_{\U{sift}},(\vec{x}
    (\tilde{\mathcal{C}}^{-1}_{\U{PA}})^{\U{T}})_X}\bra{N_{\U{sift}},(\vec{x})_X}_{A_{\U{sift}}(B_{\U{sift}})}
    .
    \label{eq:UPACPA}
\end{align}
For a vector $\vec{a}$ of length $l$, let $\vec{a}_{>i}$ be
\begin{equation}
    \vec{a}_{> i} := (a_{i+1},a_{i+2},\cdots,a_{l}).
\end{equation}
For state $\mathcal{E}^{\U{verify},\U{vir}}\circ\mathcal{E}^{\U{EC},\U{vir}}\circ\mathcal{E}^{\U{sift}}(\hat{\rho}_{\U{QC,vir}})$, the operation performed by Alice and Bob at Steps~\ref{step:pa}, \ref{step:pabitflip} and \ref{virbob_xB} is described by the CPTP map
\begin{align}
\mathcal{E}^{\U{PA,vir}}: \U{Im}(\mathcal{E}^{\U{verify},\U{vir}}\circ\mathcal{E}^{\U{EC},\U{vir}}\circ\mathcal{E}^{\U{sift}})
\to A_{\U{sift}}B_{\U{sift}} C^{\U{Hash}}_AC^{\U{HashResult}}_BC_{\U{EC}}C^{\U{Length}}_{\U{Judge}}C^{\U{Length}}_{\U{Key}}.
\end{align}Note that the $X$-basis measurement outcomes $\vec{x}_A^{\U{PA}}$ and $\vec{x}_B^{\U{PA}}$ obtained respectively in Steps~\ref{step:pa} and \ref{step:pabitflip} are not included in the output of this CPTP map, and hence these pieces of information do not enter Eve's CPTP input (namely, Eve cannot access these pieces of information). 
In the virtual protocol, we consider that Alice and Bob are in the same location and can secretly exchange 
$\vec{x}_A^{\U{PA}}$ and $\vec{x}_B^{\U{PA}}$. The virtual protocol introduces additional information that does not appear in the actual protocol, but this information is not given to Eve when constructing the virtual protocol.
\\
The CPTP map $\mathcal{E}^{\U{PA,vir}}$ acting on state $\hat{\rho}_{A_{\U{sift}}B_{\U{sift}}C^{\U{Length}}_{\U{Key}}C^{\U{Length}}_{\U{Judge}}}$ of systems $A_{\U{sift}}B_{\U{sift}}C^{\U{Length}}_{\U{Key}}C^{\U{Length}}_{\U{Judge}}$ can be written as
\begin{align}
     &\mathcal{E}^{\U{PA,vir}}(\hat{\rho}_{A_{\U{sift}}B_{\U{sift}}C^{\U{Length}}_{\U{Key}}C^{\U{Length}}_{\U{Judge}}})
     :=  \sum_{N_{\U{sift}},N_{\U{fin}}=0:N_{\U{sift}}\ge N_{\U{fin}}}^N
     \sum_{\vec{x}_A^{\U{PA}}\in\{0,1\}^{N_{\U{sift}}-N_{\U{fin}}}}\sum_{\vec{x}_B^{\U{PA}}\in\{0,1\}
    ^{N_{\U{sift}}}}\notag\\
    &\mathcal{E}_{N_{\U{sift}},N_{\U{fin}}}^{\U{PA4,vir}}
    \circ\mathcal{E}_{N_{\U{sift}},N_{\U{fin}},\vec{x}_A^{\U{PA}},\vec{x}_B^{\U{PA}}}^{\U{PA3,vir}}
    \circ\mathcal{E}_{N_{\U{sift}},N_{\U{fin}},\vec{x}_A^{\U{PA}},\vec{x}_B^{\U{PA}}}^{\U{PA2,vir}}
    \circ\mathcal{E}_{N_{\U{sift}}}^{\U{PA1,vir}}(\hat{\rho}_{A_{\U{sift}}B_{\U{sift}}C^{\U{Length}}_{\U{Key}}C^{\U{Length}}_{\U{Judge}}}).
            \label{eq:CPTPPAvir}
\end{align}
Here, $\mathcal{E}_{N_{\U{sift}}}^{\U{PA1,vir}}$ is defined by
\begin{align}
&    \mathcal{E}^{\U{PA1,vir}}_{N_{\U{sift}}}(\hat{\rho}_{A_{\U{sift}}B_{\U{sift}}C^{\U{Length}}_{\U{Key}}C^{\U{Length}}_{\U{Judge}}})
    \notag\\
    :=
&\frac{1}{2^{N_{\U{hashPA}}}}
\sum_{r_{\U{hashPA}}\in\{0,1\}^{N_{\U{hashPA}}}}
            (\hat{U}_{\U{PA}}\otimes \hat{U}_{\U{PA}})
            \hat{P}[\ket{1}_{C_{\U{Judge}}^{\U{Length}}}]
            \hat{\rho}_{A_{\U{sift}}B_{\U{sift}}C^{\U{Length}}_{\U{Key}}C^{\U{Length}}_{\U{Judge}}}
            \hat{P}[\ket{1}_{C_{\U{Judge}}^{\U{Length}}}]
            \notag\\
&            (\hat{U}_{\U{PA}}\otimes \hat{U}_{\U{PA}})^{\dagger}
+            \hat{P}[\ket{0}_{C_{\U{Judge}}^{\U{Length}}}]
            \hat{\rho}_{A_{\U{sift}}B_{\U{sift}}C^{\U{Length}}_{\U{Key}}C^{\U{Length}}_{\U{Judge}}}
            \hat{P}[\ket{0}_{C_{\U{Judge}}^{\U{Length}}}].
       \label{eq:CPTPPAvir1}
\end{align}
As for $\mathcal{E}_{N_{\U{sift}},N_{\U{fin}},\vec{x}_A^{\U{PA}},\vec{x}_B^{\U{PA}}}^{\U{PA2,vir}}$, it is defined for the input $\hat{\sigma}_1:=\mathcal{E}^{\U{PA1,vir}}_{N_{\U{sift}}}(\hat{\rho}_{A_{\U{sift}}B_{\U{sift}}C^{\U{Length}}_{\U{Key}}C^{\U{Length}}_{\U{Judge}}})$ as
\begin{align} &\mathcal{E}^{\U{PA2,vir}}_{N_{\U{sift}},N_{\U{fin}},\vec{x}_A^{\U{PA}},\vec{x}_B^{\U{PA}}}(\hat{\sigma}_1)
\notag\\
    :=&
            \left(
          \sum_{
\vec{x}\in\{0,1\}^{N_{\U{sift}}}: \vec{x}_{\le N_{\U{sift}}-N_{\U{fin}}}=
\vec{x}_A^{\U{PA}}}
\hat{P}[\ket{N_{\U{sift}},(\vec{x})_X}_{A_{\U{sift}}}]           \right)
          \hat{P}[\ket{N_{\U{sift}},(\vec{x}_B^{\U{PA}})_X}_{B_{\U{sift}}}]
                       \notag\\
            &    
             \hat{P}[\ket{N_{\U{fin}}}_{C_{\U{Key}}^{\U{Length}}}]
             \hat{P}[\ket{1}_{C_{\U{Judge}}^{\U{Length}}}]
            \hat{\sigma}_1
            \hat{P}[\ket{1}_{C_{\U{Judge}}^{\U{Length}}}]
            \hat{P}[\ket{N_{\U{fin}}}_{C_{\U{Key}}^{\U{Length}}}]
            \notag\\
          &\left(
          \sum_{
\vec{x}\in\{0,1\}^{N_{\U{sift}}}: \vec{x}_{\le N_{\U{sift}}-N_{\U{fin}}}=
\vec{x}_A^{\U{PA}}}
\hat{P}[\ket{N_{\U{sift}},(\vec{x})_X}_{A_{\U{sift}}}]           \right)
\hat{P}[\ket{N_{\U{sift}},(\vec{x}_B^{\U{PA}})_X}_{B_{\U{sift}}}] 
            \notag\\
+            &\hat{P}[\ket{0}_{C_{\U{Judge}}^{\U{Length}}}]
            \hat{\sigma}_1
            \hat{P}[\ket{0}_{C_{\U{Judge}}^{\U{Length}}}].
          \label{eq:CPTPPAvir2}
\end{align}
As for $\mathcal{E}_{N_{\U{sift}},N_{\U{fin}},\vec{x}_A^{\U{PA}},\vec{x}_B^{\U{PA}}}^{\U{PA3,vir}}$, it is defined for the input $\hat{\sigma}_2:=\mathcal{E}^{\U{PA2,vir}}_{N_{\U{sift}},N_{\U{fin}},\vec{x}_A^{\U{PA}},\vec{x}_B^{\U{PA}}}(\hat{\sigma}_1)$ as
\begin{align}
        \mathcal{E}^{\U{PA3,vir}}_{N_{\U{sift}},N_{\U{fin}},\vec{x}_A^{\U{PA}},\vec{x}_B^{\U{PA}}}(\hat{\sigma}_2)
        :=&\hat{U}_Z(\vec{x}_A^{\U{PA}},\vec{x}_B^{\U{PA}})
             \hat{P}[\ket{1}_{C_{\U{Judge}}^{\U{Length}}}]
            \hat{\sigma}_2
            \hat{P}[\ket{1}_{C_{\U{Judge}}^{\U{Length}}}]
         \hat{U}_Z(\vec{x}_A^{\U{PA}},\vec{x}_B^{\U{PA}})^{\dagger}  \notag\\
+            &\hat{P}[\ket{0}_{C_{\U{Judge}}^{\U{Length}}}]
            \hat{\sigma}_2
            \hat{P}[\ket{0}_{C_{\U{Judge}}^{\U{Length}}}].
                   \label{eq:CPTPPAvir3}
\end{align}
Here, the unitary operation $\hat{U}_Z(\vec{x}^{\U{PA}}_A,\vec{x}_B^{\U{PA}})$, which depends on Alice's and Bob's $X$-basis measurement outcomes ($\vec{x}_A^{\U{PA}},\vec{x}_B^{\U{PA}}$), is performed by Alice at Step~\ref{step:pabitflip} to correct phase errors in the $Z$-basis. This implies that this unitary operation is composed of a tensor product of Pauli-$Z$ operators.
\\\\
As for $ \mathcal{E}^{\U{PA4,vir}}_{N_{\U{sift}},N_{\U{fin}}}$, it is defined for the input $\hat{\sigma}_3:=\mathcal{E}^{\U{PA3,vir}}_{N_{\U{sift}},N_{\U{fin}},\vec{x}_A^{\U{PA}},\vec{x}_B^{\U{PA}}}(\hat{\sigma}_2)$ as
\begin{align}
        &\mathcal{E}^{\U{PA4,vir}}_{N_{\U{sift}},N_{\U{fin}}}(\hat{\sigma}_3)
        :=
         \tr_{B_{\U{sift}}C_{\U{Length}}^{\U{Key}}}
          \left( \sum_{\vec{a}\in\{0,1\}^{N_{\U{sift}}}}
    \ket{N_{\U{fin}},\vec{a}_{>N_{\U{sift}}-N_{\U{fin}}}}\bra{N_{\U{sift}},\vec{a}}_{A_{\U{sift}}}\right)
    \hat{P}[\ket{1}_{C_{\U{Judge}}^{\U{Length}}}]]
            \hat{\sigma}_3
            \notag\\
            &\hat{P}[\ket{1}_{C_{\U{Judge}}^{\U{Length}}}]]
    \left( \sum_{\vec{a}\in\{0,1\}^{N_{\U{sift}}}}
    \ket{N_{\U{fin}},\vec{a}_{>N_{\U{sift}}-N_{\U{fin}}}}\bra{N_{\U{sift}},\vec{a}}_{A_{\U{sift}}}\right)^\dagger
    +\hat{P}[\ket{0}_{C_{\U{Judge}}^{\U{Length}}}]
            \hat{\sigma}_3
            \hat{P}[\ket{0}_{C_{\U{Judge}}^{\U{Length}}}].
        \label{eq:CPTPPAvir4}
\end{align}
Note that this operation is essentially tracing out the first $N_{\U{sift}}-N_{\U{fin}}$ qubits in  $A_{\U{sift}}$.

The final state in the virtual protocol is described by
\begin{equation}
  \hat{\rho}_{\U{PA,vir}}:= 
  \tr_{R_1...R_N}\mathcal{E}^{\rm final}\circ
{\mathcal{E}^{\U{PA},\U{vir}}}\circ {\mathcal{E}^{\U{verify,vir}}} \circ {\mathcal{E}^{\U{EC,\vir}}} \circ {\mathcal{E}^{\U{sift}}}(\hat{\rho}_{\U{QC,vir}}).
\label{eq:rhoPAvir_1}
\end{equation}
For the state $\hat{\rho}_{\U{PA,vir}}$, let $\hat{\rho}_{|N_{\U{fin}}}^{\U{PA,vir}}$ be the state when the secret key length is $N_{\U{fin}}$:
\begin{equation}
    \hat{\rho}_{|N_{\U{fin}}}^{\U{PA,vir}}:= 
    \frac{\hat{E}_{N_{\U{fin}}} \hat{\rho}_{\U{PA,vir}}
        \hat{E}_{N_{\U{fin}}}}
        {\Pr_{\U{PA}}(N_{\U{fin}})}.
        \label{eq:rhovirPANfin}
\end{equation}
Here, the projector $\hat{E}_{N_{\U{fin}}}$ is 
defined in Eq.~(\ref{eq:eqENfinPro}). 
\\
From the definitions of ${\mathcal{E}^{\U{sift}}},{\mathcal{E}^{\U{EC,vir}}},{\mathcal{E}^{\U{verify,vir}}}, {\mathcal{E}^{\U{PA,vir}}}$ and ${\mathcal{E}^{\U{final}}}$, the equation
\begin{equation}
    \hat{\rho}_{\U{PA,vir}} = \sum_{N_{\U{fin}}=0}^{N} \Pr_{\U{PA}}(N_{\U{fin}}) 
    \hat{\rho}_{|N_{\U{fin}}}^{\U{PA,vir}}
\end{equation}
holds. 
\\\\
{\bf Step~\ref{step:virAliceZkey}}
\\
This step is described by the following CPTP map
$$
{\mathcal{E}^{Z}_{A_{\U{sift}}}}:A_{\U{sift}}\to A_{\U{sift}},
$$
\begin{equation}
    {\mathcal{E}^{Z}_{A_{\U{sift}}}}(\hat{\rho}):= \sum_{N_{\U{fin}=0}}^{N}\sum_{
    \vec{a}\in\{0,1\}^{N_{\U{fin}}}}
     \hat{E}_{N_{\U{fin}},\vec{a}}^Z \hat{\rho}\hat{E}_{N_{\U{fin}},\vec{a}}^Z
     \label{mathcalEZ}
\end{equation}
with
\begin{equation}
    \hat{E}_{N_{\U{fin}},\vec{a}}^Z:=
    \hat{P}\left[\ket{N_{\U{fin}},\vec{a}}_{A_{\U{sift}}}\right].
\end{equation}

\subsubsection{Equivalence of the states of Alice's secret key and Eve's system in actual and virtual protocols}
\label{subsec:equivalence_actual_virtual}
\begin{proposition}
\label{prop_jituvireq}
For the final state in the virtual protocol when the secret key length is $N_{\U{fin}}$ [namely, $\hat{\rho}_{|N_{\U{fin}}}^{\U{PA,vir}}$ in Eq.~(\ref{eq:rhovirPANfin})] and the actual state of Alice's and Eve's systems when the secret key length is $N_{\U{fin}}$ [namely, $\hat{\rho}^{AE}_{\U{PA}|N_{\U{fin}}}$ in Eq.~(\ref{eq:rhoPAAENfinactual})], 
\begin{equation}
\label{eq-fid-trace-E-vir-PA-actual}
    {\mathcal{E}^{Z}_{A_{\U{sift}}}}(\hat{\rho}_{|N_{\U{fin}}}^{\U{PA,vir}})
    = \hat{\rho}_{\U{PA}|N_{\U{fin}}}^{AE}
\end{equation}
holds.
\\
Also, let
\begin{equation}
    \hat{\rho}^{\U{ideal,vir}}_{|N_{\U{fin}}}:=
   \hat{P}\left[\ket{N_{\U{fin}},+^{N_{\U{fin}}}}_{A_{\U{sift}}}
    \right]\otimes \tr_{A_{\U{sift}}} \hat{\rho}_{\U{PA}|N_{\U{fin}}}^{AE}
\end{equation}
be the ideal final state of Alice's and Eve's systems in the virtual protocol when the secret key length is $N_{\U{fin}}$. Here, 
\begin{equation}
    \ket{N_{\U{fin}},+^{N_{\rm fin}}}_{A_{\U{sift}}}
    := 2^{-N_{\U{fin}}/2}\sum_{\vec{k}_A\in\{0,1\}^{N_{\U{fin}}}}
    \ket{N_{\U{fin}},\vec{k}_A}_{A_{\U{sift}}}
\label{eq:ketplusX}
\end{equation}
denotes the $X$-basis eigenstate of system $A_{\U{sift}}$. 
\\
From the definition of ${\mathcal{E}^{Z}_{A_{\U{sift}}}}$ and $\hat{\rho}^{AE}_{\U{ideal}|N_{\U{fin}}}$ defined in Eq.~(\ref{eq:rhoAEidealNfin}), 
\begin{equation}
\label{eq-fid-trace-E-vir-PA-ideal}
 {\mathcal{E}^{Z}_{A_{\U{sift}}}}(\hat{\rho}^{\U{ideal,vir}}_{|N_{\U{fin}}})
 = \hat{\rho}^{AE}_{\U{ideal}|N_{\U{fin}}}
\end{equation}
holds.
\end{proposition}

{\bf Proof of Proposition~\ref{prop_jituvireq}}
\\
Since the final state in the virtual protocol $\hat{\rho}_{\U{PA,vir}}$ is written as shown in Eq.~(\ref{eq:rhoPAvir_1}), each map can be considered separately.
\\
As for the map $\tr_{R_1...R_N}\mathcal{E}^{\rm final}$, it does not affect the state of system $A_{\U{sift}}$. This implies
\begin{equation}
    {\mathcal{E}^{Z}_{A_{\U{sift}}}}\circ (\tr_{R_1...R_N}\mathcal{E}^{\rm final})
    =(\tr_{R_1...R_N}\mathcal{E}^{\rm final})\circ {\mathcal{E}^{Z}_{A_{\U{sift}}}}.
\end{equation}
Next, the map $\mathcal{E}^{\U{PA4,vir}}_{N_{\U{sift}},N_{\U{fin}}}$ in Eq.~(\ref{eq:CPTPPAvir4}) simply traces out the first $N_{\U{sift}}-N_{\U{fin}}$ qubits of system $A_{\U{sift}}$. Therefore, nothing changes if these qubits are measured in the $Z$ basis beforehand. This implies
\begin{equation}
    {\mathcal{E}^{Z}_{A_{\U{sift}}}}\circ \mathcal{E}^{\U{PA4,vir}}_{N_{\U{sift}},N_{\U{fin}}}
    =\mathcal{E}^{\U{PA4,vir}}_{N_{\U{sift}},N_{\U{fin}}}\circ {\mathcal{E}^{Z}_{A_{\U{sift}}}}.
    \label{eq:EZEPA4}
\end{equation}
Regarding the map $\mathcal{E}_{N_{\U{sift}},N_{\U{fin}},\vec{x}_A^{\U{PA}},\vec{x}_B^{\U{PA}}}^{\U{PA3,vir}}$ in Eq.~(\ref{eq:CPTPPAvir3}), since $\hat{U}_Z(\vec{x}_A^{\U{PA}},\vec{x}_B^{\U{PA}})$ is composed of a tensor product of 
Pauli-$Z$ operators and does not affect the subsequent $Z$-basis measurement,
\begin{equation}
    {\mathcal{E}^{Z}_{A_{\U{sift}}}}\circ \mathcal{E}_{N_{\U{sift}},N_{\U{fin}},\vec{x}_A^{\U{PA}},\vec{x}_B^{\U{PA}}}^{\U{PA3,vir}}
    = {\mathcal{E}^{Z}_{A_{\U{sift}}}}
        \label{eq:EZEPA3}
\end{equation}
holds. Combining Eqs.~(\ref{eq:EZEPA4}) and (\ref{eq:EZEPA3}) leads to
\begin{align}
{\mathcal{E}^{Z}_{A_{\U{sift}}}}
    \circ \mathcal{E}_{N_{\U{sift}},N_{\U{fin}}}^{\U{PA4,vir}}
    \circ \mathcal{E}_{N_{\U{sift}},N_{\U{fin}},\vec{x}_A^{\U{PA}},\vec{x}_B^{\U{PA}}}^{\U{PA3,vir}}
    \circ \mathcal{E}_{N_{\U{sift}},N_{\U{fin}},\vec{x}_A^{\U{PA}},\vec{x}_B^{\U{PA}}}^{\U{PA2,vir}}
    ={\mathcal{E}^{Z}_{A_{\U{sift}}}}
    \circ \mathcal{E}_{N_{\U{sift}},N_{\U{fin}}}^{\U{PA4,vir}}
    \circ \mathcal{E}_{N_{\U{sift}},N_{\U{fin}},\vec{x}_A^{\U{PA}},\vec{x}_B^{\U{PA}}}^{\U{PA2,vir}}.
\end{align}
Since the map $\mathcal{E}_{N_{\U{sift}},N_{\U{fin}}}^{\U{PA4,vir}}$ simply traces out the first $N_{\U{sift}}-N_{\U{fin}}$ qubits of system $A_{\U{sift}}$ and all the qubits of system $B_{\U{sift}}$, measurement on these systems, which is performed by $\mathcal{E}_{N_{\U{sift}},N_{\U{fin}},\vec{x}_A^{\U{PA}},\vec{x}_B^{\U{PA}}}^{\U{PA2,vir}}$ in Eq.~(\ref{eq:CPTPPAvir2}), does affect the output. This means
\begin{equation}
\begin{split}
    & \sum_{\vec{x}_A^{\U{PA}}\in\{0,1\}^{N_{\U{sift}}-N_{\U{fin}}}}\sum_{\vec{x}_B^{\U{PA}}\in\{0,1\}
    ^{N_{\U{sift}}}}
    {\mathcal{E}^{Z}_{A_{\U{sift}}}}
    \circ \mathcal{E}_{N_{\U{sift}},N_{\U{fin}}}^{\U{PA4,vir}}
    \circ \mathcal{E}_{N_{\U{sift}},N_{\U{fin}},\vec{x}_A^{\U{PA}},\vec{x}_B^{\U{PA}}}^{\U{PA2,vir}}\\
    =&
    \mathcal{E}_{N_{\U{sift}},N_{\U{fin}}}^{\U{PA4,vir}}
    \circ {\mathcal{E}^{Z}_{A_{\U{sift}}}} \circ \mathcal{E}_{B_{\U{sift}}}^Z
\end{split}
\label{eq:EZEPA4PA2}
\end{equation}
with 
\begin{equation}
    \mathcal{E}_{B_{\U{sift}}}^Z(\hat{\rho}):= \sum_{N'_{\U{sift}}=0}^{N}
    \sum_{\vec{b}\in\{0,1\}^{N'_{\U{sift}}}}
     \hat{P}\left[\ket{N'_{\U{sift}},\vec{b}}_{B_{\U{sift}}}\right]
     \hat{\rho}
     \hat{P}\left[\ket{N'_{\U{sift}},\vec{b}}_{B_{\U{sift}}}\right].
\end{equation}
Finally, the remaining maps $\mathcal{N}:=\mathcal{E}_{N_{\U{sift}}}^{\U{PA1,vir}}\circ\mathcal{E}^{\U{verify,vir}}\circ\mathcal{E}^{\U{EC,vir}}\circ\mathcal{E}^{\U{sift}}$, performing the $Z$-basis measurement ${\mathcal{E}^{Z}_{A_{\U{sift}}}} \circ \mathcal{E}_{B_{\U{sift}}}^Z$ in Eq.~(\ref{eq:EZEPA4PA2}) after applying these maps is equivalent to performing the $Z$-basis measurement first and then applying these maps:
\begin{align}
    {\mathcal{E}^{Z}_{A_{\U{sift}}}} \circ \mathcal{E}_{B_{\U{sift}}}^Z\circ\mathcal{N}=
    \mathcal{N}\circ
    {\mathcal{E}^{Z}_{A_{\U{sift}}}} \circ \mathcal{E}_{B_{\U{sift}}}^Z.
\end{align}
This equation holds because the unitary operation $\hat{U}_{\U{PA}}$ within $\mathcal{E}_{N_{\U{sift}}}^{\U{PA1,vir}}$, $\hat{U}_{\U{verify}}(\tilde{\mathcal{C}}_{\U{verify}})$ within $\mathcal{E}^{\U{verify,vir}}$, and $\hat{U}_{\U{synd}}(\tilde{\mathcal{C}}_{\U{synd}})$ within $\mathcal{E}^{\U{EC},A,\U{vir}}$ are all binary matrices in the $Z$ basis. They simply transform a $Z$-basis eigenstate (i.e., a bit value) into another bit value. Therefore, when considering the $Z$-basis measurement in Step~\ref{step:virAliceZkey}, Alice can measure her system $A_{\U{sift}}$ in the $Z$ basis at the beginning of Step~\ref{step3avirtual}. 
\\
The above discussions imply that the statistics of Alice's secret key in the virtual protocol is exactly the same as that in the actual protocol, and hence, Eq.~(\ref{eq-fid-trace-E-vir-PA-actual}) holds.
\\\\
Regarding Eq.~(\ref{eq-fid-trace-E-vir-PA-ideal}), it follows directly from the fact that a uniform and random $Z$-basis measurement outcome is obtained by measuring an $X$-basis eigenstate in the $Z$-basis.

\subsubsection{Main propositions to derive $\epsilon_{\U{secrecy}}$ in Eq.~(\ref{eq:secrecy})}
\label{subsec:mainprops}
To state the main propositions needed for deriving the secrecy parameter, we introduce the following definition. 

\begin{definition}
\label{def:ENWABEX}
Let 
\begin{align}
\hat{E}_{\vec{n},\vec{\omega},\vec{\alpha},\vec{\beta},\vec{e}_{X}}&:=
\bigotimes^{N}_{i=1}
\Bigg(\delta(e_{X,i},\U{Noclick})\sum_{a_i\in\{0,1\}}
\hat{P}\left[
\ket{\omega_i,\alpha_i,a_i}_{A_i^{\U{CR}}}\ket{n_i}_{R_i}
\ket{\beta_i}_{B_i^{\U{basis}}}\ket{\U{Noclick}}_{B_i^{\U{bit}}}
\right] \notag\\
&+\frac{\delta(e_{X,i},0)+\delta(e_{X,i},1)}{4}\times\notag\\
&
 \sum_{x\in\{0,1\}}\hat{P}\left[\sum_{a_i,b_i\in\{0,1\}}(-1)^{a_ix+b_i(x+e_{X,i})}
 \ket{\omega_i,\alpha_i,a_i}_{A_i^{\U{CR}}}\ket{n_i}_{R_i}
\ket{\beta_i}_{B_i^{\U{basis}}}\ket{b_i}_{B_i^{\U{bit}}}
\right]\Bigg)
\label{eq:ENWABEX}
\end{align}
with $\vec{n}:=n_1...n_N\in[0,\infty)^N,\vec{\omega}:=\omega_1...\omega_N\in\{S,D,V\}^N,\vec{\alpha}:=\alpha_1...\alpha_N\in\{Z,X\}^N,\vec{\beta}:=\beta_1...\beta_N\in\{Z,X\}^N,\vec{e}_X:=e_{X,1}...e_{X,N}\in\{0,1,\U{Noclick}\}^N$ denote the POVM element acting on systems $A_1^{\U{CR}}...A_N^{\U{CR}}R_1...R_NB_1^{\U{bit}}...B_N^{\U{bit}}B_1^{\U{basis}}...B_N^{\U{basis}}$. Then, the probability of obtaining the outcomes $\vec{n},\vec{\omega},\vec{\alpha},\vec{\beta},\vec{e}_{X}$ when state $\hat{\rho}_{\U{QC,vir}}$ in Eq.~(\ref{eq:rhoQCvir}) is measured by this POVM is defined by
\begin{equation}
\label{state:definition}
 \begin{split}
 \Pr_{\U{QC}}(\vec{n},\vec{\omega},\vec{\alpha},\vec{\beta},\vec{e}_{X})
:=\tr \(\hat{E}_{\vec{n},\vec{\omega},\vec{\alpha},\vec{\beta},\vec{e}_{X}}
\hat{\rho}_{\U{QC,vir}}\).
 \end{split}
\end{equation}
\end{definition}

\begin{proposition}
\label{prop:markov}
    {\bf Markov property about intensity choices}\\
    Let 
    \begin{align}
    F_i:=\vec{n}_{\le i},\vec{\omega}_{\le i},\vec{\alpha}_{\le i},\vec{\beta}_{\le i},\vec{e}_{X,\le i}
    \label{eq:Fi}
    \end{align} 
be the collection up to the $i$th elements of each of $\vec{n},\vec{\omega},\vec{\alpha},\vec{\beta},\vec{e}_{X}$. Then, for the probability defined in Eq.~(\ref{state:definition}), 
\begin{equation}
\Pr_{\U{QC}}(n_i,\omega_i,\alpha_i,\beta_i, e_{X,i}|F_{i-1})
=p^{\U{int}}_{\omega_i|n_i}\Pr_{\U{QC}}(n_i,\alpha_i,\beta_i,e_{X,i} |F_{i-1})
\label{eq:omegaatodasi}
\end{equation}
is satisfied. Here, $p^{\U{int}}_{\omega_i|n_i}$ is defined by
\begin{equation}
    p^{\U{int}}_{\omega | n} := \frac{p_{\omega}p_{\mu_\omega,n}^{\U{CS}}}{\sum_{\omega\in\{S,D,V\}}p_{\omega}p_{\mu_\omega,n}^{\U{CS}}}
    \label{eq:def-p-omega-n_2}
\end{equation}
with $p_{\mu,n}^{\U{CS}}$ defined in Eq.~(\ref{eq:poisson}).
\end{proposition}

{\bf Proof of Proposition~\ref{prop:markov}}
\\
Applying Bayes' theorem yields
\begin{align} 
\Pr_{\U{QC}}(n_i,\omega_i,\alpha_i,\beta_i,e_{X,i}|F_{i-1})=\Pr_{\U{QC}}(n_i,\alpha_i,\beta_i,e_{X,i}|F_{i-1})
  \Pr_{\U{QC}}(\omega_i|n_i,\alpha_i,\beta_i,e_{X,i},F_{i-1}).
 \label{eq:omegaato1}
\end{align}
The crux of the proof of this proposition is to show
\begin{align}
\Pr_{\U{QC}}(\alpha_i,\beta_i,e_{X,i},F_{i-1}|n_i,\omega_i)
=
\Pr_{\U{QC}}(\alpha_i,\beta_i,e_{X,i},F_{i-1}|n_i).
\label{eq:omegaini}
\end{align}
This equation holds for the following reasons. First, the left-hand-side is equal to
\begin{align}
&\Pr_{\U{QC}}(\alpha_i,\beta_i,e_{X,i},F_{i-1}|n_i,\omega_i)
\notag\\
=&\Pr_{\U{QC}}(
\vec{\alpha}_{\le i},\vec{\beta}_{\le i},\vec{e}_{X,\le i},\vec{n}_{\le i-1},\vec{\omega}_{\le i-1}
|n_i,\omega_i)
\notag\\
=&
\Pr_{\U{QC}}(
\vec{e}_{X,\le i}
|\vec{\alpha}_{\le i},\vec{\beta}_{\le i},\vec{n}_{\le i-1},\vec{\omega}_{\le i-1},n_i,\omega_i)
\Pr_{\U{QC}}(
\vec{\alpha}_{\le i},\vec{\beta}_{\le i},\vec{n}_{\le i-1},\vec{\omega}_{\le i-1}
|n_i,\omega_i)
\notag\\
=&
\Pr_{\U{QC}}(
\vec{e}_{X,\le i}
|\vec{\alpha}_{\le i},\vec{\beta}_{\le i},\vec{n}_{\le i-1},\vec{\omega}_{\le i-1},n_i,\omega_i)
\Pr_{\U{QC}}(
\vec{\alpha}_{\le i},\vec{\beta}_{\le i},\vec{n}_{\le i-1},\vec{\omega}_{\le i-1})
\notag\\
=&\Pr_{\U{QC}}(
\vec{e}_{X,\le i}
|\vec{\alpha}_{\le i},\vec{\beta}_{\le i},\vec{n}_{\le i-1},\vec{\omega}_{\le i-1},n_i)
\Pr_{\U{QC}}(
\vec{\alpha}_{\le i},\vec{\beta}_{\le i},\vec{n}_{\le i-1},\vec{\omega}_{\le i-1}).
\label{eq:PrQCLEFT}
\end{align}
The first equation follows by Eq.~(\ref{eq:Fi}). The second equation is from the Bayes' theorem. The third equation comes from the fact that $\vec{\alpha}_{\le i},\vec{\beta}_{\le i},\vec{n}_{\le i-1},\vec{\omega}_{\le i-1}$ are independent of $n_i,\omega_i$. The fourth equation, where $\omega_i$ is omitted from the condition of the probability, holds for the following reasons.
\begin{itemize}
\item
$\Pr_{\U{QC}}(
\vec{e}_{X,\le i}
|\vec{\alpha}_{\le i},\vec{\beta}_{\le i},\vec{n}_{\le i-1},\vec{\omega}_{\le i-1},n_i,\omega_i)$ represents the probability of obtaining the outcome $\vec{e}_{X,\le i}$ when the $i$th emitted state is characterized by $n_i,\omega_i$. From Eq.~(\ref{eq:virtualstateAlice}), once $n_i$ is fixed, the $i$th emitted state of system $A_i^{\U{sig}}$ is given by $\ket{\psi_{n_i,\theta_{a_i,\alpha_i}}}_{A_i^{\U{sig}}}$, which is independent of $\omega_i$. 
\item
From the information exchanging and processing flowchart Fig.~\ref{fig:flowsift}, 
the information of $\omega_i$ becomes public after Bob has completed his measurement up to the $i$th received pulse. Therefore, $\vec{e}_{X,\le i}$, which is also determined by Bob's $i$th measurement outcome, is independent of $\omega_i$.
\end{itemize}
Next, the right-hand side of Eq.~(\ref{eq:omegaini}) is equal to
\begin{align}
\Pr_{\U{QC}}(\alpha_i,\beta_i,e_{X,i},F_{i-1}|n_i)
=&\Pr_{\U{QC}}(
\vec{\alpha}_{\le i},\vec{\beta}_{\le i},\vec{e}_{X,\le i},\vec{n}_{\le i-1},\vec{\omega}_{\le i-1}
|n_i)
\notag\\
=&
\Pr_{\U{QC}}(
\vec{e}_{X,\le i}
|\vec{\alpha}_{\le i},\vec{\beta}_{\le i},\vec{n}_{\le i-1},\vec{\omega}_{\le i-1},n_i)
\Pr_{\U{QC}}(
\vec{\alpha}_{\le i},\vec{\beta}_{\le i},\vec{n}_{\le i-1},\vec{\omega}_{\le i-1}
|n_i)
\notag\\
=&
\Pr_{\U{QC}}(
\vec{e}_{X,\le i}
|\vec{\alpha}_{\le i},\vec{\beta}_{\le i},\vec{n}_{\le i-1},\vec{\omega}_{\le i-1},n_i)
\Pr_{\U{QC}}(
\vec{\alpha}_{\le i},\vec{\beta}_{\le i},\vec{n}_{\le i-1},\vec{\omega}_{\le i-1})
.
\label{eq:PrQCRIGHT}
\end{align}
The first equation follows by Eq.~(\ref{eq:Fi}). The second equation is from the Bayes' theorem. The third equation comes from the fact that $\vec{\alpha}_{\le i},\vec{\beta}_{\le i},\vec{n}_{\le i-1},\vec{\omega}_{\le i-1}$ are independent of $n_i$. Equation~(\ref{eq:PrQCLEFT}) is equal to Eq.~(\ref{eq:PrQCRIGHT}), which ends the proof of Eq.~(\ref{eq:omegaini}).
\\
It is straightforward to show that Eq.~(\ref{eq:omegaini}) is equivalent to 
\begin{align}
\Pr_{\U{QC}}(\omega_i|n_i,\alpha_i,\beta_i,e_{X,i},F_{i-1})
=
p_{\omega_i|n_i}^{\U{int}},
\label{eq:markovderiv}
\end{align}
and substituting this equation to the right-hand side of Eq.~(\ref{eq:omegaato1}) results in Eq.~(\ref{eq:omegaatodasi}). 
\\
Note that Eq.~(\ref{eq:markovderiv}) is equivalent to stating that the following stochastic process 
$$
\vec{\alpha}_{\le i},\vec{\beta}_{\le i},\vec{e}_{X,\le i},\vec{n}_{\le i-1},\vec{\omega}_{\le i-1}~~\to~~n_i~~\to~~\omega_i
$$
is a Markov process.

\begin{proposition}
\label{meidai:decoy1}
    {\bf Decoy-state method: lower bound on the detection probability from single-photon emissions}\\
Let 
\begin{align}
 p_{n_i,Z,i}^{\U{det}}
	&:=\sum_{e_{X,i}\in\{0,1\}}\Pr_{\U{QC}}(n_i,\alpha_i=\beta_i=Z,e_{X,i}|F_{i-1}),
         \label{eq:defsProdecoyZ1}
\\
  {p_{\omega_i,Z,i}^{\U{det}}}
	 &:=
     \sum_{e_{X,i}\in\{0,1\}}
     \Pr_{\U{QC}}(\omega_i,\alpha_i=\beta_i=Z,e_{X,i}|F_{i-1})
     \label{eq:defsProdecoyZ2}
\end{align}
be sums of the probabilities with respect to the probability 
\[
\Pr_{\U{QC}}(\vec{n},\vec{\omega},\vec{\alpha},\vec{\beta},\vec{e}_{X})
\]
defined in Eq.~(\ref{state:definition}). Then, the lower bound on the detection probability from single-photon emission with the $Z$ basis is given as
\begin{equation}
 p_{n_i=1,Z,i}^{\U{det}}\geq \lambda {p_{S,Z,i}^{\U{det}}}+\zeta {p_{D,Z,i}^{\U{det}}}+\gamma {p_{V,Z,i}^{\U{det}}},
 \end{equation}
 where
 \begin{align}
 \lambda:=-\left(
-\frac{\mu_D^2}{\mu_S^2}
\frac{p_{S|1}^{\U{int}}}{p_{S,0}^{\U{int}}}+\frac{p_{D|1}^{\U{int}}}{p_{D,0}^{\U{int}}}-\frac{p_{V|1}^{\U{int}}}{p_{V,0}^{\U{int}}}
\right)^{-1}\frac{1}{p_{S,0}^{\U{int}}}\frac{\mu_D^2}{\mu_S^2}
\le0,
\end{align}
 \begin{align}
\zeta:=\left(-
\frac{\mu_D^2}{\mu_S^2}\frac{p_{S|1}^{\U{int}}}{p_{S,0}^{\U{int}}}+\frac{p_{D|1}^{\U{int}}}{p_{D,0}^{\U{int}}}-\frac{p_{V|1}^{\U{int}}}{p_{V,0}^{\U{int}}}
\right)^{-1}\frac{1}{p_{D,0}^{\U{int}}}\ge0,
\end{align}
 \begin{align}
\gamma:=-\left(
-\frac{\mu_D^2}{\mu_S^2}\frac{p_{S|1}^{\U{int}}}{p_{S,0}^{\U{int}}}+\frac{p_{D|1}^{\U{int}}}{p_{D,0}^{\U{int}}}-\frac{p_{V|1}^{\U{int}}}{p_{V,0}^{\U{int}}}
\right)^{-1}\frac{1}{p_{V,0}^{\U{int}}}\le0
\end{align}
with 
\begin{align}
p_{\omega,n}^{\U{int}}:=p_{\omega}p_{\mu_\omega,n}^{\U{CS}}.
\end{align}
\end{proposition}

{\bf Proof of Proposition~\ref{meidai:decoy1}}\\
Note that this proposition follows from the standard decoy-state method, as shown in~\cite{practicaldecoy2005}.

From Eqs.~(\ref{eq:omegaatodasi}), (\ref{eq:defsProdecoyZ1}) and (\ref{eq:defsProdecoyZ2}), 
\begin{align}
{p_{\omega_i,Z,i}^{\U{det}}}&=\sum_{n_i=0}^\infty\sum_{e_{X,i}\in\{0,1\}}
     \Pr_{\U{QC}}(n_i,\omega_i,\alpha_i=\beta_i=Z,e_{X,i}|F_{i-1})
    \notag\\
    &=\sum_{n_i=0}^\infty p_{\omega_i|n_i}^{\U{int}}\sum_{e_{X,i}\in\{0,1\}}
     \Pr_{\U{QC}}(n_i,\alpha_i=\beta_i=Z,e_{X,i}|F_{i-1})
     \notag\\
     &=\sum_{n_i=0}^\infty p_{\omega_i|n_i}^{\U{int}}p_{n_i,Z,i}^{\U{det}}
     \label{eq:decoymaineq}
     \end{align}
holds. Let
\begin{align}
\kappa_n:=\frac{\mu_D^n}{\mu_S^n}\ge0
\end{align}
for $n\ge2$, and as $\mu_D/\mu_S\le1$, 
\begin{align}
\kappa_n\le \kappa_2
\label{eq:kappa}
\end{align}
holds. Using Eq.~(\ref{eq:decoymaineq}) for ${p_{S,Z,i}^{\U{det}}},{p_{D,Z,i}^{\U{det}}},{p_{V,Z,i}^{\U{det}}}$ gives
\begin{align}
&-\kappa_2\frac{{p_{S,Z,i}^{\U{det}}}}{p_{S,0}^{\U{int}}}
+\frac{{p_{D,Z,i}^{\U{det}}}}{p_{D,0}^{\U{int}}}
-\frac{{p_{V,Z,i}^{\U{det}}}}{p_{V,0}^{\U{int}}}
=
\left(
-\frac{\kappa_2p_{S|1}^{\U{int}}}{p_{S,0}^{\U{int}}}+\frac{p_{D|1}^{\U{int}}}{p_{D,0}^{\U{int}}}-\frac{p_{V|1}^{\U{int}}}{p_{V,0}^{\U{int}}}
\right)
p_{1,Z,i}^{\U{det}}
\notag\\
+&
\underbrace{
\left(
-\frac{\kappa_2p_{S|0}^{\U{int}}}{p_{S,0}^{\U{int}}}+\frac{p_{D|0}^{\U{int}}}{p_{D,0}^{\U{int}}}-\frac{p_{V|0}^{\U{int}}}{p_{V,0}^{\U{int}}}
\right)}_{=:C_{0,Z,i}}p_{0,Z,i}^{\U{det}}
+
\sum^{\infty}_{n=2}\underbrace{
\left(
-\frac{\kappa_2p_{S|n}^{\U{int}}}{p_{S,0}^{\U{int}}}+\frac{p_{D|n}^{\U{int}}}{p_{D,0}^{\U{int}}}-\frac{p_{V|n}^{\U{int}}}{p_{V,0}^{\U{int}}}
\right)}_{=:C_{n,Z,i}}
p_{n,Z,i}^{\U{det}}.
\label{eq:decoylower}
\end{align}
Here, $C_{0,Z,i}$ and $C_{n,Z,i}$ are non-positive because of Eq.~(\ref{eq:kappa}): 
\begin{align}
C_{0,Z,i}=-\frac{\kappa_2p_{S|0}^{\U{int}}}{p_{S,0}^{\U{int}}}+\frac{p_{D|0}^{\U{int}}}{p_{D,0}^{\U{int}}}-\frac{p_{V|0}^{\U{int}}}{p_{V,0}^{\U{int}}}=-\frac{\kappa_2}{p_0^{\U{int}}}\le0,
\end{align}
\begin{align}
C_{n,Z,i}=\frac{\mu_S^n}{n!p_n^{\U{int}}}(-\kappa_2+\kappa_n)\le0,
\end{align}
where
\begin{align}
    p_n^{\U{int}}:=\sum_{\omega\in\{S,D,V\}}p_{\omega,n}^{\U{int}}.
\end{align}
$C_{0,Z,i},C_{n,Z,i}\le0$ implies
\begin{align}
-\kappa_2\frac{{p_{S,Z,i}^{\U{det}}}}{p_{S,0}^{\U{int}}}
+\frac{{p_{D,Z,i}^{\U{det}}}}{p_{D,0}^{\U{int}}}
-\frac{{p_{V,Z,i}^{\U{det}}}}{p_{V,0}^{\U{int}}}
\le
\left(
-\frac{\kappa_2p_{S|1}^{\U{int}}}{p_{S,0}^{\U{int}}}+\frac{p_{D|1}^{\U{int}}}{p_{D,0}^{\U{int}}}-\frac{p_{V|1}^{\U{int}}}{p_{V,0}^{\U{int}}}
\right)p_{1,Z,i}^{\U{det}}.
\end{align}
The prefactor of $p_{1,Z,i}^{\U{det}}$ is non-negative because of 
\begin{align}
-\frac{\kappa_2p_{S|1}^{\U{int}}}{p_{S,0}^{\U{int}}}+\frac{p_{D|1}^{\U{int}}}{p_{D,0}^{\U{int}}}-\frac{p_{V|1}^{\U{int}}}{p_{V,0}^{\U{int}}}
=-\frac{\mu^2_D}{\mu_S}+\mu_D=\frac{\mu_D(\mu_S-\mu_D)}{\mu_S}\ge0,
\end{align}
which leads to
\begin{align}
p_{1,Z,i}^{\U{det}}\ge\left(
-\frac{\kappa_2p_{S|1}^{\U{int}}}{p_{S,0}^{\U{int}}}+\frac{p_{D|1}^{\U{int}}}{p_{D,0}^{\U{int}}}-\frac{p_{V|1}^{\U{int}}}{p_{V,0}^{\U{int}}}
\right)^{-1}\left(
-\kappa_2\frac{{p_{S,Z,i}^{\U{det}}}}{p_{S,0}^{\U{int}}}
+\frac{{p_{D,Z,i}^{\U{det}}}}{p_{D,0}^{\U{int}}}
-\frac{{p_{V,Z,i}^{\U{det}}}}{p_{V,0}^{\U{int}}}
\right).
\end{align}

\begin{proposition}
\label{meidai:kato1}
    {\bf Lower bound on the number of detections from single-photon emissions}\\
The measurement outcomes $\vec{n}:=n_1...n_N\in[0,\infty)^N, \vec{\alpha}:=\alpha_1...\alpha_N\in\{Z,X\}^N, \vec{\beta}:=\beta_1...\beta_N\in\{Z,X\}^N,\vec{e}_X:=e_{X,1}...e_{X,N}\in\{0,1,\U{Noclick}\}^N$ when the state $\hat{\rho}_{\U{QC,vir}}$ defined in Eq.~(\ref{eq:rhoQCvir}) is measured with the POVM in Definition~\ref{def:ENWABEX} satisfy
\begin{align}
 \Pr\( 
\sum_{i=1}^N\sum_{e=0,1}\delta(n_i,1)\delta(\alpha_i,Z)\delta(\beta_i,Z)\delta(e_{X,i},e) 
 < \underline{N}_{1,Z}\) < 
 \frac{1}{8}\epsilon_{\U{secrecy}}^2.
 \label{eq:meidai:kato1}
\end{align}
Here, $\underline{N}_{1,Z}$ is defined in Eq.~(\ref{eq:LowerN1zsection25}).
\end{proposition}

{\bf Proof of Proposition~\ref{meidai:kato1}}\\
From Kato's inequality, Theorem~1 of~\cite{Katoinequality}, the following inequality holds for any $a_{\U{K}},b_{\U{K}}\in \mathbb{R}$ satisfying $b_{\U{K}}\ge|a_{\U{K}}|$:
\begin{align}
\label{eq:epavirverify}
\Pr\left[
\sum_{i=1}^Np_{n_i=1,Z,i}^{\U{det}}\ge N_{1,Z}+\left[b_{\U{K}}+a_{\U{K}}\left(\frac{2N_{1,Z}}{N}-1\right)
\right]\sqrt{N}
\right]\le 
\exp\left[-\frac{2b_{\U{K}}^2-2a_{\U{K}}^2}{(1+\frac{4a_{\U{K}}}{3\sqrt{N}})^2}\right],
\end{align}
where
\begin{align}
N_{1,Z}:=\sum_{i=1}^N\sum_{e=0,1}\delta(n_i,1)\delta(\alpha_i,Z)\delta(\beta_i,Z)\delta(e_{X,i},e).
\label{eq:definitionN1Z}
\end{align}
The following optimization problem:
\begin{align}
\min
\left[b_{\U{K}}+a_{\U{K}}\(\frac{2\tilde{N}_{1,Z}}{N}-1\)
\right]\sqrt{N}
\end{align}
such that
\begin{align}
\exp\left[-\frac{2b_{\U{K}}^2-2a_{\U{K}}^2}{(1+\frac{4a_{\U{K}}}{3\sqrt{N}})^2}\right]=\frac{\epsilon_{\U{secrecy}}^2}{32}~~\U{and}~~b_{\U{K}}\ge|a_{\U{K}}|
\end{align}
is analytically solved in~\cite{tightnpj}\footnote{This can be seen from Eq.~(31) on page 8 of~\cite{tightnpj}.}. Here, $\tilde{N}_{1,Z}$ is an estimated value of $N_{1,Z}$. The optimal values of $a_{\U{K}}$ and $b_{\U{K}}$ are
\begin{align}
a_{\U{K}}^{1,Z}:=
a_{\U{K}}\(N,\tilde{N}_{1,Z},\frac{\epsilon_{\U{secrecy}}^2}{32}\)  ~~\U{and}~~
b_{\U{K}}^{1,Z}:=b_{\U{K}}\(N,\tilde{N}_{1,Z},\frac{\epsilon_{\U{secrecy}}^2}{32}\),
\end{align}
respectively, where these functions are defined by
\begin{align}
a_{\U{K}}(s,t,\epsilon):=
&\frac{216\sqrt{s}t(s-t)\ln\epsilon-48s^{\frac{3}{2}}(\ln\epsilon)^2
+27\sqrt{2}(s-2t)\sqrt{-s^2(\ln\epsilon)[9t(s-t)-2s\ln\epsilon]}}
{4(9s-8\ln\epsilon)[9t(s-t)-2s\ln\epsilon]
}
\label{kato-ineq-def-a}
\end{align}
\begin{align}
b_{\U{K}}(s,t,\epsilon):=
\frac{\sqrt{18a_{\U{K}}(s,t,\epsilon)^2s-[16a_{\U{K}}(s,t,\epsilon)^2+24a_{\U{K}}(s,t,\epsilon)\sqrt{s}+9s]\ln\epsilon}}{3\sqrt{2s}}.
\label{kato-ineq-def-b}
\end{align}
If $a_{\U{K}}^{1,Z}>-\frac{\sqrt{N}}{2}$ , by setting $a_{\U{K}}$ and $b_{\U{K}}$ as $a_{\U{K}}^{1,Z}$ and $b_{\U{K}}^{1,Z}$, 
respectively, Eq.~(\ref{eq:epavirverify}) implies that
\begin{align}
N_{1,Z}\ge\(1+a_{\U{K}}^{1,Z}\frac{2}{\sqrt{N}}\)^{-1}\left[\sum_{i=1}^Np^{\U{det}}_{1,Z,i}-(b_{\U{K}}^{1,Z}-a_{\U{K}}^{1,Z})\sqrt{N}\right]
\label{eq:N}
\end{align}
holds except with probability $\frac{\epsilon_{\U{secrecy}}^2}{32}$. 
On the other hand, if $a_{\U{K}}^{1,Z}\le-\frac{\sqrt{N}}{2}$, $N_{1,Z}\ge0$.

From proposition~\ref{meidai:decoy1}, $p_{1,Z,i}^{\U{det}}$ is lower-bounded by the linear combination of ${p_{S,Z,i}^{\U{det}}},{p_{D,Z,i}^{\U{det}}}$ and ${p_{V,Z,i}^{\U{det}}}$ as
$$
p_{1,Z,i}^{\U{det}}\ge \lambda {p_{S,Z,i}^{\U{det}}}+\zeta {p_{D,Z,i}^{\U{det}}}+\gamma {p_{V,Z,i}^{\U{det}}}~~
(\lambda\le0, \zeta\ge0, \gamma\le0).
$$
Applying this to Eq.~(\ref{eq:N}) leads to
\begin{align}
N_{1,Z}\ge\(1+a_{\U{K}}^{1,Z}\frac{2}{\sqrt{N}}\)^{-1}
\left[\lambda\sum_{i=1}^N {p_{S,Z,i}^{\U{det}}}+\zeta\sum_{i=1}^N {p_{D,Z,i}^{\U{det}}}+\gamma\sum_{i=1}^N {p_{V,Z,i}^{\U{det}}}-(b_{\U{K}}^{1,Z}-a_{\U{K}}^{1,Z})\sqrt{N}\right].
\label{main}
\end{align}
In the following, we evaluate the upper bounds on $\sum_{i=1}^N {p_{S,Z,i}^{\U{det}}}$ and $\sum_{i=1}^N {p_{V,Z,i}^{\U{det}}}$, and the lower bound on $\sum_{i=1}^N {p_{D,Z,i}^{\U{det}}}$. 
\\\\
{\bf 1: Upper bound on $\sum_{i=1}^N {p_{S,Z,i}^{\U{det}}}$}
\\
From Kato's inequality, Theorem 1 of~\cite{Katoinequality}, the following inequality holds for any $a_{\U{K}},b_{\U{K}}\in \mathbb{R}$ satisfying $b_{\U{K}}\ge|a_{\U{K}}|$ except with probability $\exp\left[-\frac{2b_{\U{K}}^2-2a_{\U{K}}^2}{(1+\frac{4a_{\U{K}}}{3\sqrt{N}})^2}\right]$:
\begin{align}
\sum_{i=1}^N{p_{S,Z,i}^{\U{det}}}\le 
N_S^{\U{sift}}\(1+a_{\U{K}}\frac{2}{\sqrt{N}}\)+\(b_{\U{K}}-a_{\U{K}}\)\sqrt{N}.
\end{align}
By setting $a_{\U{K}}$ and $b_{\U{K}}$ as
\begin{align}
a_{\U{K}}^{S}:=a_{\U{K}}\(N,\tilde{N}_S^{\U{sift}},\frac{\epsilon_{\U{secrecy}}^2}{32}\)~~\U{and}~~
b_{\U{K}}^{S}:=b_{\U{K}}\(N,\tilde{N}_S^{\U{sift}},\frac{\epsilon_{\U{secrecy}}^2}{32}\),
\end{align}
respectively, 
\begin{align}
\sum_{i=1}^N{p_{S,Z,i}^{\U{det}}}\le 
N_S^{\U{sift}}\(1+a_{\U{K}}^{S}\frac{2}{\sqrt{N}}\)+(b_{\U{K}}^{S}-a_{\U{K}}^{S})\sqrt{N}
\label{1}
\end{align}
holds except with probability $\frac{\epsilon_{\U{secrecy}}^2}{32}$. Here, $\tilde{N}_S^{\U{sift}}$ denotes an estimated value of $N_S^{\U{sift}}$.
\\\\
{\bf 2: Upper bound on $\sum_{i=1}^N {p_{V,Z,i}^{\U{det}}}$}
\\
From Kato's inequality, Theorem 1 of~\cite{Katoinequality}, the following inequality holds for any $a_{\U{K}},b_{\U{K}}\in \mathbb{R}$ satisfying $b_{\U{K}}\ge|a_{\U{K}}|$ except with probability $\exp\left[-\frac{2b_{\U{K}}^2-2a_{\U{K}}^2}{(1+\frac{4a_{\U{K}}}{3\sqrt{N}})^2}\right]$:
\begin{align}
\sum_{i=1}^N{p_{V,Z,i}^{\U{det}}}\le 
N_V^{\U{sift}}\(1+a_{\U{K}}\frac{2}{\sqrt{N}}\)+\(b_{\U{K}}-a_{\U{K}}\)\sqrt{N}.
\end{align}
By setting $a_{\U{K}}$ and $b_{\U{K}}$ as
\begin{align}
a_{\U{K}}^{V}:=a_{\U{K}}\(N,\tilde{N}_V^{\U{sift}},\frac{\epsilon_{\U{secrecy}}^2}{32}\)~~\U{and}~~b_{\U{K}}^{V}:=b_{\U{K}}\(N,\tilde{N}_V^{\U{sift}},\frac{\epsilon_{\U{secrecy}}^2}{32}\),
\end{align}
respectively,
\begin{align}
\sum_{i=1}^N{p_{V,Z,i}^{\U{det}}}\le 
N_V^{\U{sift}}\(1+a_{\U{K}}^{V}\frac{2}{\sqrt{N}}\)+(b_{\U{K}}^{V}-a_{\U{K}}^{V})\sqrt{N}
\label{2}
\end{align}
holds except with probability $\frac{\epsilon_{\U{secrecy}}^2}{32}$. Here, $\tilde{N}_V^{\U{sift}}$ denotes an estimated value of $N_V^{\U{sift}}$.
\\\\
{\bf 3: Lower bound on $\sum_{i=1}^N {p_{D,Z,i}^{\U{det}}}$}
\\
From Kato's inequality, Theorem 1 of~\cite{Katoinequality}, the following inequality holds for any $a_{\U{K}},b_{\U{K}}\in \mathbb{R}$ satisfying $b_{\U{K}}\ge|a_{\U{K}}|$:
\begin{align}
\Pr\left[
N_D^{\U{sift}}\ge \sum_{i=1}^N{p_{D,Z,i}^{\U{det}}}+\left[b_{\U{K}}+a_{\U{K}}\left(\frac{2N_D^{\U{sift}}}{N}-1\right)
\right]\sqrt{N}
\right]\le 
\exp\left[-\frac{2b_{\U{K}}^2-2a_{\U{K}}^2}{(1-\frac{4a_{\U{K}}}{3\sqrt{N}})^2}\right].
\end{align}
By setting $a_{\U{K}}$ and $b_{\U{K}}$ as
\begin{align}
a_{\U{K}}^{D}:=a'_{\U{K}}\(N,\tilde{N}_D^{\U{sift}},\frac{\epsilon_{\U{secrecy}}^2}{32}\)~~\U{and}~~
b_{\U{K}}^{D}:=b'_{\U{K}}\(N,\tilde{N}_D^{\U{sift}},\frac{\epsilon_{\U{secrecy}}^2}{32}\),
\end{align}
respectively, with
\begin{align}
a'_{\U{K}}(s,t,\epsilon):=
&\frac{-216\sqrt{s}t(s-t)\ln\epsilon+48s^{\frac{3}{2}}(\ln\epsilon)^2
+27\sqrt{2}(s-2t)\sqrt{-s^2(\ln\epsilon)[9t(s-t)-2s\ln\epsilon]}}
{4(9s-8\ln\epsilon)[9t(s-t)-2s\ln\epsilon]
}
\end{align}
and
\begin{align}
b'_{\U{K}}(s,t,\epsilon):=
\frac{\sqrt{18a_{\U{K}}(s,t,\epsilon)^2s-[16a_{\U{K}}(s,t,\epsilon)^2-24a_{\U{K}}(s,t,\epsilon)\sqrt{s}+9s]\ln\epsilon}}{3\sqrt{2s}},
\end{align}

\begin{align}
\sum_{i=1}^N{p_{D,Z,i}^{\U{det}}}\ge 
N_D^{\U{sift}}-
\left[b_{\U{K}}^{D}+a_{\U{K}}^{D}\left(\frac{2N_D^{\U{sift}}}{N}-1\right)
\right]\sqrt{N}
\label{3}
\end{align}
holds except with probability $\frac{\epsilon_{\U{secrecy}}^2}{32}$. Here, $\tilde{N}_D^{\U{sift}}$ denotes an estimated value of $N_D^{\U{sift}}$.
\\\\
Substituting the bounds in Eqs.~(\ref{1}),(\ref{2}), and (\ref{3}) to Eq.~(\ref{main}), 
\begin{align}
&N_{1,Z}\ge 
\(1+a_{\U{K}}^{1,Z}\frac{2}{\sqrt{N}}\)^{-1}\notag\\
&\Bigg\{\lambda
\left[
N_S^{\U{sift}}(1+a_{\U{K}}^{S}\frac{2}{\sqrt{N}})+(b_{\U{K}}^{S}-a_{\U{K}}^{S})\sqrt{N}
\right]
+\gamma
\left[
N_V^{\U{sift}}(1+a_{\U{K}}^{V}\frac{2}{\sqrt{N}})+(b_{\U{K}}^{V}-a_{\U{K}}^{V})\sqrt{N}
\right]
\notag\\
&+\zeta
\left[
N_D^{\U{sift}}-
\left[b_{\U{K}}^{D}+a_{\U{K}}^{D}\left(\frac{2N_D^{\U{sift}}}{N}-1\right)
\right]\sqrt{N}
\right]
-(b_{\U{K}}^{1,Z}-a_{\U{K}}^{1,Z})\sqrt{N}
\Bigg\}
\end{align}
holds except with probability $\frac{\epsilon_{\U{secrecy}}^2}{8}$.

\begin{proposition}
\label{decoy2meidai}
       {\bf Decoy-state method: upper bound on the single-photon error detection probability}\\
       Let 
\begin{align}
p_{n_i,X,i}^{\U{Error}}
	&:=\Pr_{\U{QC}}(n_i,\alpha_i=\beta_i=X,e_{X,i}=1|F_{i-1}),
    \label{eq:qnixi}
\\
  p_{\omega_i,X,i}^{\U{Error}}
	 &:=
     \Pr_{\U{QC}}(\omega_i,\alpha_i=\beta_i=X,e_{X,i}=1|F_{i-1})
\end{align}
be sums of the probabilities with respect to the probability 
    $$
 \Pr_{\U{QC}}(\vec{n},\vec{\omega},\vec{\alpha},\vec{\beta},\vec{e}_{X})
$$
defined in Eq.~(\ref{state:definition}). Then, the upper bound on the single-photon error detection probability with the $X$ basis is given as
\begin{equation}
 p_{n_i=1,X,i}^{\U{Error}} \le\frac{p_{D,X,i}^{\U{Error}}}{p_{D|1}^{\U{int}}}-\frac{p_{D|0}^{\U{int}}}{p_{D|1}^{\U{int}}}
 \frac{p_{V,X,i}^{\U{Error}}}{p_{V|0}^{\U{int}}}. 
 \end{equation}
\end{proposition}

{\bf Proof of Proposition~\ref{decoy2meidai}}\\
Note that this proposition follows from the standard decoy-state method, as shown in~\cite{practicaldecoy2005}.

Doing a similar argument to derive Eq.~(\ref{eq:decoymaineq}) gives
 \begin{align}
p_{\omega_i,X,i}^{\U{Error}}=\sum_{n_i=0}^\infty p_{\omega_i|n_i}^{\U{int}}p_{n_i,X,i}^{\U{Error}}.
     \end{align}
Using this equation for $p_{D,X,i}^{\U{Error}}$ and $p_{V,X,i}^{\U{Error}}$ leads to
 \begin{align}
\frac{e^{\mu_D}}{p_D}p_{D,X,i}^{\U{Error}}-\frac{1}{p_V}p_{V,X,i}^{\U{Error}}
=\frac{\mu_D}{p_1^{\U{int}}}p_{1,X,i}^{\U{Error}}+\sum_{n=2}^{\infty}\frac{e^{\mu_D}p_{n|D}^{\U{int}}}{p_n^{\U{int}}}p_{n,X,i}^{\U{Error}}
\ge\frac{\mu_D}{p_1^{\U{int}}}p_{1,X,i}^{\U{Error}},
\end{align}
which results in
\begin{align}
p_{1,X,i}^{\U{Error}}\le \frac{p_1^{\U{int}}}{\mu_D}\left[
\frac{e^{\mu_D}}{p_D}p_{D,X,i}^{\U{Error}}-\frac{1}{p_V}p_{V,X,i}^{\U{Error}}
\right]=\frac{p_{D,X,i}^{\U{Error}}}{p_{D|1}^{\U{int}}}-\frac{p_{D|0}^{\U{int}}}{p_{D|1}^{\U{int}}} \frac{p_{V,X,i}^{\U{Error}}}{p_{V|0}^{\U{int}}}.
\end{align}

\begin{proposition}
    \label{meidai:ZXtouka}
    {\bf Equivalence of error detection probabilities in single-photon emissions in the $X$ and $Z$ bases}
    \\
    Let 
\begin{equation}
 p_{1,Z,i}^{\U{Error}}
	:=\Pr_{\U{QC}}
    (n_i=1,\alpha_i=\beta_i=Z, e_{X,i}=1|F_{i-1})
\end{equation}
be the probability of obtaining the $X$-basis error from single-photon emissions when Alice's and Bob's basis choices are $Z$, and let $p_{1,X,i}^{\U{Error}}$ be the probability defined in Eq.~(\ref{eq:qnixi}). Then, 
\begin{equation}
p_{1,Z,i}^{\U{Error}}=\frac{p^2_Z}{p^2_X}p_{1,X,i}^{\U{Error}}
\label{eq:prop9main}
\end{equation}
holds.
\end{proposition}
{\bf Proof of Proposition~\ref{meidai:ZXtouka}}
\\
Consider that
\begin{align}
    p_{1,X,i}^{\U{Error}}&=\Pr_{\U{QC}}(\beta_i=X, e_{X,i}=1|n_i=1,\alpha_i=X,F_{i-1})
    \Pr_{\U{QC}}(n_i=1,\alpha_i=X|F_{i-1})\notag\\
    &=\Pr_{\U{QC}}(\beta_i=X, e_{X,i}=1|n_i=1,\alpha_i=X,F_{i-1})p_{n_i=1}^{\U{int}}p_{\alpha_i=X}\notag\\
        &=\Pr_{\U{QC}}(e_{X,i}=1|n_i=1,\alpha_i=X,F_{i-1})p_{n_i=1}^{\U{int}}p_{\alpha_i=X}p_{\beta_i=X}.
            \label{eq:q1xiprop9}
\end{align}
The first equation follows from Bayes' theorem. The second equation is obtained by using the perfect-state-preparation assumption in Table~\ref{table:assumption}. The third equation comes from the fact that Bob measures the incoming system in the $X$ basis independently of $\beta_i$, meaning that the choice of $\beta_i$ has no influence on the value of $e_{X,i}$. Therefore, $\beta_i$ is independent of all the random variables that appear in the probability.
\\
By making a similar argument for $p_{1,Z,i}^{\U{Error}}$, 
\begin{align}
    p_{1,Z,i}^{\U{Error}}=\Pr_{\U{QC}}(e_{X,i}=1|n_i=1,\alpha_i=Z,F_{i-1})p_{n_i=1}^{\U{int}}p_{\alpha_i=Z}p_{\beta_i=Z}
    \label{eq:q1ziprop9}
\end{align}
holds. The only difference between the first factors in Eqs.~(\ref{eq:q1xiprop9}) and (\ref{eq:q1ziprop9}) is whether $\alpha_i=X$ or $\alpha_i=Z$. However, this difference does not affect the probability of $e_{X,i}=1$, namely, 
\begin{align}
\Pr_{\U{QC}}(e_{X,i}=1|n_i=1,\alpha_i=X,F_{i-1})=\Pr_{\U{QC}}(e_{X,i}=1|n_i=1,\alpha_i=Z,F_{i-1}).
\label{eq:proxexiZX}
\end{align}
This is because, from Eq.~(\ref{eq:virtualstateAlice}), the $i$th emitted states by Alice with $n_i=1,\alpha_i=Z$ and with $n_i=1,\alpha_i=X$ are the same as
\begin{align}
\sum_{a_i\in\{0,1\}}p_{a_i}\hat{P}[\ket{\psi_{1,\theta_{a_i},Z}}_{A_i^{\U{sig}}}]
=\sum_{a_i\in\{0,1\}}p_{a_i}\hat{P}[\ket{\psi_{1,\theta_{a_i},X}}_{A_i^{\U{sig}}}]
=\frac{\hat{P}[\ket{0}_{{A^{\U{sig}1}_i}}\ket{1}_{{A^{\U{sig}2}_i}}]+\hat{P}[\ket{1}_{{A^{\U{sig}1}_i}}\ket{0}_{{A^{\U{sig}2}_i}}]}{2}.
\end{align}
Combining Eq.~(\ref{eq:proxexiZX}) with Eqs.~(\ref{eq:q1xiprop9}) and (\ref{eq:q1ziprop9}) results in Eq.~(\ref{eq:prop9main}).

\begin{proposition}
\label{meidai:kato2}
    {\bf Upper bound on the number of phase errors from single-photon emissions}\\
The measurement outcomes $\vec{n}:=n_1...n_N\in[0,\infty)^N, \vec{\alpha}:=\alpha_1...\alpha_N\in\{Z,X\}^N, \vec{\beta}:=\beta_1...\beta_N\in\{Z,X\}^N$ and $\vec{e}_X:=e_{X,1}...e_{X,N}\in\{0,1,\U{Noclick}\}^N$ when the state $\hat{\rho}_{\U{QC,vir}}$ defined in Eq.~(\ref{eq:rhoQCvir}) is measured with the POVM in Definition~\ref{def:ENWABEX} satisfy
\begin{equation}
 \Pr\( 
\sum_{i=1}^N\delta(n_i,1)\delta(\alpha_i,Z)\delta(\beta_i,Z)\delta(e_{X,i},1) 
 < \overline{N}_{\U{ph}}\) < 
 \frac{1}{8}\epsilon_{\U{secrecy}}^2.
 \label{eq:meidai:kato2}
\end{equation}
Here, $\overline{N}_{\U{ph}}$ is defined in Eq.~(\ref{eq:UpperN1zsection25}).
\end{proposition}
{\bf Proof of Proposition~\ref{meidai:kato2}}
\\
From Kato's inequality, Theorem 1 of~\cite{Katoinequality}, the following inequality holds for any $a_{\U{K}},b_{\U{K}}\in \mathbb{R}$ satisfying $b_{\U{K}}\ge|a_{\U{K}}|$:
\begin{align}
\Pr\left[
N_{\U{ph}}\ge \sum_{i=1}^Np_{1,Z,i}^{\U{Error}}+\left[b_{\U{K}}+a_{\U{K}}\left(\frac{2N_{\U{ph}}}{N}-1\right)
\right]\sqrt{N}
\right]\le 
\exp\left[-\frac{2b_{\U{K}}^2-2a_{\U{K}}^2}{(1-\frac{4a_{\U{K}}}{3\sqrt{N}})^2}\right],
\label{eq:NphKato}
\end{align}
where 
\begin{align}
N_{\U{ph}}:=\sum_{i=1}^N\delta(n_i,1)\delta(\alpha_i,Z)\delta(\beta_i,Z)\delta(e_{X,i},1).
\label{eq:definitionNph}
\end{align}
If
\begin{align}
a_{\U{K}}^{\U{ph}}:=a'_{\U{K}}(N,\tilde{N}_{\U{ph}},\frac{\epsilon_{\U{secrecy}}^2}{24})<\frac{\sqrt{N}}{2},
\end{align}
by setting $a_{\U{K}}$ and $b_{\U{K}}$ as $a_{\U{K}}^{\U{ph}}$ and 
$b_{\U{K}}^{\U{ph}}:=b'_{\U{K}}(N,\tilde{N}_{\U{ph}},\frac{\epsilon_{\U{secrecy}}^2}{24})$, respectively, Eq.~(\ref{eq:NphKato}) implies that
\begin{align}
N_{\U{ph}}\le 
\left(1-\frac{2a_{\U{K}}^{\U{ph}}}{\sqrt{N}}\right)^{-1}
\left[\sum_{i=1}^Np_{1,Z,i}^{\U{Error}}+(b_{\U{K}}^{\U{ph}}-a_{\U{K}}^{\U{ph}})\sqrt{N}
\right]
\end{align}
holds except with probability $\frac{\epsilon_{\U{secrecy}}^2}{24}$. Here, $\tilde{N}_{\U{ph}}$ is an estimated value of $N_{\U{ph}}$. 
On the other hand, if $a_{\U{K}}^{\U{ph}}\ge\frac{\sqrt{N}}{2}$, 
$N_{\U{ph}}\le N$.

Combining this inequality and propositions~\ref{decoy2meidai} and \ref{meidai:ZXtouka} leads to
\begin{align}
N_{\U{ph}}
&\le 
\left(1-\frac{2a_{\U{K}}^{\U{ph}}}{\sqrt{N}}\right)^{-1}
\left[\frac{p^2_Z}{p^2_X}\sum_{i=1}^Np_{1,X,i}^{\U{Error}}+(b_{\U{K}}^{\U{ph}}-a_{\U{K}}^{\U{ph}})\sqrt{N}
\right]\notag\\
&\le 
\left(1-\frac{2a_{\U{K}}^{\U{ph}}}{\sqrt{N}}\right)^{-1}
\left[\frac{p^2_Z}{p^2_X}\sum_{i=1}^N
\left(
\frac{p_{D,X,i}^{\U{Error}}}{p_{D|1}^{\U{int}}}-\frac{p_{D|0}^{\U{int}}}{p_{D|1}^{\U{int}}p_{V|0}^{\U{int}}}p_{V,X,i}^{\U{Error}}\right)
+(b_{\U{K}}^{\U{ph}}-a_{\U{K}}^{\U{ph}})\sqrt{N}
\right].
\label{eq:katophN}
\end{align}
In the following, we evaluate the upper bound on $\sum_{i=1}^N p_{D,X,i}^{\U{Error}}$ and the lower bound on $\sum_{i=1}^N p_{V,X,i}^{\U{Error}}$. 
\\\\
{\bf 1: Upper bound on $\sum_{i=1}^N p_{D,X,i}^{\U{Error}}$}
\\
From Kato's inequality, Theorem 1 of~\cite{Katoinequality}, the following inequality holds for any $a_{\U{K}},b_{\U{K}}\in \mathbb{R}$ satisfying $b_{\U{K}}\ge|a_{\U{K}}|$ except with probability $\exp\left[-\frac{2b_{\U{K}}^2-2a_{\U{K}}^2}{(1+\frac{4a_{\U{K}}}{3\sqrt{N}})^2}\right]$:
\begin{align}
\sum_{i=1}^Np_{D,X,i}^{\U{Error}}\le 
N_{D,X}^{\U{Error}}\(1+a_{\U{K}}\frac{2}{\sqrt{N}}\)+(b_{\U{K}}-a_{\U{K}})\sqrt{N}.
\end{align}
By setting $a_{\U{K}}$ and $b_{\U{K}}$ as
\begin{align}
a_{\U{K}}^{D,X}:=a_{\U{K}}\(N,\tilde{N}_{D,X}^{\U{Error}},\frac{\epsilon_{\U{secrecy}}^2}{24}\)~~\U{and}~~b_{\U{K}}^{D,X}:=b_{\U{K}}\(N,\tilde{N}_{D,X}^{\U{Error}},\frac{\epsilon_{\U{secrecy}}^2}{24}\),
\end{align}
respectively,
\begin{align}
\sum_{i=1}^Np_{D,X,i}^{\U{Error}}\le 
N_{D,X}^{\U{Error}}\(1+a_{\U{K}}^{D,X}\frac{2}{\sqrt{N}}\)+(b_{\U{K}}^{D,X}-a_{\U{K}}^{D,X})\sqrt{N}
\label{eq:KatoXBED}
\end{align}
holds except with probability $\frac{\epsilon_{\U{secrecy}}^2}{24}$. Here, $\tilde{N}_{D,X}^{\U{Error}}$ denotes an estimated value of $N_{D,X}^{\U{Error}}$.
\\\\
{\bf 2: Lower bound on $\sum_{i=1}^N p_{V,X,i}^{\U{Error}}$}
\\
From Kato's inequality, Theorem 1 of~\cite{Katoinequality}, the following inequality holds for any $a_{\U{K}},b_{\U{K}}\in \mathbb{R}$ satisfying $b_{\U{K}}\ge|a_{\U{K}}|$:
\begin{align}
\Pr\left[
N_{V,X}^{\U{Error}}\ge \sum_{i=1}^Np_{V,X,i}^{\U{Error}}+\left[b_{\U{K}}+a_{\U{K}}\left(\frac{2N_{V,X}^{\U{Error}}}{N}-1\right)
\right]\sqrt{N}
\right]\le 
\exp\left[-\frac{2b_{\U{K}}^2-2a_{\U{K}}^2}{(1-\frac{4a_{\U{K}}}{3\sqrt{N}})^2}\right].
\end{align}
By setting $a_{\U{K}}$ and $b_{\U{K}}$ as
\begin{align}
a_{\U{K}}^{V,X}:=a'_{\U{K}}\(N,\tilde{N}_{V,X}^{\U{Error}},\frac{\epsilon_{\U{secrecy}}^2}{24}\)~~\U{and}~~
b_{\U{K}}^{V,X}:=b'_{\U{K}}\(N,\tilde{N}_{V,X}^{\U{Error}},\frac{\epsilon_{\U{secrecy}}^2}{24}\),
\end{align}
respectively, 
\begin{align}
\sum_{i=1}^Np_{V,X,i}^{\U{Error}}\ge 
N_{V,X}^{\U{Error}}-
\left[b_{\U{K}}^{V,X}+a_{\U{K}}^{V,X}\left(\frac{2N_{V,X}^{\U{Error}}}{N}-1\right)
\right]\sqrt{N}
\label{eq:KatoXBEV}
\end{align}
holds except with probability $\frac{\epsilon_{\U{secrecy}}^2}{24}$. Here, $\tilde{N}_{V,X}^{\U{Error}}$ denotes an estimated value of $N_{V,X}^{\U{Error}}$.
\\
Substituting the upper bound in Eq.~(\ref{eq:KatoXBED}) and the lower bound in Eq.~(\ref{eq:KatoXBEV}) to Eq.~(\ref{eq:katophN}) results in
\begin{align}
&N_{\U{ph}}
\le 
\left(1-\frac{2a_{\U{K}}^{\U{ph}}}{\sqrt{N}}\right)^{-1}
\Bigg\{
\frac{p^2_Z}{p^2_Xp_{D|1}^{\U{int}}}
\left[N_{D,X}^{\U{Error}}(1+a_{\U{K}}^{D,X}\frac{2}{\sqrt{N}})+(b_{\U{K}}^{D,X}-a_{\U{K}}^{D,X})\sqrt{N}
\right]
\notag\\
&-\frac{p^2_Zp_{D|0}^{\U{int}}}{p^2_Xp_{D|1}^{\U{int}}p_{V|0}^{\U{int}}}
\left[
N_{V,X}^{\U{Error}}-
\left[b_{\U{K}}^{V,X}+a_{\U{K}}^{V,X}\left(\frac{2N_{V,X}^{\U{Error}}}{N}-1\right)
\right]\sqrt{N}
\right]
+(b_{\U{K}}^{\U{ph}}-a_{\U{K}}^{\U{ph}})\sqrt{N}
\Bigg\},
\end{align}
which holds except with probability $\frac{\epsilon_{\U{secrecy}}^2}{8}$. 

\begin{proposition}
    \label{meidai:errorXPQC}
    {\bf  The number of phase error patterns after Step~2 in the virtual protocol}\\
    For the probability 
\begin{equation}
 \Pr_{\U{QC}}(\vec{n},\vec{\omega},\vec{\alpha},\vec{\beta},\vec{e}_{X})
\end{equation}
defined in Eq.~(\ref{state:definition}), there exists a set 
\[
\left\{\Omega_{\U{QC},N_{\U{sift}}}\subset\{0,1\}^{N_{\U{sift}}}\right\}_{N_{\U{sift}}}
\]
with its cardinality satisfying
\begin{align}
     |\Omega_{\U{QC},N_{\U{sift}}}|\leq 2^{N_{\U{sift}}-\underline{N}_{1,Z}}\times 2^{\underline{N}_{1,Z}}h(\overline{N}_{\U{ph}}/\underline{N}_{1,Z}),
     \label{eq:NXpatternQC}
\end{align}
such that
\begin{equation}
 \sum_{N_{\U{sift}}=1}^{N}\Pr_{\U{QC}}\(N_{\U{sift}},(e_{X,i})_{i\in S_{\U{sift}}} \in \Omega_{\U{QC},N_{\U{sift}}}\)
 \geq 1-\frac{1}{4}\epsilon_{\U{secrecy}}^2
 \label{eq:prop11prob}
\end{equation}
holds. Here, $\underline{N}_{1,Z}$ and $\overline{N}_{\U{ph}}$ are defined in 
Eqs.~(\ref{eq:LowerN1zsection25}) and (\ref{eq:UpperN1zsection25}), respectively.
\end{proposition}
{\bf Proof of Proposition~\ref{meidai:errorXPQC}}
\\
The following set
\begin{equation}
\begin{split}
 \Omega_{\U{QC},N_{\U{sift}}}:=
 \bigg\{(e_{X,i})_{i\in S_{\U{sift}}} 
 \mathrel{}\bigg|\mathrel{}
|S_{\U{sift}}|=N_{\U{sift}},~N_{1,Z} \geq \underline{N}_{1,Z},~N_{\U{ph}}\leq \overline{N}_{\U{ph}} \bigg\}
\end{split}
\label{eq:omegaQCNsift}
\end{equation}
satisfies Eq.~(\ref{eq:NXpatternQC}). Recall that $N_{1,Z}$ and $N_{\U{ph}}$ are respectively defined in Eqs.~(\ref{eq:definitionN1Z}) and (\ref{eq:definitionNph}). This is because there is no information about $e_{X,i}$ with $n_i=0$ and $n_i\ge2$, and hence $e_{X,i}$ can take all possible values. For $n_i=1$, as the number of phase errors is at most $\overline{N}_{\U{ph}}$, the number of phase error patterns is upper-bounded by $2^{\underline{N}_{1,Z}h(\overline{N}_{\U{ph}}/\underline{N}_{1,Z})}$. 
\\ 
The probability in Eq.~(\ref{eq:prop11prob}) with the set $\Omega_{\U{QC},N_{\U{sift}}}$ given in Eq.~(\ref{eq:omegaQCNsift}) is evaluated as follows.
\begin{align}
 &\sum_{N_{\U{sift}}=1}^{N}\Pr_{\U{QC}}\(N_{\U{sift}},(e_{X,i})_{i\in S_{\U{sift}}} \notin \Omega_{\U{QC},N_{\U{sift}}}\)\notag\\
 =&\sum_{N_{\U{sift}}=1}^{N}\Pr_{\U{QC}}\Bigg(N_{\U{sift}},N_{1,Z} < \underline{N}_{1,Z}~\U{or}~
 N_{\U{ph}} > \overline{N}_{\U{ph}}\Bigg)\notag\\
 \le&\sum_{N_{\U{sift}}=0}^{N}\Pr_{\U{QC}}\Bigg(N_{\U{sift}},N_{1,Z} < \underline{N}_{1,Z}~\U{or}~
 N_{\U{ph}} > \overline{N}_{\U{ph}}\Bigg)\notag\\
 =&\Pr_{\U{QC}}\Bigg(N_{1,Z} <\underline{N}_{1,Z}~\U{or}~N_{\U{ph}} > \overline{N}_{\U{ph}}\Bigg)\notag\\
 \le&\Pr_{\U{QC}}\(N_{1,Z} < \underline{N}_{1,Z}\)+
 \Pr_{\U{QC}}\(N_{\U{ph}} > \overline{N}_{\U{ph}}\)\notag\\
\le&\frac{1}{4}\epsilon_{\U{secrecy}}^2
\end{align}
The first equation follows from the definition of $\Omega_{\U{QC},N_{\U{sift}}}$. 
The second equation is obtained by marginalizing over $N_{\U{sift}}$. 
The second inequality follows from the union bound. 
The third inequality holds from Eqs.~(\ref{eq:meidai:kato1}) and (\ref{eq:meidai:kato2}).

\begin{proposition}
    \label{meidai:errorXPAmaePQC}
    {\bf The number of phase-error patterns before privacy amplification}
   \\
Let 
\begin{equation}
 \hat{E}_{N_{\U{sift}},\vec{e}_{X}} 
:= \sum_{\vec{x}\in\{0,1\}^{N_{\U{sift}}}} \hat{P}\left[
\ket{N_{\U{sift}},(\vec{x})_X}_{A_{\U{sift}}}\ket{N_{\U{sift}},(\vec{x}\oplus\vec{e}_X)_X}_{B_{\U{sift}}}
\right]
\label{prop9:Eprojector}
\end{equation}
denote the projector onto the subspace where the $X$-basis measurement outcomes performed on $A_{\U{sift}}$ and $B_{\U{sift}}$ differ by $\vec{e}_{X}$ when the sifted key length is $N_{\U{sift}}$. For the state immediately after completing error verification $\hat{\rho}_{\U{verify,vir}}$, defined in Eq.~(\ref{eq:rhoverifyvir}), let
\begin{equation}
 \begin{split}
 \Pr_{\U{verify}}(N_{\U{sift}},\vec{e}_{X})
:=\tr \(\hat{E}_{N_{\U{sift}},\vec{e}_{X}}\hat{\rho}_{\U{verify,vir}}\)
 \end{split}
\end{equation}
represent the probability that the sifted key length is $N_{\U{sift}}$ and that Alice's and Bob's $X$-basis measurement outcomes differ by $\vec{e}_X$. Then, there exists a set
\[
\{\Omega_{\U{verify},N_{\U{sift}}}\subset\{0,1\}^{N_{\U{sift}}}\}_{N_{\U{sift}}}
\]
with
\begin{align}
|\Omega_{\U{verify},N_{\U{sift}}}| &\leq 2^{N_{\U{EC}}+N_{\U{verify}}}2^{N_{\U{sift}}-\underline{N}_{1,Z}} 2^{\underline{N}_{1,Z}h(\overline{N}_{\U{ph}}/\underline{N}_{1,Z})}
=:\overline{|\Omega_{\U{verify},N_{\U{sift}}}|}
\label{eq:meidai12cardOmega}
\end{align}
such that
\begin{equation} \sum_{N_{\U{sift}}=1}^N\Pr_{\U{verify}}(N_{\U{sift}},\vec{e}_{X} \in \Omega_{\U{verify},N_{\U{sift}}}  )\geq 1-\frac{1}{4}\epsilon_{\U{secrecy}}^2.
\end{equation}
\end{proposition}

{\bf Proof of Proposition~\ref{meidai:errorXPAmaePQC}}
\\
From proposition~\ref{meidai:errorXPQC}, there exists a set $\{\Omega_{\U{QC},N_{\U{sift}}}\subset\{0,1\}^{N_{\U{sift}}}\}_{N_{\U{sift}}}$ with $|\Omega_{\U{QC},N_{\U{sift}}}|\leq 2^{N_{\U{sift}}-\underline{N}_{1,Z}} 2^{\underline{N}_{1,Z}h(\overline{N}_{\U{ph}}/\underline{N}_{1,Z})}$ satisfying
\begin{equation}
    \sum_{N_{\U{sift}}=1}^{N}
    \sum_{\vec{e}_X\in \Omega_{\U{QC},N_{\U{sift}}}}
    \tr\(
    \hat{E}_{N_{\U{sift}},\vec{e}_X}(\hat{\rho}_{\U{QC,vir}})
    \) \geq 1-\frac{1}{4}\epsilon_{\U{secrecy}}^2.
\end{equation}
From Eq.~(\ref{eq:kraususiftoperation}), the sifting operation $\mathcal{E}^{\U{sift}}$ does not alter the states of systems $A_i^{\U{CR}}$ and $B_i^{\U{CR}}$ with $i\in S_{\U{sift}}$, and hence $e_{X,i}$ for $i\in S_{\U{sift}}$ remains unchanged by performing $\mathcal{E}^{\U{sift}}$. This implies
\begin{equation}
    \sum_{N_{\U{sift}}=1}^{N}
    \sum_{\vec{e}_X\in \Omega_{\U{QC},N_{\U{sift}}}}
    \tr\(
    \hat{E}_{N_{\U{sift}},\vec{e}_X} {\mathcal{E}^{\U{sift}}}(\hat{\rho}_{\U{QC,vir}})
    \) \geq 1-\frac{1}{4}\epsilon_{\U{secrecy}}^2.
    \label{eq:pro9sift}
\end{equation}
In the proof of this proposition, we first discuss the increase in the number of phase-error patterns $\vec{e}_X$ due to the bit-error correction operation $\mathcal{E}^{\U{EC,vir}}$ in Steps~\ref{step:ec} and \ref{step3bvirtualbob}. We then discuss the increase in the number of phase-error patterns $\vec{e}_X$ due to the error-verification operation $\mathcal{E}^{\U{verify,vir}}$ in Steps~\ref{step:verify} and \ref{stepBobvirverify}.
\\\\
{\bf 1. Increase in the number of $\vec{e}_X$ through $\mathcal{E}^{\U{EC,vir}}$}
\\
The goal here is to prove that, for the set 
\begin{equation}
    \Omega_{\U{EC},N_{\U{sift}}}:=
    \{\vec{e}_X\oplus\vec{b}\tilde{\mathcal{C}}_{\U{synd}}^{\U{T}}
    \mid \forall \vec{e}_X\in \Omega_{\U{QC},N_{\U{sift}}}, \forall \vec{b} \in \{0,1\}^{N_{\U{sift}}}, \forall i\in [N_{\U{sift}}]\setminus [N_{\U{EC}}], b_i=0\},
    \label{def:omega-EC}
\end{equation}
and for the state immediately after performing $\mathcal{E}^{\U{EC,vir}}$, 
\begin{equation}
    \sum_{N_{\U{sift}}=1}^{N} \sum_{\vec{e}_X\in \Omega_{\U{EC},N_{\U{sift}}}}
    \tr\(\hat{E}_{N_{\U{sift}},\vec{e}_X}{\mathcal{E}^{\U{EC,vir}}}\circ{\mathcal{E}^{\U{sift}}}(\hat{\rho}_{\U{QC,vir}})\)
    \geq 1-\frac{1}{4}\epsilon_{\U{secrecy}}^2
\end{equation}
holds. To prove this inequality, we first show that the following three equations [Eqs.~(\ref{eq:prop9Usynd1})-(\ref{eq:prop9Usynd3})] hold for the sets $\Omega_{\U{EC},N_{\U{sift}}}$ and 
\begin{equation}
    \Omega'_{\U{EC},N_{\U{sift}}}:=
    \{\vec{e}_X(\tilde{\mathcal{C}}_{\U{synd}}^{\U{T}})^{-1}\oplus\vec{b}
    \mid \forall \vec{e}_X\in \Omega_{\U{QC},N_{\U{sift}}}, \forall \vec{b} \in \{0,1\}^{N_{\U{sift}}}, \forall i\in [N_{\U{sift}}]\setminus [N_{\U{EC}}], b_i=0\}.
\end{equation}
\begin{enumerate}
    \item 
    Using the definition of unitary operation $\hat{U}_{\U{synd}}(\tilde{\mathcal{C}}_{\U{synd}})$ in Eq.~(\ref{eq:unitarysyndtildeCec}) and the fact that this unitary operation transforms $\vec{e}_X\oplus\vec{b}\tilde{\mathcal{C}}_{\U{synd}}^{\U{T}}$ to $\vec{e}_X(\tilde{\mathcal{C}}_{\U{synd}}^{\U{T}})^{-1}\oplus\vec{b}$, 
\begin{equation}
    \hat{U}_{\U{synd}}(\tilde{\mathcal{C}}_{\U{synd}})
    \(\sum_{\vec{e}_X\in  \Omega_{\U{EC},N_{\U{sift}}}}
    \hat{E}_{N_{\U{sift}},\vec{e}_X}
    \)
    \hat{U}_{\U{synd}}^{\dagger}(\tilde{\mathcal{C}}_{\U{synd}})
    =   \sum_{\vec{e}_X\in  \Omega'_{\U{EC},N_{\U{sift}}}}
     \hat{E}_{N_{\U{sift}},\vec{e}_X}.
     \label{eq:prop9Usynd1}
\end{equation}
\item 
Since the set $\Omega'_{\U{EC},N_{\U{sift}}}$ already covers all the possible patterns of the $X$-basis measurement outcomes for $N_{\U{EC}}$ qubits that are measured with Kraus operators $\hat{K}^{\U{EC,vir}}_{N_{\U{sift}},\vec{a}_{N_{\U{EC}}}}$ in Eq.~(\ref{eq:krausECvir}), the set of $\vec{e}_X$ remains unchanged under this measurement. This implies
\begin{align}
\sum_{\vec{a}\in 
    \{\vec{a}\in\{0,1\}^{N_{\U{sift}}}\mid 
    \vec{a}_{\leq N_{\U{EC}}} = \vec{a}_{N_\U{EC}}\}}\hat{K}^{\U{EC,vir}\dagger}_{N_{\U{sift}},\vec{a}_{N_{\U{EC}}}} 
     \(\sum_{\vec{e}_X\in  \Omega'_{\U{EC},N_{\U{sift}}}}
     \hat{E}_{N_{\U{sift}},\vec{e}_X}
     \)
    \hat{K}^{\U{EC,vir}}_{N_{\U{sift}},\vec{a}_{N_{\U{EC}}}} 
    =\sum_{\vec{e}_X\in  \Omega'_{\U{EC},N_{\U{sift}}}}
    \hat{E}_{N_{\U{sift}},\vec{e}_X}.
   \label{eq:prop9Usynd2}
\end{align}
\item 
Multiplying $\hat{U}_{\U{synd}}^{\dagger}(\tilde{\mathcal{C}}_{\U{synd}})$ by both sides of Eq.~(\ref{eq:prop9Usynd1}) gives
\begin{equation}
    \hat{U}_{\U{synd}}^{\dagger}(\tilde{\mathcal{C}}_{\U{synd}})
    \(\sum_{\vec{e}_X\in  \Omega'_{\U{EC},N_{\U{sift}}}}
    \hat{E}_{N_{\U{sift}},\vec{e}_X}
    \)
    \hat{U}_{\U{synd}}(\tilde{\mathcal{C}}_{\U{synd}})
    =  
     \sum_{\vec{e}_X\in  \Omega_{\U{EC},N_{\U{sift}}}}
     \hat{E}_{N_{\U{sift}},\vec{e}_X}.
          \label{eq:prop9Usynd3}
\end{equation}
\end{enumerate}
Employing Eqs.~(\ref{eq:prop9Usynd1})-(\ref{eq:prop9Usynd3}) in a step by step manner, as for the adjoint map $\mathcal{E}^{\U{EC},A,\U{vir}\dagger}$ of $\mathcal{E}^{\U{EC},A,\U{vir}}$ defined in Eq.~(\ref{eq:EECAvir}),
\begin{equation}
     \mathcal{E}^{\U{EC},A,\U{vir}\dagger}\(\sum_{\vec{e}_X\in  \Omega_{\U{EC},N_{\U{sift}}}}\hat{E}_{N_{\U{sift}},\vec{e}_X} \)
    =\sum_{\vec{e}_X\in  \Omega_{\U{EC},N_{\U{sift}}}}\hat{E}_{N_{\U{sift}},\vec{e}_X}
    \label{eq:ec-A-dist}
\end{equation}
holds. As for Bob's CPTP map $\mathcal{E}^{\U{EC},B,\U{vir}}$ defined in Eq.~(\ref{eq:EECBvir}), since the projection operator in the $X$ basis is invariant under the Pauli $X$ operation:
\[
\hat{X}_{B,N_{\U{sift}},i}\hat{E}_{N_{\U{sift}},\vec{e}_X} \hat{X}_{B,N_{\U{sift}},i} = \hat{E}_{N_{\U{sift}},\vec{e}_X},
\]
\begin{equation}
     \mathcal{E}^{\U{EC},B,\U{vir},\dagger}\(
     \sum_{\vec{e}_X\in  \Omega_{\U{EC},N_{\U{sift}}}}
     \hat{E}_{N_{\U{sift}},\vec{e}_X}\)
    =\sum_{\vec{e}_X\in  \Omega_{\U{EC},N_{\U{sift}}}}\hat{E}_{N_{\U{sift}},\vec{e}_X}
    \label{eq:ec-B-dist}
\end{equation}
holds. From Eqs.~(\ref{eq:EecvirAB}), (\ref{eq:ec-A-dist}) and (\ref{eq:ec-B-dist}), 
\begin{equation}
{\mathcal{E}^{\U{EC,vir}}}^\dagger
\(
\sum_{\vec{e}_X\in  \Omega_{\U{EC},N_{\U{sift}}}}
\hat{E}_{N_{\U{sift}},\vec{e}_X}
        \) 
        = \sum_{\vec{e}_X\in  \Omega_{\U{EC},N_{\U{sift}}}}\hat{E}_{N_{\U{sift}},\vec{e}_X}
        \label{eq:ec-dist}
\end{equation}
is satisfied. 
\\
Then, consider that
\begin{equation}
    \begin{split}
    &\sum_{N_{\U{sift}}=1}^{N} \sum_{\vec{e}_X\in \Omega_{\U{EC},N_{\U{sift}}}}
    \tr\(\hat{E}_{N_{\U{sift}},\vec{e}_X}{\mathcal{E}^{\U{EC,vir}}}\circ{\mathcal{E}^{\U{sift}}}(\hat{\rho}_{\U{QC,vir}})\)
    \\
     =&\sum_{N_{\U{sift}}=1}^{N} \sum_{\vec{e}_X\in \Omega_{\U{EC},N_{\U{sift}}}}
    \tr\({\mathcal{E}^{\U{EC,vir}}}^\dagger(\hat{E}_{N_{\U{sift}},\vec{e}_X}){\mathcal{E}^{\U{sift}}}(\hat{\rho}_{\U{QC,vir}})\)\\
    =&\sum_{N_{\U{sift}}=0}^{N} \sum_{\vec{e}_X\in \Omega_{\U{EC},N_{\U{sift}}}}
    \tr\(\hat{E}_{N_{\U{sift}},\vec{e}_X}{\mathcal{E}^{\U{sift}}}(\hat{\rho}_{\U{QC,vir}})\)\\
        \geq&\sum_{N_{\U{sift}}=0}^{N}
    \sum_{\vec{e}_X\in \Omega_{\U{QC},N_{\U{sift}}}}
    \tr\(
    \hat{E}_{N_{\U{sift}},\vec{e}_X} {\mathcal{E}^{\U{sift}}}(\hat{\rho}_{\U{QC,vir}})
    \) \\
    \geq & 1-\frac{1}{4}\epsilon_{\U{secrecy}}^2.
    \end{split}
    \label{eq:EC-prob-bound}
\end{equation}
The first equation follows from the fact that
\begin{align}
\tr\(\hat{E}\mathcal{E}(\hat{\rho})\)=
\tr\(\mathcal{E}^{\dagger}(\hat{E})\hat{\rho}\)
\label{eq:derivationfirstequation}
\end{align}
holds for any operator $\hat{E}$, any CPTP map $\mathcal{E}$, and any density operator $\hat{\rho}$
    \footnote{
Note that Eq.~(\ref{eq:derivationfirstequation}) can be proven using the Kraus representation of the CPTP map, along with the cyclic property and the linearity of the trace.}. The second equation follows by Eq.~(\ref{eq:ec-dist}). The first inequality follows by 
\begin{equation}
    \sum_{\vec{e}_X\in  \Omega_{\U{EC},N_{\U{sift}}}}\hat{E}_{N_{\U{sift}},\vec{e}_X} \geq 
    \sum_{\vec{e}_X\in  \Omega_{\U{QC},N_{\U{sift}}}}\hat{E}_{N_{\U{sift}},\vec{e}_X},
    \label{eq:EC-triv-ineq}
\end{equation}
which is obtained from Eq.~(\ref{def:omega-EC}). The second inequality follows from Eq.~(\ref{eq:pro9sift}). 
\\\\
{\bf 2. Increase in the number of $\vec{e}_X$ through $\mathcal{E}^{\U{verify,vir}}$}
\\
Next, we discuss the increase in the number of phase-error patterns $\vec{e}_X$ due to the error-verification $\mathcal{E}^{\U{verify,vir}}$ in Steps~\ref{step:verify} and \ref{stepBobvirverify}. The goal here is to prove that, for the set
\begin{equation}
\begin{split}
    & \Omega_{\U{verify},N_{\U{sift}}}
     :=
    \{\vec{e}_X\oplus\vec{b}\tilde{\mathcal{C}}_{\U{synd}}^{\U{T}}
    \oplus\vec{c}\tilde{\mathcal{C}}_{\U{verify}}^{\U{T}}
    \mid \forall \vec{e}_X\in \Omega_{\U{QC},N_{\U{sift}}}, \forall \vec{b} \in \{0,1\}^{N_{\U{sift}}}, \forall i\in [N_{\U{sift}}]\setminus [N_{\U{EC}}], b_i=0,\\
    &\forall \vec{c} \in \{0,1\}^{N_{\U{sift}}}, \forall i\in [N_{\U{sift}}]\setminus [N_{\U{verify}}], b_i=0
    \},
\end{split}
\label{def:omega-verify}
\end{equation}
and for the state immediately after performing $\mathcal{E}^{\U{verify,vir}}$, 
\begin{equation}
    \sum_{N_{\U{sift}}=0}^{N} \sum_{\vec{e}_X\in \Omega_{\U{verify},N_{\U{sift}}}}
    \tr\(\hat{E}_{N_{\U{sift}},\vec{e}_X}{\mathcal{E}^{\U{verify,vir}}}\circ{\mathcal{E}^{\U{EC,vir}}}\circ{\mathcal{E}^{\U{sift}}}(\hat{\rho}_{\U{QC,vir}})\)
    \geq 1-\frac{1}{4}\epsilon_{\U{secrecy}}^2
\end{equation}
holds. To prove this inequality, we first show that the following three equations [Eqs.~(\ref{eq:prop9Uverify1})-(\ref{eq:prop9Uverify3})]  hold for the sets $\Omega_{\U{verify},N_{\U{sift}}}$ and 
\begin{equation}
\begin{split}
    & \Omega'_{\U{verify},N_{\U{sift}}}
     :=
    \{\vec{e}_X(\tilde{\mathcal{C}}_{\U{verify}}^{\U{T}})^{-1}\oplus\vec{b}\tilde{\mathcal{C}}_{\U{synd}}^{\U{T}}(\tilde{\mathcal{C}}_{\U{verify}}^{\U{T}})^{-1}
    \oplus\vec{c}
    \mid \forall \vec{e}_X\in \Omega_{\U{QC},N_{\U{sift}}}, \\&
    \forall \vec{b} \in \{0,1\}^{N_{\U{sift}}}, \forall i\in [N_{\U{sift}}]\setminus [N_{\U{EC}}], b_i=0,\forall \vec{c} \in \{0,1\}^{N_{\U{sift}}}, \forall i\in [N_{\U{sift}}]\setminus [N_{\U{verify}}], b_i=0
    \}.
\end{split}
\end{equation}
\begin{enumerate}
    \item 
    Using the definition of the unitary operation $\hat{U}_{\U{verify}}(\tilde{\mathcal{C}}_{\U{verify}})$ in Eq.~(\ref{eq:unitaryverify}) and the fact that this unitary operation transforms $\vec{e}_X\oplus\vec{b}\tilde{\mathcal{C}}_{\U{synd}}^{\U{T}}
    \oplus\vec{c}\tilde{\mathcal{C}}_{\U{verify}}^{\U{T}}$ to $\vec{e}_X(\tilde{\mathcal{C}}_{\U{verify}}^{\U{T}})^{-1}\oplus\vec{b}\tilde{\mathcal{C}}_{\U{synd}}^{\U{T}}(\tilde{\mathcal{C}}_{\U{verify}}^{\U{T}})^{-1}
    \oplus\vec{c}$, 
\begin{equation}
        \hat{U}_{\U{verify}}(\tilde{\mathcal{C}}_{\U{verify}})
\(
    \sum_{\vec{e}_X\in \Omega_{\U{verify},N_{\U{sift}}}}
    \hat{E}_{N_{\U{sift}},\vec{e}_X}
 \)
    \hat{U}_{\U{verify}}^{\dagger}(\tilde{\mathcal{C}}_{\U{verify}})
    = \sum_{\vec{e}_X\in \Omega'_{\U{verify},N_{\U{sift}}}}
    \hat{E}_{N_{\U{sift}},\vec{e}_X}
    \label{eq:prop9Uverify1}
\end{equation}
holds.
\item 
Since the set $\Omega'_{\U{verify},N_{\U{sift}}}$ already covers all possible patterns of the $X$-basis measurement outcomes for $N_{\U{verify}}$ qubits that are measured with Kraus operators $\hat{K}^{\U{verify,vir}}_{N_{\U{sift}},\vec{a}_{N_{\U{verify}}},\vec{b}_{N_{\U{verify}}}}$ defined in Eq.~(\ref{eq:krausverify}), the set of $\vec{e}_X$ remains unchanged under this measurement. This implies
\begin{align}
   & 
   \sum_{
   \vec{a}_{N_{\U{verify}}},\vec{b}_{N_{\U{verify}}}\in\{0,1\}^{N_{\U{verify}}}}
\hat{K}^{\U{verify,vir}\dagger}_{N_{\U{sift}},\vec{a}_{N_{\U{verify}}},\vec{b}_{N_{\U{verify}}}}\(\sum_{\vec{e}_X\in \Omega'_{\U{verify},N_{\U{sift}}}}
\hat{E}_{N_{\U{sift}},\vec{e}_X}
\)\hat{K}^{\U{verify,vir}}_{N_{\U{sift}},\vec{a}_{N_{\U{verify}}},\vec{b}_{N_{\U{verify}}}} 
\notag\\
    =& \sum_{\vec{e}_X\in \Omega'_{\U{verify},N_{\U{sift}}}}
    \hat{\Pi}^{\U{verifyPass}}_{N_{\U{sift}}}\hat{E}_{N_{\U{sift}},\vec{e}_X} \hat{\Pi}^{\U{verifyPass}}_{N_{\U{sift}}}
    \label{eq:prop9Uverify2}
\end{align}
where $\hat{\Pi}^{\U{verifyPass}}_{N_{\U{sift}}}$ is a projector defined by
\begin{equation}
\begin{split}
     \hat{\Pi}^{\U{verifyPass}}_{N_{\U{sift}}}
     :=& \sum_{\vec{a}_{N_{\U{verify}}}\in\{0,1\}^{N_{\U{verify}}}}
    \sum_{\vec{a}\in \{0,1\}^{N_{\U{sift}}}, \vec{a}_{\leq N_{\U{verify}}}=\vec{a}_{N_{\U{verify}}}}
    \hat{P}[\ket{N_{\U{sift}},\vec{a}}_{A_{\U{sift}}}]\\
    &\otimes\sum_{\vec{b}\in \{0,1\}^{N_{\U{sift}}}, \vec{b}_{\leq N_{\U{verify}}}=\vec{a}_{N_{\U{verify}}}}\hat{P}[\ket{N_{\U{sift}},\vec{b}}_{B_{\U{sift}}}]\otimes \hat{P}[\ket{1}_{C^{\U{Length}}_{\U{Judge}}}].
\end{split}
\end{equation}
This projector  $\hat{\Pi}^{\U{verifyPass}}_{N_{\U{sift}}}$ commutes with $\sum_{\vec{e}_X\in \Omega'_{\U{verify},N_{\U{sift}}}} \hat{E}_{N_{\U{sift}},\vec{e}_X}$  because $\sum_{\vec{e}_X\in \Omega'_{\U{verify},N_{\U{sift}}}} \hat{E}_{N_{\U{sift}},\vec{e}_X}$ acts as an identity operator for first $N_{\U{verify}}$ bits of $\vec{a}$ and $\vec{b}$ in 
    $\ket{N_{\U{sift}},\vec{a}}_{A_{\U{sift}}}$ and $\ket{N_{\U{sift}},\vec{a}}_{B_{\U{sift}}}$, and $\hat{\Pi}^{\U{verifyPass}}$ acts as an identity operator for last $N_{\U{sift}}-N_{\U{verify}}$ bits of them.

\item 
Multiplying $\hat{U}_{\U{verify}}^{\dagger}(\tilde{\mathcal{C}}_{\U{verify}})$ by both sides of Eq.~(\ref{eq:prop9Uverify1}) gives
\begin{equation}
        \hat{U}_{\U{verify}}^{\dagger}(\tilde{\mathcal{C}}_{\U{verify}})  
    \(
   \sum_{\vec{e}_X\in \Omega'_{\U{verify},N_{\U{sift}}}}\hat{E}_{N_{\U{sift}},\vec{e}_X}\)
    \hat{U}_{\U{verify}}(\tilde{\mathcal{C}}_{\U{verify}})
    = \sum_{\vec{e}_X\in \Omega_{\U{verify},N_{\U{sift}}}}
    \hat{E}_{N_{\U{sift}},\vec{e}_X}.
        \label{eq:prop9Uverify3}
\end{equation}
\end{enumerate}
Employing Eqs.~(\ref{eq:prop9Uverify1})-(\ref{eq:prop9Uverify3}) in a step by step manner, as for the adjoint map $\mathcal{E}^{\U{verify,vir}\dagger}$ of $\mathcal{E}^{\U{verify,vir}}$ in Eq.~(\ref{eq:mapverifyvir}),
\begin{equation}
         \mathcal{E}^{\U{verify,vir}\dagger}
    \(\sum_{\vec{e}_X\in \Omega_{\U{verify},N_{\U{sift}}}}
    \hat{E}_{N_{\U{sift}},\vec{e}_X}\)
    =\sum_{\vec{e}_X\in \Omega_{\U{verify},N_{\U{sift}}}}
    \hat{\Pi}^{\U{verifyPass}}_{N_{\U{sift}}}\hat{E}_{N_{\U{sift}},\vec{e}_X}\hat{\Pi}^{\U{verifyPass}}_{N_{\U{sift}}}
    \label{eq:pro9verifysecond}
\end{equation}
holds. 
\\
Let $\mathbf{1}_{N_{\U{sift}}}$ be denoted by $\sum_{\vec{a},\vec{b}\in\{0,1\}^{N_{\U{sift}}}}P[\ket{N_{\U{sift}},\vec{a}}_{A_{\U{sift}}}\ket{N_{\U{sift}},\vec{a}}_{B_{\U{sift}}}]$. From the discussions so far, the target probability in this proposition is calculated as follows. 
\begin{equation}
    \begin{split}
    &\sum_{N_{\U{sift}}=1}^{N} \sum_{\vec{e}_X\not\in \Omega_{\U{verify},N_{\U{sift}}}}
    \tr\(\hat{E}_{N_{\U{sift}},\vec{e}_X}{\mathcal{E}^{\U{verify,vir}}}\circ{\mathcal{E}^{\U{EC,vir}}}\circ{\mathcal{E}^{\U{sift}}}(\hat{\rho}_{\U{QC,vir}})\)\\
    =&\sum_{N_{\U{sift}}=1}^{N} 
    \tr\((\mathbf{1}_{N_{\U{sift}}} -\sum_{\vec{e}_X\in \Omega_{\U{verify},N_{\U{sift}}}}\hat{E}_{N_{\U{sift}},\vec{e}_X}){\mathcal{E}^{\U{verify,vir}}}\circ{\mathcal{E}^{\U{EC,vir}}}\circ{\mathcal{E}^{\U{sift}}}(\hat{\rho}_{\U{QC,vir}})\)\\
    =&\sum_{N_{\U{sift}}=1}^{N} 
    \tr\({\mathcal{E}^{\U{verify,vir}}}^\dagger(\mathbf{1}_{N_{\U{sift}}}-\sum_{\vec{e}_X\in \Omega_{\U{verify},N_{\U{sift}}}}\hat{E}_{N_{\U{sift}},\vec{e}_X}){\mathcal{E}^{\U{EC,vir}}}\circ{\mathcal{E}^{\U{sift}}}(\hat{\rho}_{\U{QC,vir}})\)\\
    =&\sum_{N_{\U{sift}}=1}^{N} 
    \tr\(
    \(\hat{\Pi}^{\U{verifyPass}}_{N_{\U{sift}}}-\sum_{\vec{e}_X\in \Omega_{\U{verify},N_{\U{sift}}}}\hat{\Pi}^{\U{verifyPass}}_{N_{\U{sift}}}\hat{E}_{N_{\U{sift}},\vec{e}_X}\hat{\Pi}^{\U{verifyPass}}_{N_{\U{sift}}}\)\mathcal{E}^{\U{EC,vir}}\circ{\mathcal{E}^{\U{sift}}}(\hat{\rho}_{\U{QC,vir}})\)\\
    \leq &\sum_{N_{\U{sift}}=1}^{N} 
    \tr\(
    \(\mathbf{1}_{N_{\U{sift}}}-\sum_{\vec{e}_X\in \Omega_{\U{verify},N_{\U{sift}}}}\hat{E}_{N_{\U{sift}},\vec{e}_X}\)\mathcal{E}^{\U{EC,vir}}\circ{\mathcal{E}^{\U{sift}}}(\hat{\rho}_{\U{QC,vir}})\)\\
    = &1-\sum_{N_{\U{sift}}=1}^{N} \sum_{\vec{e}_X\in \Omega_{\U{verify},N_{\U{sift}}}}
    \tr\(    \hat{E}_{N_{\U{sift}},\vec{e}_X}\mathcal{E}^{\U{EC,vir}}\circ{\mathcal{E}^{\U{sift}}}(\hat{\rho}_{\U{QC,vir}})\)\\
    \leq &1-\sum_{N_{\U{sift}}=1}^{N} \sum_{\vec{e}_X\in  \Omega_{\U{EC},N_{\U{sift}}}}
    \tr\(    \hat{E}_{N_{\U{sift}},\vec{e}_X}\mathcal{E}^{\U{EC,vir}}\circ{\mathcal{E}^{\U{sift}}}(\hat{\rho}_{\U{QC,vir}})\)\\
    \leq& 1 - \(1-\frac{1}{4}\epsilon_{\U{secrecy}}^2 \) = \frac{1}{4}\epsilon_{\U{secrecy}}^2
    \end{split}   
\end{equation}
The first equation is the decomposition of $\mathbf{1}_{A_{\U{sift}}B_{\U{sift}}}$ to $\mathbf{1}_{N_{\U{sift}}}$.
The second equation is derived using the same reasoning as the one in Eq.~(\ref{eq:EC-prob-bound}), as explained in Eq.~(\ref{eq:derivationfirstequation}). The third equation follows by Eq.~(\ref{eq:pro9verifysecond}). The first inequality follows from the commutativity of $\hat{\Pi}^{\U{verifyPass}}_{N_{\U{sift}}}$ and $\sum_{\vec{e}_X\in \Omega'_{\U{verify},N_{\U{sift}}}} \hat{E}_{N_{\U{sift}},\vec{e}_X}$, the property of projectors $(\hat{\Pi}^{\U{verifyPass}}_{N_{\U{sift}}})^2 = \hat{\Pi}^{\U{verifyPass}}_{N_{\U{sift}}}$, which implies
\begin{equation}
\begin{split}
     &\hat{\Pi}^{\U{verifyPass}}_{N_{\U{sift}}}-\sum_{\vec{e}_X\in \Omega_{\U{verify},N_{\U{sift}}}}\hat{\Pi}^{\U{verifyPass}}_{N_{\U{sift}}}\hat{E}_{N_{\U{sift}},\vec{e}_X}\hat{\Pi}^{\U{verifyPass}}_{N_{\U{sift}}}\\
    =&  \mathbf{1}_{N_{\U{sift}}}-\sum_{\vec{e}_X\in \Omega_{\U{verify},N_{\U{sift}}}}\hat{E}_{N_{\U{sift}},\vec{e}_X}\\
    & - \(\mathbf{1}_{N_{\U{sift}}}-\hat{\Pi}^{\U{verifyPass}}_{N_{\U{sift}}}\) \(\mathbf{1}_{N_{\U{sift}}}-\sum_{\vec{e}_X\in \Omega_{\U{verify},N_{\U{sift}}}}\hat{E}_{N_{\U{sift}},\vec{e}_X}\)\(\mathbf{1}_{N_{\U{sift}}}-\hat{\Pi}^{\U{verifyPass}}_{N_{\U{sift}}}\),
\end{split}
\end{equation}
where the third term of the right-hand side is a positive operator. The fourth equality comes from  the decomposition of $\mathbf{1}_{A_{\U{sift}}B_{\U{sift}}}$.
The second inequality follows from
\begin{equation}
    \sum_{\vec{e}_X\in  \Omega_{\U{verify},N_{\U{sift}}}}\hat{E}_{N_{\U{sift}},\vec{e}_X} \geq 
    \sum_{\vec{e}_X\in  \Omega_{\U{EC},N_{\U{sift}}}}\hat{E}_{N_{\U{sift}},\vec{e}_X}, 
    \label{eq:verify-triv-ineq}
\end{equation}
which is obtained from Eqs.~(\ref{def:omega-EC}) and (\ref{def:omega-verify}). The third inequality follows by Eq.~(\ref{eq:EC-prob-bound}).
\\\\
Since the cardinality of set $\{\vec{b} \in \{0,1\}^{N_{\U{sift}}}\mid \forall i\in [N_{\U{sift}}]\setminus [N_{\U{EC}}], b_i=0\}$ is $2^{N_{\U{EC}}}$, cardinality of set $\{\vec{c} \in \{0,1\}^{N_{\U{sift}}}\mid \forall i\in [N_{\U{sift}}]\setminus [N_{\U{verify}}], c_i=0\}$ is $2^{N_{\U{verify}}}$, and from Eq.~(\ref{eq:NXpatternQC}), the cardinality of $\Omega_{\U{verify},N_{\U{sift}}}$ is upper-bounded as
\begin{equation}
    |\Omega_{\U{verify},N_{\U{sift}}}|
    \leq 2^{N_{\U{EC}}+N_{\U{verify}}}2^{N_{\U{sift}}-\underline{N}_{1,Z}} 2^{\underline{N}_{1,Z}h(\overline{N}_{\U{ph}}/\underline{N}_{1,Z})}.
\end{equation}

\begin{proposition}
    \label{meidai:errorXPAatoPQC}
    {\bf The number of phase error patterns after privacy amplification}
    \\
        Let 
    \begin{equation}
\begin{split}
\hat{E}_{N_{\U{fin}},\vec{e}_{X}}
:=
\hat{P}\left[\sum_{\vec{k}_A\in \{0,1\}^{N_{\U{fin}}}}
2^{-N_{\U{fin}}}
(-1)^{-\vec{k}_A\cdot\vec{e}_X}
\ket{N_{\U{fin}},\vec{k}_A}_{A_{\U{sift}}}
\right]
\end{split}
\end{equation}
denote   the projector onto the subspace where the $X$-basis measurement outcome performed on $A_{\U{sift}}$ is $\vec{e}_X\in\{0,1\}^{N_{\U{fin}}}$ when the secret key length is $N_{\U{fin}}$. 
\\
For the final state in the virtual protocol:
\begin{equation}
  \hat{\rho}_{\U{PA,vir}}:= 
  \tr_{R_1...R_N}[\mathcal{E}^{\rm final}\circ
{\mathcal{E}^{\U{PA},\U{vir}}}\circ {\mathcal{E}^{\U{verify,vir}}} \circ {\mathcal{E}^{\U{EC,\vir}}} \circ {\mathcal{E}^{\U{sift}}}(\hat{\rho}_{\U{QC,vir}})],
\label{eq:rhoPAvir}
\end{equation}
let
\begin{equation}
 \begin{split}
 \Pr_{\U{PA}}(N_{\U{fin}},\vec{e}_{X}):=\tr \(\hat{E}_{N_{\U{fin}},\vec{e}_{X}}
\hat{\rho}_{\U{PA,vir}}
\)
 \end{split}
\label{eq:PrPANfinex}
\end{equation}
represent the probability that the sifted key length is $N_{\U{sift}}$ and that Alice obtains the $X$-basis measurement outcome $\vec{e}_X\in\{0,1\}^{N_{\U{fin}}}$. Then, 
\begin{equation}
\begin{split}
    &\sum_{N_{\U{fin}}=1}^{N} \Pr_{\U{PA}}(N_{\U{fin}})F\({\tr}_E\hat{\rho}_{|N_{\U{fin}}}^{\U{PA,vir}},\hat{P}[\ket{N_{\U{fin}},+^{N_{\rm fin}}}_{A_{\U{sift}}}]\)\\
 = &\sum_{N_{\U{fin}}=1}^{N} \Pr_{\U{PA}}(N_{\U{fin}})
    \tr \(\hat{E}_{N_{\U{fin}},\vec{e}_{X} = 0^{\otimes N_{\U{fin}}}}
\tr_E\hat{\rho}_{|N_{\U{fin}}}^{\U{PA,vir}}
\)\\   
   =&
  \sum_{N_{\U{fin}}=1}^N\Pr_{\U{PA}}(N_{\U{fin}},\vec{e}_X
  = 0^{\otimes N_{\U{fin}}}) 
  \\
  \ge& 1-\frac{1}{2}\epsilon_{\U{secrecy}}^2
\end{split}
\label{eq:meidai13profinal}
\end{equation}
holds. Here, $\ket{N_{\U{fin}},+^{N_{\rm fin}}}_{A_{\U{sift}}}$ is defined in Eq.~(\ref{eq:ketplusX}).
    \end{proposition}
{\bf Proof of Proposition~\ref{meidai:errorXPAatoPQC}}
\\
Let
\begin{align}
\hat{\rho}_{\U{verify,vir}}
:=
{\mathcal{E}^{\U{verify,vir}}}\circ{\mathcal{E}^{\U{EC,vir}}}\circ{\mathcal{E}^{\U{sift}}}(\hat{\rho}_{\U{QC,vir}})
\end{align}
be the state of Alice's, Bob's and Eve's systems immediately after completing error verification. To calculate the target probability in Eq.~(\ref{eq:meidai13profinal}), let 
$$
\hat{D}_X: A_{\U{sift}}\to A_{\U{sift}}
$$
denote an operation that preserves the diagonal elements of the input state in the $X$ basis while setting the off-diagonal elements to 0. This operation classicalizes the input state, resulting in a classical bit string in the $X$ basis. 
\\
Also, using definitions of $\hat{E}_{N_{\U{sift}},\vec{e}_{X}}$ and $\Omega_{\U{verify},N_{\U{sift}}}$ in Eqs.~(\ref{prop9:Eprojector}) and (\ref{def:omega-verify}), let
\begin{align}
\hat{E}_{\U{good}}:=\sum_{N_{\U{sift}}=1}^N
\sum_{\vec{e}_X\in\Omega_{\U{verify},N_{\U{sift}}}}\hat{E}_{N_{\U{sift}},\vec{e}_{X}}
\end{align}
denote the projector onto a ``good space" in the sense that the number of phase error patterns immediately before privacy amplification is upper bounded as specified in Eq.~(\ref{eq:meidai12cardOmega}). Then, the probability in Eq.~(\ref{eq:meidai13profinal}) is calculated as follows.
\begin{align}
&\sum_{N_{\U{fin}}=1}^N\Pr_{\U{PA}}(N_{\U{fin}},\vec{e}_X= 0^{\otimes N_{\U{fin}}})
\notag\\
=&\sum_{N_{\U{fin}}=1}^N\tr \left(\hat{E}_{N_{\U{fin}},\vec{e}_{X}=0}\hat{\rho}_{\U{PA,vir}}\right)
\notag\\
=&\sum_{N_{\U{fin}}=1}^N\tr \left(\hat{E}_{N_{\U{fin}},\vec{e}_{X}=0}
{\mathcal{E}^{\U{final}}}\circ
{\mathcal{E}^{\U{PA},\U{vir}}}
(\hat{\rho}_{\rm verify,vir})
\right)
\notag\\
=&\sum_{N_{\U{fin}}=1}^N\tr \left(\hat{E}_{N_{\U{fin}},\vec{e}_{X}=0}
{\mathcal{E}^{\U{final}}}\circ
{\mathcal{E}^{\U{PA},\U{vir}}}\hat{D}_X
(\hat{\rho}_{\rm verify,vir})
\right)
\notag\\
=&
\sum_{N_{\U{fin}}=1}^N\tr \left(\hat{E}_{N_{\U{fin}},\vec{e}_{X}=0}
{\mathcal{E}^{\U{final}}}\circ
{\mathcal{E}^{\U{PA},\U{vir}}}\hat{D}_X
(\hat{E}_{\U{good}}\hat{\rho}_{\rm verify,vir}\hat{E}_{\U{good}})\right)
\notag\\
+&
\sum_{N_{\U{fin}}=1}^N\tr \left(\hat{E}_{N_{\U{fin}},\vec{e}_{X}=0}
{\mathcal{E}^{\U{final}}}\circ
{\mathcal{E}^{\U{PA},\U{vir}}}\hat{D}_X
(\hat{I}-\hat{E}_{\U{good}})\hat{\rho}_{\rm verify,vir}(\hat{I}-\hat{E}_{\U{good}})\right)
\notag\\
\ge&
\sum_{N_{\U{fin}}=1}^N\tr \left(\hat{E}_{N_{\U{fin}},\vec{e}_{X}=0}
{\mathcal{E}^{\U{final}}}\circ
{\mathcal{E}^{\U{PA},\U{vir}}}\hat{D}_X
(\hat{E}_{\U{good}}\hat{\rho}_{\rm verify,vir}\hat{E}_{\U{good}})\right)
\notag\\
=&
\tr[\hat{E}_{\U{good}}\hat{\rho}_{\U{verify,vir}}]
\sum_{N_{\U{fin}}=1}^N\tr \left(\hat{E}_{N_{\U{fin}},\vec{e}_{X}=0}
{\mathcal{E}^{\U{final}}}\circ
{\mathcal{E}^{\U{PA},\U{vir}}}\hat{D}_X
(\hat{\sigma}_{\rm verify,vir})\right)
\notag\\
\ge&\left(1-\frac{1}{4}\epsilon_{\U{secrecy}}^2\right)
\sum_{N_{\U{fin}}=1}^N\tr \left(\hat{E}_{N_{\U{fin}},\vec{e}_{X}=0}
{\mathcal{E}^{\U{final}}}\circ{\mathcal{E}^{\U{PA},\U{vir}}}\hat{D}_X
(\hat{\sigma}_{\rm verify,vir})\right)
\label{eq:meidai13main}
\end{align}
The first equation follows by Eq.~(\ref{eq:PrPANfinex}). The second equation comes from Eq.~(\ref{eq:rhoPAvir}). The third equality, multiplying by $\hat{D}_X$, means that $\hat{\rho}_{\U{verify, vir}}$ is classicalized in the $X$ basis. The reasons why multiplying by $\hat{D}_X$ is allowed are as follows: 
\begin{enumerate}
\item
The unitary operation $\hat{U}_{\U{PA}}$ within $\mathcal{E}^{\U{PA1,vir}}_{N_{\U{sift}}}$ in Eq.~(\ref{eq:CPTPPAvir1}) is a binary matrix in the $X$ basis, according to Eq.~(\ref{eq:UPACPA}), and it does not create superpositions of states in the $X$ basis.
\item
The unitary operation $\hat{U}_Z(\vec{x}_A^{\U{PA}},\vec{x}_B^{\U{PA}})$ within $\mathcal{E}^{\U{PA3,vir}}_{N_{\U{sift}},N_{\U{fin}},\vec{x}_A^{\U{PA}},\vec{x}_B^{\U{PA}}}$ in Eq.~(\ref{eq:CPTPPAvir3}) is a bit-flip operation in the $X$ basis. 
\end{enumerate}
In the fourth equation, the terms with $(\hat{I}-\hat{E}_{\U{good}})\hat{\rho}_{\rm verify,vir}\hat{E}_{\U{good}}$ and $\hat{E}_{\U{good}}\hat{\rho}_{\rm verify,vir}(\hat{I}-\hat{E}_{\U{good}})$ do not appear because these are off-diagonal elements in the $X$ basis, which vanish upon applying the operation $\hat{D}_X$. The first inequality comes from the non-negativity of the second term in the fourth equation. In the fifth equation, we defined
\begin{align}
\hat{\sigma}_{\U{verify,vir}}
:=\frac{\hat{E}_{\U{good}}(\hat{\rho}_{\U{verify, vir}})\hat{E}_{\U{good}}}
{\tr\left(\hat{E}_{\U{good}}\hat{\rho}_{\U{verify, vir}}\right)}.
\end{align}
The second inequality follows by proposition~\ref{meidai:errorXPAmaePQC}. 
\\
From Eq.~(\ref{eq:CPTPPAvir})

\begin{equation}
\begin{split}
     &\mathcal{E}^{\U{PA,vir}}\(\hat{D}_X(\hat{\sigma}_{\U{verify,vir}})\) 
     =\sum_{N_{\U{sift}},N_{\U{fin}}=1:N_{\U{sift}}\ge N_{\U{fin}}}^N
     \sum_{\vec{x}_B^{\U{PA}}\in\{0,1\}
    ^{N_{\U{sift}}}}\\
    &
    \underbrace{
    \sum_{\vec{x}_A^{\U{PA}}\in\{0,1\}^{N_{\U{sift}}-N_{\U{fin}}}}\mathcal{E}_{N_{\U{sift}},N_{\U{fin}}}^{\U{PA4,vir}}
    \circ\mathcal{E}_{N_{\U{sift}},N_{\U{fin}},\vec{x}_A^{\U{PA}},\vec{x}_B^{\U{PA}}}^{\U{PA3,vir}}
    \circ\mathcal{E}_{N_{\U{sift}},N_{\U{fin}},\vec{x}_A^{\U{PA}},\vec{x}_B^{\U{PA}}}^{\U{PA2,vir}}
    \circ\mathcal{E}_{N_{\U{sift}}}^{\U{PA1,vir}}}
    _{=:\mathcal{E}^{\U{PA,vir}}_{N_{\U{sift}},N_{\U{fin}},\vec{x}_B^{\U{PA}}}}
\(\hat{D}_X(\hat{\sigma}_{\U{verify,vir}})\),
\end{split}
\end{equation}
and using this results in
\begin{align}
&\sum_{N_{\U{fin}}=1}^N\tr \left(\hat{E}_{N_{\U{fin}},\vec{e}_{X}=0}
{\mathcal{E}^{\U{final}}}\circ{\mathcal{E}^{\U{PA},\U{vir}}}\hat{D}_X
(\hat{\sigma}_{\rm verify,vir})\right)
\notag\\
=&
\sum_{N_{\U{sift}},N_{\U{fin}}=1:N_{\U{sift}}\ge N_{\U{fin}}}^N
     \sum_{\vec{x}_B^{\U{PA}}\in\{0,1\}
    ^{N_{\U{sift}}}}
    \tr\left[\mathcal{E}^{\U{PA,vir}}_{N_{\U{sift}},N_{\U{fin}},\vec{x}_B^{\U{PA}}}
\hat{D}_X(\hat{\sigma}_{\U{verify,vir}})\right]
\notag\\
&\frac{
    \expect{N_{\U{fin}},+^{N_{\U{fin}}}|{\mathcal{E}^{\U{final}}}
\circ
\mathcal{E}^{\U{PA,vir}}_{N_{\U{sift}},N_{\U{fin}},\vec{x}_B^{\U{PA}}}
\hat{D}_X(\hat{\sigma}_{\U{verify,vir}})
|N_{\U{fin}},+^{N_{\U{fin}}}}
}
{    \tr\left[\mathcal{E}^{\U{PA,vir}}_{N_{\U{sift}},N_{\U{fin}},\vec{x}_B^{\U{PA}}}
\hat{D}_X(\hat{\sigma}_{\U{verify,vir}})\right]
}.
\label{eq:meidai13middle}
\end{align}
Once $\vec{x}_B^{\U{PA}}$ is fixed, by Alice measuring $N_{\U{sift}}-N_{\U{fin}}$ qubits of system $A_{\U{sift}}$ in the $X$ basis, the failure probability to uniquely identify the $X$-basis measurement outcome after $\mathcal{E}^{\U{final}}$ is upper-bounded by
\begin{align}
\overline{|\Omega_{\U{verify},N_{\U{sift}}}|}
\times
2^{N_{\U{sift}}-N_{\U{fin}}}
&=2^{N_{\U{EC}}+N_{\U{verify}}}2^{N_{\U{sift}}-\underline{N}_{1,Z}} 2^{\underline{N}_{1,Z}h(\overline{N}_{\U{ph}}/\underline{N}_{1,Z})}\times
2^{N_{\U{sift}}-N_{\U{fin}}}
\notag\\
&=2^{-\log_2\frac{\epsilon_{\U{secrecy}}^2}{4}}=\frac{\epsilon_{\U{secrecy}}^2}{4}.
\end{align}
Therefore, Eq.~(\ref{eq:meidai13middle}) leads to
\begin{align}
&\sum_{N_{\U{fin}}=1}^N\tr \left(\hat{E}_{N_{\U{fin}},\vec{e}_{X}=0}
{\mathcal{E}^{\U{final}}}\circ{\mathcal{E}^{\U{PA},\U{vir}}}\hat{D}_X
(\hat{\sigma}_{\rm verify,vir})\right)
\notag\\
\ge&
\left(1-\frac{1}{4}\epsilon_{\U{secrecy}}^2\right)
\sum_{N_{\U{sift}},N_{\U{fin}}=1:N_{\U{sift}}\ge N_{\U{fin}}}^N
     \sum_{\vec{x}_B^{\U{PA}}\in\{0,1\}
    ^{N_{\U{sift}}}}
    \tr\left[\mathcal{E}^{\U{PA,vir}}_{N_{\U{sift}},N_{\U{fin}},\vec{x}_B^{\U{PA}}}
\hat{D}_X(\hat{\sigma}_{\U{verify,vir}})\right]
\notag\\
=&
1-\frac{1}{4}\epsilon_{\U{secrecy}}^2.
\end{align}
Combining this with Eq.~(\ref{eq:meidai13main}) results in
\begin{align}
\sum_{N_{\U{fin}}=1}^N\Pr_{\U{PA}}(N_{\U{fin}},\vec{e}_X= 0^{\otimes N_{\U{fin}}})
\ge\left(1-\frac{1}{4}\epsilon_{\U{secrecy}}^2\right)^2\ge1-\frac{1}{2}\epsilon_{\U{secrecy}}^2,
\end{align}
which ends the proof.

\begin{proposition}
\label{meidai:secrecy}
{\bf $\epsilon_{\U{secrecy}}\U{-secrecy}$}
\\
Equation~(\ref{eq:secrecy}) holds, that is, 
    \begin{align}
\frac{1}{2}\sum_{N_{\U{fin}}=1}^N\Pr_{\U{PA}}(N_{\U{fin}})
||\hat{\rho}^{AE}_{\U{PA}|N_{\U{fin}}}-\hat{\rho}^{AE}_{\U{ideal}|N_{\U{fin}}}||\le\epsilon_{\U{secrecy}}.
\end{align}
\end{proposition}
{\bf Proof of Proposition~\ref{meidai:secrecy}}
\\
First, we introduce the following lemma.
\begin{lemma}
\label{lemma:prop14}
For any state $\hat{\rho}_{AE}$ of systems $A$ and $E$, let	
\begin{equation}
 \hat{\rho}_{E} = \tr_{A}(\hat{\rho}_{AE})
\end{equation}
and
\begin{equation}
 \hat{\rho}_{A} = \tr_{E}(\hat{\rho}_{AE})
\end{equation}
be the marginal states of $\hat{\rho}_{AE}$. Then, for any state $\ket{\phi}_{A}$ of system $A$, 
\begin{equation}
 F(\hat{\rho}_{AE},\ket{\phi}\bra{\phi}_{A}\otimes\hat{\rho}_E) \geq \left[F(\hat{\rho}_{A},\ket{\phi}\bra{\phi}_{A})\right]^2
\end{equation}
holds. Here, the fidelity $F$ is defined in Eq.~(\ref{eq:fidelity}).
\end{lemma}
{\bf Proof of Lemma~\ref{lemma:prop14}}
\\
Let $W$ be a reference system, $\ket{\psi}_{AEW}$ be the purification of $\hat{\rho}_{AE}$, and $\{\ket{e_i}\}_i$ be an orthonormal basis of system $A$ that includes $\ket{e_0}=\ket{\phi}$. Using the following definitions: 
\begin{equation}
 p_i := \tr_{AEW}\(\ket{e_i}\bra{e_i}_{A}\ket{\psi}\bra{\psi}_{AEW}\)
\end{equation}
\begin{equation}
 \sqrt{p_i}\ket{\psi_i}_{EW} := \bra{e_i}_A\ket{\psi}_{AEW},
\end{equation}
$\ket{\psi}_{AEW}$ is written as
\begin{equation}
\begin{split}
 \ket{\psi}_{AEW} =&
 \sum_{i} \ket{e_i}\bra{e_i}_{A} \ket{\psi}_{AEW}\\
=&\sqrt{p_0}\ket{\phi}_A\ket{\psi_0}_{EW}+\sum_{i\neq 0}\sqrt{p_i}\ket{e_i}_A\ket{\psi_i}_{EW}.
\end{split}
\end{equation}
The definition of the fidelity gives
\begin{equation}
\begin{split}
 F(\hat{\rho}_{A},\ket{\phi}\bra{\phi}_{A}) 
 =& \max_{\ket{\chi}_{EW}}\left|\bra{\psi}_{AEW}\ket{\phi}_A\ket{\chi}_{EW}  \right|^2\\
 =&p_0\max_{\ket{\chi}_{ER}}\left|\bra{\psi_0}_{EW}\ket{\chi}_{EW}  \right|^2\\
 =& p_0.
\end{split}
\end{equation}
Next, system $S$ is introduced with the same orthonormal basis as system $A$, and the following state is defined.
\begin{equation}
 \ket{\xi}_{AEWS}:= \ket{\psi}_{AEW}\ket{\phi}_{S}
\end{equation}
This state is a a purification of $\hat{\rho}_{AE}$, which can be seen from
\begin{equation}
 \tr_{WS}\(\ket{\xi}\bra{\xi}_{AEWS}\) = \tr_W\(\ket{\psi}\bra{\psi}_{AEW}\) = \hat{\rho}_{AE}.
\end{equation}
Furthermore, the following state
\begin{equation}
 \ket{\omega}_{EWS}:= \sum_{i}\sqrt{p_i}\ket{\psi_i}_{EW}\ket{e_i}_S
\end{equation}
is a purification of $\hat{\rho}_E$ because 
\begin{equation}
 \tr_{WS}\(\ket{\omega}\bra{\omega}_{EWS}\) = \sum_{i}p_i\tr_W\(\ket{\psi_i}\bra{\psi_i}_{EW}\)
= \tr_{AW}\(\ket{\psi}\bra{\psi}_{AEW}\) = \hat{\rho}_E.
\end{equation}
Using the definition of the fidelity leads to
\begin{equation}
 \begin{split}
	F(\hat{\rho}_{AE},\ket{\phi}\bra{\phi}_{A}\otimes\hat{\rho}_E)
=& \max_{\ket{\chi'}_{EWS}} \left| \bra{\xi}_{AEWS} \ket{\phi}_A\ket{\chi'}_{EWS}  \right|^2\\
\geq & \left| \bra{\xi}_{AEWS} \ket{\phi}_A\ket{\omega}_{EWS}  \right|^2\\
	=& p_0^2\\
	=& \left[F(\hat{\rho}_{A},\ket{\psi}\bra{\psi}_{A})\right]^2,
 \end{split}
\end{equation}
which ends the proof of Lemma~\ref{lemma:prop14}.
\\\\
Then, we have
\begin{equation}
\begin{split}
&\frac{1}{2}\sum_{N_{\U{fin}}=0}^N\Pr_{\U{PA}}(N_{\U{fin}})
||\hat{\rho}^{AE}_{\U{PA}|N_{\U{fin}}}-\hat{\rho}^{AE}_{\U{ideal}|N_{\U{fin}}}|| \\
=&\frac{1}{2}\sum_{N_{\U{fin}}=1}^N\Pr_{\U{PA}}(N_{\U{fin}})
||\hat{\rho}^{AE}_{\U{PA}|N_{\U{fin}}}-\hat{\rho}^{AE}_{\U{ideal}|N_{\U{fin}}}|| \\
=&\frac{1}{2}\sum_{N_{\U{fin}}=1}^N\Pr_{\U{PA}}(N_{\U{fin}})
\|{\mathcal{E}^{Z}_{A_{\U{sift}}}}(\hat{\rho}_{|N_{\U{fin}}}^{\U{PA,vir}})
-{\mathcal{E}^{Z}_{A_{\U{sift}}}}(\hat{\rho}^{\U{ideal,vir}}_{|N_{\U{fin}}}) \| \\
\leq & \frac{1}{2}\sum_{N_{\U{fin}}=1}^N\Pr_{\U{PA}}(N_{\U{fin}})
\|\hat{\rho}_{|N_{\U{fin}}}^{\U{PA,vir}}
-\hat{\rho}^{\U{ideal,vir}}_{|N_{\U{fin}}}\|\\
\leq & \sum_{N_{\U{fin}}=1}^N\Pr_{\U{PA}}(N_{\U{fin}})
\sqrt{1-F(\hat{\rho}_{|N_{\U{fin}}}^{\U{PA,vir}},\hat{\rho}^{\U{ideal,vir}}_{|N_{\U{fin}}} )}\\
\leq & \sum_{N_{\U{fin}}=1}^N\Pr_{\U{PA}}(N_{\U{fin}})
\sqrt{1- \(F(\tr_E\hat{\rho}_{|N_{\U{fin}}}^{\U{PA,vir}}, \hat{P}[\ket{N_{\U{fin}},+^{N_{\U{fin}}}}])\)^2} \\
\leq & \sum_{N_{\U{fin}}=1}^N\Pr_{\U{PA}}(N_{\U{fin}})
\sqrt{2\(1- F(\tr_E \hat{\rho}_{|N_{\U{fin}}}^{\U{PA,vir}}, \hat{P}[\ket{N_{\U{fin}},+^{N_{\U{fin}}}}])\)} \\
\leq & 
\sqrt{2\(1- \sum_{N_{\U{fin}}=1}^N\Pr_{\U{PA}}(N_{\U{fin}}) F(\tr_E \hat{\rho}_{|N_{\U{fin}}}^{\U{PA,vir}}, \hat{P}[\ket{N_{\U{fin}},+^{N_{\U{fin}}}}])\)} \\
\leq & \epsilon_{\U{secrecy}}.
\end{split}
\end{equation}
The first equality comes from Eqs.~(\ref{eq-fid-trace-E-vir-PA-actual}) and (\ref{eq-fid-trace-E-vir-PA-ideal}). The second equality follows from the fact that the trace distance is zero when $N_{\U{fin}}=0$. The first inequality follows from the CPTP monotonicity of the trace distance. The second inequality comes from the relation between the trace distance and the fidelity. The third inequality follows by Lemma~\ref{lemma:prop14} with $\ket{\phi} = \ket{N_{\U{fin}},+^{N_{\rm fin}}}_{A_{\U{sift}}} $. The fourth inequality comes from the inequality $1-x^2 = 2(1-x)-(1-x)^2 \leq 2(1-x)$ for any real number $x$. The fifth inequality follows from the concavity of $\sqrt{2(1-x)}$ for $x$. The sixth inequality follows from Proposition~\ref{meidai:errorXPAatoPQC}.

\section{Conclusion and Discussion}
In this paper, we explicitly describe the decoy-state BB84 QKD protocol using a flowchart format and represent the 
actual operations performed by Alice and Bob with completely positive trace-preserving (CPTP) maps. 
Based on this concrete mathematical description of the protocol, we provide an explicit and self-contained security proof for this protocol based on the phase error correction approach. 

As for future work, rather than giving individual security proofs for each QKD protocol, we believe it will become increasingly important to modularize security proof techniques so that common components can be extracted and reused across different protocols. 
Furthermore, since our goal in this work was to give an explicit security proof of the de facto standard QKD protocol, we assumed a somewhat idealized device model involving a coherent light source with no modulation errors and threshold detectors with a fixed detection efficiency and a dark count probability. However, for the purpose of certifying real-world QKD systems, it will be important to develop security proofs based on more relaxed and realistic device assumptions, such as incorporating imperfections in sources (see, for example, \cite{GLLP2004,margarida2023,gile2023}) 
and detectors~(see, for example, \cite{yanbao2021,Phasenorbert2024,Ale2025}).

\section*{Acknowledgements}
This work is based on the outcomes of an activity aimed at developing a security proof document for quantum key distribution.
During this process, we received numerous valuable comments and suggestions from the QKD Technical Review Committee of the Quantum Forum. In particular, we would like to express our sincere gratitude to Prof.~Tamaki, Dr.~Tsurumaru, Prof.~Matsumoto, Prof. Takeoka, Dr.~Honjo, Mr.~Hideshima, and Mr.~Saito for their significant contributions during the review process. We also extend our heartfelt thanks to the QKD Technical Review Committee as a whole for dedicating their time and expertise, which greatly contributed to enhancing the quality of this work.
Furthermore, we are especially grateful to Dr.~Sasaki, Chair of the Quantum Key Distribution Technology Promotion Committee of the Quantum Forum, for his leadership and commitment in steering the effort toward the development of this document. His guidance was instrumental in coordinating this initiative and driving it forward. 
This work was partly supported by the following national projects: 
``Research and Development for Construction of a Global Quantum Cryptography Network  (JPJ008957)" 
in ``R\&D of ICT Priority Technology (JPMI00316)" of Ministry of Internal Affairs and Communication (MIC), Japan. 
We also acknowledge support from JST CREST (Grant Number JPMJCR2113, Japan) and JSPS KAKENHI 
(Grant Number JP23K25793), which provided important additional resources that contributed to this work.


\begin{thebibliography}{99}
\bibitem{bb84}
C. H. Bennett and G. Brassard. Quantum cryptography: Public key distribution and coin tossing. 
Theoretical Computer Science {\bf 560}, 7-11 (2014).
\bibitem{kiyoNphoto}
H.-K. Lo, M. Curty, and K. Tamaki. Secure quantum key distribution. Nature Photonics {\bf 8}, 595 (2014).
\bibitem{feihureview}
F. Xu, X. Ma, Q. Zhang, H.-K. Lo, and J.-W. Pan. Secure quantum key distribution with realistic devices. 
Reviews of Modern Physics {\bf 92}, 025002 (2020).
\bibitem{Pirandola2020}
S. Pirandola, U. L. Andersen, L. Banchi, M. Berta, D. Bunandar, R. Colbeck, D. Englund, T. Gehring, C. Lupo, C, 
C. Ottaviani, et al. Advances in quantum cryptography. Adv. Opt. Photonics {\bf 12}, 1012 (2020).
\bibitem{imple1}
M. Peev, et al, The SECOQC quantum key distribution network in Vienna. New J. Phys. {\bf 11}, 075001 (2009).
\bibitem{imple2}
M. Sasaki, et al, Field test of quantum key distribution in the Tokyo QKD Network. Opt. Express, {\bf 19}, 10387 (2011).  
\bibitem{imple3}
J. F. Dynes, et al, Cambridge quantum network. npj Quantum Inf. {\bf 5}, 101 (2019).
\bibitem{imple4}
S. K. Joshi, et al, A trusted node-free eight-user metropolitan quantum communication network. Sci. Adv. {\bf 6}, 
eaba0959 (2020).
\bibitem{imple5}
M. Avesani, et al, Resource-effective quantum key distribution: A field trial in Padua city center. 
Opt. Lett. {\bf 46}, 2848-2851 (2021).
\bibitem{imple6}
E. Bersin, et al,  Development of a Boston-area 50-Km Fiber Quantum Network Testbed. 
Phys. Rev. App. {\bf 21}, 014024 (2024).
\bibitem{decoy1}
W. Y. Hwang, Quantum key distribution with high loss: toward global secure communication. 
Phys. Rev. Lett. {\bf 91}, 057901 (2003).
\bibitem{decoy2}
X.-B. Wang, Beating the Photon-Number-Splitting Attack in Practical Quantum Cryptography. 
Phys. Rev. Lett. {\bf 94}, 230503 (2005).
\bibitem{decoy2005HoiKwong}
H.-K. Lo, X. Ma, and K. Chen. Decoy State Quantum Key Distribution. Phys. Rev. Lett. {\bf 94}, 230504 (2005).
\bibitem{concise2014}
C. C. W. Lim, M. Curty, N. Walenta, F. Xu, and H. Zbinden. 
Concise security bounds for practical decoy-state quantum key distribution. Physical Review A {\bf 89}, 022307 (2014).
\bibitem{nakayamahayashi2014}
M. Hayashi and R. Nakayama. Security analysis of the decoy method with the Bennett-Brassard 1984 
protocol for finite key lengths. New Journal of Physics {\bf 16}, 063009 (2014).
\bibitem{rusca2018}
D. Rusca, A. Boaron, F. Gr\"{u}nenfelder, A. Martin, and H. Zbinden. Finite-key analysis on the 1-decoy state QKD protocol. 
arXiv:1801.03443 (2018).
\bibitem{Tupkary2024}
D. Tupkary, S. Nahar, P. Sinha, and N. L\"{u}tkenhaus. 
Phase error rate estimation with basis-efficiency mismatch for decoy-state BB84, arXiv:2408.17349 [quant-ph] (2024).
\bibitem{Wiesemann2024}
J. Wiesemann, J. Krause, D. Tupkary, N. L\"utkenhaus, D. Rusca, N. Walenta. 
A consolidated and accessible security proof for finite-size decoy-state quantum key distribution. 
arXiv:2405.16578 (2024).
\bibitem{kamin2024}
L. Kamin, A. Arqand, I. George, N. L\"utkenhaus, and E. Y.-Z. Tan. Finite-size analysis of prepare-and-measure and 
decoy-state QKD via entropy accumulation. arXiv:2406.10198 (2024).
\bibitem{Tupkary2025}
D. Tupkary, E. Y.-Z. Tan, S. Nahar, L. Kamin, N. L\"utkenhaus. 
QKD security proofs for decoy-state BB84: protocol variations, proof techniques, gaps and limitations. 
arXiv:2502.10340 (2025).
\bibitem{ben2005}
M. Ben-Or, M. Horodecki, D. W. Leung, D. Mayers, and J. Oppenheim. The universal 
composable security of quantum key distribution. Theory of Cryptography 3378, 386-406 (2005). 
\bibitem{renner2009}
J. Muller-Quade and R. Renner. Composability in quantum cryptography. New J. Phys. {\bf 11}, 085006 (2009).
\bibitem{renner2022}
C. Portmann and R. Renner. Security in quantum cryptography. Reviews of Modern Physics {\bf 94}, 025008 (2022).
\bibitem{hoiknowng2003}
H. K. Lo. Method for decoupling error correction from privacy amplification, New J. Phys. {\bf 5}, 36 (2003).
\bibitem{koashi2009njp}
M. Koashi. Simple security proof of quantum key distribution based on complementarity, 
New Journal of Physics {\bf 11}, 045018 (2009).
\bibitem{tomamichel2017}
M. Tomamichel and A. Leverrier, 
A largely self-contained and complete security proof for quantum key distribution, Quantum {\bf 1}, 14 (2017).
\bibitem{EUP1}
M. Tomamichel and R. Renner. Uncertainty Relation for Smooth Entropies. 
Physical Review Letters, {\bf 106}, 110506 (2011).
\bibitem{koashi2005}
M. Koashi. Simple security proof of quantum key distribution via uncertainty principle. arXiv:quant-ph/0505108 (2005).
\bibitem{tsurumaru2010}
T. Tsurumaru and M. Hayashi, Dual universality of hash functions and its applications to quantum cryptography. 
IEEE Transactions on Information Theory {\bf 59}, 7 (2013).
\bibitem{wegmancarter}
M. N. Wegman and J. L. Carter, New hash functions and their use in authentication and set equality. 
Journal of Computer and System Sciences {\bf 22}, 265-279 (1981).
\bibitem{Katoinequality}
G. Kato. Concentration inequality using unconfirmed knowledge. arXiv:2002.04357 (2020).
\bibitem{tightnpj}
G. Curr\'{a}s-Lorenzo, \'{A}. Navarrete, K. Azuma, et al. Tight finite-key security for twin-field quantum key distribution. npj Quantum Inf {\bf 7}, 22 (2021). 
\bibitem{Beaudry2008}
N. J. Beaudry, T. Moroder, and N. L\"{u}tkenhaus. Squashing Models for Optical Measurements in Quantum Communication, 
Physical Review Letters {\bf 101}, 093601 (2008).
\bibitem{tsurumaru2008}
T. Tsurumaru and K. Tamaki. Security proof for quantum-key-distribution systems with threshold detectors. 
Physical Review A {\bf 78}, 032302 (2008).
\bibitem{Tsurumaru2010}
T. Tsurumaru. Squash operator and symmetry. Phys. Rev. {\bf A 81}, 012328 (2010).
\bibitem{fung2011}
C.-H. F. Fung, H. F. Chau, and H.-K. Lo. Universal squash model for optical communications using linear optics and 
threshold detectors. Physical Review A {\bf 84}, 020303 (2011).
\bibitem{Gittsovich2014}
O. Gittsovich, N. J. Beaudry, V. Narasimhachar, R. R. Alvarez, T. Moroder, and N. L\"{u}tkenhaus. Squashing model for 
detectors and applications to quantum-key-distribution protocols. Physical Review A {\bf 89}, 012325 (2014).
\bibitem{Zhang2021}
Y. Zhang, P. J. Coles, A. Winick, J. Lin, and N. L\"{u}tkenhaus. Security proof of practical quantum key distribution with 
detection-efficiency mismatch. Physical Review Research {\bf 3}, 013076 (2021).
\bibitem{Upadhyaya2021}
T. Upadhyaya, T. Van Himbeeck, J. Lin, and N. L\"{u}tkenhaus. Dimension Reduction in Quantum Key Distribution for 
Continuous- and Discrete-Variable Protocols. PRX Quantum {\bf 2}, 020325 (2021).
\bibitem{Nahar2025}
S. Nahar, and N. L\"{u}tkenhaus. Imperfect detectors for adversarial tasks with applications to quantum key distribution. 
arXiv:2503.06328 (2025).
\bibitem{practicaldecoy2005}
X. Ma, B. Qi, Y. Zhao, and H.-K. Lo. Practical decoy state for quantum key distribution. 
Phys. Rev. A {\bf 72}, 012326 (2005).
\bibitem{GLLP2004}
D. Gottesman, H.-K. Lo, N. L\"{u}tkenhaus, and J. Preskill. 
Security of quantum key distribution with imperfect devices. Quantum Information and Computation, {\bf 5}, 325-360 (2004).
\bibitem{margarida2023}
M. Pereira, et al. Modified BB84 quantum key distribution protocol robust to source imperfections. Physical Review Research, {\bf 5}, 023065 (2023).
\bibitem{gile2023}
G. Curr\'{a}s-Lorenzo, M. Pereira, G. Kato, M. Curty, K. Tamaki. 
A security framework for quantum key distribution implementations, arXiv:2305.05930 (2023).
\bibitem{yanbao2021}
Y. Zhang, P. J. Coles, A. Winick, J. Lin, and N. L\"{u}tkenhausl. 
Security proof of practical quantum key distribution with detection-efficiency mismatch. 
Phys. Rev. Research {\bf 3}, 013076 (2021).
\bibitem{Phasenorbert2024}
D. Tupkary, S. Nahar, P. Sinha, N. L\"{u}tkenhaus. 
Phase error rate estimation in QKD with imperfect detectors. arXiv:2408.17349 (2024).
\bibitem{Ale2025}
A. Marcomini, A. Mizutani, F. Gr\"{u}nenfelder, M. Curty, K. Tamaki. 
Loss-tolerant quantum key distribution with detection efficiency mismatch. 
Quantum Science and Technology {\bf 10}, 035002 (2025).	
\end{thebibliography}
\end{document}